%% file: main.tex
\documentclass[final,3p,times]{elsarticle}
\biboptions{compress}

\usepackage{amssymb}
\usepackage{amsmath}
\usepackage{amsthm}

\usepackage{amsfonts}
\usepackage{bm}
\usepackage{mathtools}
\usepackage{booktabs}
\usepackage{caption}
\usepackage{graphicx}
\usepackage{float}
\usepackage{url}
\usepackage[hidelinks]{hyperref}
\usepackage{array}
\usepackage{multirow}
\usepackage{enumitem}

\newcommand{\eq}[1]{{Eq.(\ref{#1})}}
\newcommand{\eqq}[2]{{Eqs.(\ref{#1},~\ref{#2})}}

\newcommand{\fig}[1]{{Fig.~\ref{#1}}}
\newcommand{\tab}[1]{{Tab.~\ref{#1}}}

\newsavebox{\sumbox}
\sbox{\sumbox}{$\displaystyle\sum$}

\journal{Computational Particle Mechanics}

\begin{document}

\begin{frontmatter}



    \title{A generalized ghost cell boundary model for incompressible SPH}


    \author[inst1]{Hiroyuki Omura}

    \affiliation[inst1]{organization={National Research Institute for Earth Science and Disaster Resilience},
        addressline={Tennodai 3-1},
        city={Tsukuba-city},
        postcode={3050006},
        state={Ibaraki},
        country={Japan}}

    \author[inst2]{Takayuki Tsunemi}

    \affiliation[inst2]{organization={University of Tsukuba},
        addressline={Tennodai 1-1-1},
        city={Tsukuba-city},
        postcode={3058573},
        state={Ibaraki},
        country={Japan}}

    \author[inst3]{Shunhua Chen}

    \affiliation[inst3]{organization={Sun Yat-sen University},
        addressline={Tangjiawan, Xiangzhou District},
        city={Zhuhai},
        postcode={519082},
        state={Guangdong},
        country={China}}

    \author[inst2]{Naoto Mitsume}

    \begin{abstract}
        This study improves and generalizes the ghost cell boundary (GCB) model for free-surface flow analysis based on particle methods.
        In the GCB framework, the wall-side contribution to particle approximations is evaluated by cell-based volume integration using Gaussian quadrature.
        The proposed formulation extends the original GCB model to an arbitrary number of quadrature points and incorporates the wall contribution into a semi-implicit particle method, such as the incompressible smoothed particle hydrodynamics (ISPH) method.
        A fixed-ghost-particle-based extrapolation is further introduced to assign velocity and pressure values at the quadrature points so that the wall boundary conditions are imposed consistently.
        In addition, the formulation is extended to fluid--rigid-body coupling by interpreting the GCB wall contribution as the force exerted on fluid particles and by applying its reaction to cells.
        The accuracy of the generalized GCB model for fixed- and moving-wall boundary problems was validated through hydrostatic pressure, dam-break, and sloshing simulations.
        Moreover, the accuracy of fluid--rigid-body coupling simulations based on the generalized GCB model was validated through falling-cylinder and wedge water-entry problems.
        These results demonstrate that the generalized GCB model can treat fixed- and moving-wall boundaries and coupling with rigid bodies in ISPH while retaining a geometrically explicit cell-based boundary representation.
    \end{abstract}



    \begin{keyword}
        Free-surface flow \sep
        Smoothed particle hydrodynamics \sep
        Ghost cell boundary model \sep
        Fixed ghost particle \sep
        Gaussian quadrature \sep
        Fluid--rigid-body coupling
    \end{keyword}

\end{frontmatter}


\section{Introduction}
\label{sec:intro}
\input{doc/1_.tex}

\section{ISPH method}
\label{sec:isph}
\input{doc/2_.tex}

\section{Generalized GCB model}
\label{sec:ggcb}
\input{doc/3_.tex}

\section{Fluid-rigid-body coupling formulation}
\label{sec:fsi}
\input{doc/4_.tex}

\section{Verification and validation}
\label{sec:vv}
\input{doc/5_.tex}

\section{Conclusions}
\label{sec:conc}
\input{doc/6_.tex}


\section*{Acknowledgment}
This work was supported by JST FOREST Program Grant Number JPMJFR215S, and JSPS KAKENHI Grant Numbers 23K26356 and 26K02921.

\appendix
\section{Computational results using the difference-based pressure gradient model}
\input{doc/app_.tex}

\bibliographystyle{elsarticle-num}
\bibliography{cas-refs}

\end{document}

%% file: doc/1_.tex
Particle methods, such as smoothed particle hydrodynamics (SPH)~\cite{Gingold1977,Lucy1977} and the moving particle semi-implicit (MPS) method~\cite{Koshizuka1996}, are meshfree approaches in which the domain is discretized by computational points, referred to as particles, rather than by a mesh.
Because these methods are generally formulated in a Lagrangian framework, they can represent free surfaces and large interface deformations in a natural manner.
Owing to these capabilities, particle methods have been extensively used to simulate incompressible free-surface flows, including wave breaking, sloshing, and fluid--structure interaction (FSI)~\cite{Monaghan1994, Liu2010, Violeau2016, Cummins1999, Shao2003, Asai2012, Bouscasse2013, Sun2015, Khayyer2018, Negishi2024, Zhang2024}.
Despite these advances, the treatment of wall boundary conditions continues to be a central challenge in SPH and MPS~\cite{Vacondio2021}.
This issue is particularly important in fixed- and moving-boundary and FSI problems because numerical errors near walls can directly affect the overall accuracy of the analysis.

The treatment of wall boundary conditions in particle methods is challenging for two main reasons.
First, because fluid particles move in a Lagrangian manner, computational points do not necessarily lie on the wall boundary.
Therefore, Dirichlet velocity and Neumann pressure conditions cannot generally be imposed directly on the wall surface.
Second, SPH differential operators are evaluated using neighboring particles within the kernel support.
Near a wall, however, this support is truncated, giving rise to kernel support deficiency.
A wall boundary model must therefore both compensate for the resulting kernel support deficiency and define wall-side pressure and velocity values consistent with the prescribed boundary conditions.

Early studies of wall boundary treatment in particle methods focused on repulsive-force models designed to suppress particle penetration into wall boundaries~\cite{Monaghan1994, Monaghan2009}.
Later work shifted toward virtual-particle approaches, in which particles placed outside the fluid domain compensate for kernel support deficiency.
This class includes wall-particle models~\cite{Koshizuka1998, Morris1997, Crespo2007, Adami2012, Rezavand2022}, ghost-particle models~\cite{Randles1996, Colagrossi2003, Bierbrauer2009, Fourtakas2019, Vela2019}, and fixed-ghost-particle (FGP) models~\cite{Fang2006, marrone2011, English2022, Antuono2023}.
Wall-particle models are straightforward to implement and generally robust.
However, the mapping of field variables outside the fluid domain often involves approximations, so the wall boundary conditions may not be satisfied exactly.
In contrast, ghost-particle models generate virtual particles dynamically by reflecting fluid particles across the wall and map field variables to them in accordance with the boundary conditions.
These models allow wall boundary conditions to be imposed more explicitly than in wall-particle models.
Extending these models to non-planar walls, however, is difficult.
Although an extension to smoothly curved boundaries has been reported~\cite{Vela2019}, applicability to more general geometries containing corners remains challenging.
FGP models avoid the dynamic generation of virtual particles in advance and retaining them at fixed locations relative to the boundary.
Field variables are first interpolated at corresponding mirror points on the fluid side of the boundary and then mapped to the fixed ghost particles.
This approach retains the robustness of wall-particle models with respect to wall geometry while improving numerical accuracy.
{
Furthermore, recent advancements in wall boundary models have sought to improve accuracy and computational efficiency through novel concepts, such as the motif-based wall boundary model~\cite{Khan2025} and the wall boundary potential model~\cite{Gotoh2026a}.
}

The approach of extending field variables across a boundary to enforce boundary conditions is not unique to particle methods.
In mesh-based methods such as finite difference methods (FDMs) and finite volume methods (FVMs), this approach has developed into a broad family of ghost-cell techniques.
When a solid boundary, interface, or free surface does not conform to grid lines or cell faces, ghost cells are introduced across the boundary.
Their values are reconstructed by interpolation or extrapolation to satisfy the prescribed boundary conditions, allowing the finite-difference stencils or finite-volume flux formulations in the fluid region to retain their original form.
The underlying concept can be traced back to the assignment of near-wall values in early incompressible-flow solvers, notably the marker-and-cell method~\cite{Harlow1965}.
Subsequent developments include the ghost fluid method~\cite{Fedkiw1999}, embedded-boundary and cut-cell methods~\cite{Johansen1998, Gibou2002}, and sharp-interface immersed-boundary methods~\cite{Ye1999, Udaykumar2001, Mittal2005}.
Ghost-cell immersed-boundary methods~\cite{Tseng2003, Ghias2007, Berthelsen2008} further reconstruct values in ghost cells located inside the solid domain using data from the boundary and the neighboring fluid.
This allows Dirichlet and Neumann conditions to be handled within a unified framework while preserving, as far as possible, the accuracy of the underlying finite-difference or finite-volume scheme.
In mesh-based methods, ghost cells thus provide physically and numerically consistent values for virtual degrees of freedom beyond the boundary, thereby completing otherwise truncated discrete stencils.
They are therefore not merely a device for preventing fluid penetration into solid boundaries, but a general numerical framework for embedding boundary conditions within discrete operators.
{
Recent developments in free-surface modeling include the introduction of space potential particles to enhance stability~\cite{Tsuruta2015} and unified semi-analytical wall boundary conditions~\cite{Ruan2024}.
Although the primary focus of the present study is on solid boundaries, treating arbitrary interfaces accurately is a shared challenge.
}

Against this background, Zheng et al. proposed the ghost cell boundary (GCB) model~\cite{zheng2020-1, zheng2020-2} for particle methods.
In the GCB model, the wall domain, including the interior of a solid structure, is discretized into ghost cells.
The wall contribution to particle approximations is evaluated through cell-based volume integration using Gaussian quadrature.
The quadrature points within the ghost cells serve as computational points in the wall domain similar to wall particles. 
The fluid field is extended to these points in accordance with the prescribed boundary conditions, thereby compensating for the missing kernel support near the wall.
The GCB model can therefore be viewed as extending the concepts underlying the wall-particle model from discrete particles to cells.
Because the framework directly employs cells and Gaussian quadrature commonly used in the finite element method (FEM), it readily accommodates complex geometries and fluid--structure interactions (e.g.,~\cite{Xie2026}).
However, the original GCB model has two principal limitations.
First, the previous studies focused primarily on explicit MPS (EMPS) formulations~\cite{Shakibaeinia2010}.
Because the EMPS method suffers from severe numerical pressure oscillations, strong artificial damping must be introduced to solve moving-boundary and FSI problems.
Therefore, the GCB model needs to be generalized so that it can be applied to semi-implicit particle methods, such as the incompressible SPH (ISPH) method~\cite{Cummins1999, Shao2003}.
Second, pressure and velocity values at the quadrature points are obtained using a simplified mapping, and the wall boundary conditions are therefore not enforced precisely.
Beyond these formulation issues, the effects of the number and spacing of quadrature points on numerical stability and accuracy require systematic assessment when the ghost-cell resolution differs from the fluid-particle resolution.

To address these limitations, this paper proposes a generalized GCB model with an improved field-mapping procedure that can also be incorporated into ISPH.
First, the formulation of the GCB wall contribution is generalized to accommodate Gaussian quadrature with an arbitrary number of quadrature points.
The formulation then specifies how the GCB wall contribution is incorporated into the pressure Poisson equation and the evaluation of the pressure gradient, velocity divergence, and Laplacian.
In addition, the FGP concept~\cite{marrone2011} is extended to the quadrature points of the GCB model.
Pressure and velocity values at these points are mapped from neighboring fluid particles while accounting for the prescribed wall boundary conditions, providing a more accurate mapping of the wall domain.
Hydrostatic pressure and dam-break benchmarks are simulated to compare the generalized and original GCB models, examine the effect of the number of quadrature points, and evaluate applicability to unstructured ghost-cell arrangements.
A fluid--rigid-body coupling formulation is also developed in which the interaction between the fluid and a cell-discretized rigid body is evaluated from the generalized GCB model.
The accuracy of this coupling formulation is assessed using falling-cylinder and wedge water-entry simulations.

The remainder of this paper is organized as follows.
Section~\ref{sec:isph} presents the governing equations for incompressible free-surface flows, the associated boundary conditions, and their discretization using the ISPH method.
Section~\ref{sec:ggcb} formulates the generalized GCB model, introduces Gaussian quadrature points, describes its incorporation into ISPH, and presents the FGP-based treatment of wall boundary conditions.
Section~\ref{sec:fsi} develops a fluid--rigid-body coupling formulation based on the generalized GCB model. 
Section~\ref{sec:vv} assesses the accuracy of the proposed method through hydrostatic pressure, dam-break, sloshing, falling-cylinder, and wedge water-entry problems.
Finally, Section~\ref{sec:conc} summarizes the conclusions of this study.

%% file: doc/2_.tex
\subsection{Governing equations}
\input{doc/2_1_.tex}

\subsection{Spatial discretization}
\input{doc/2_2_.tex}

\subsection{Time integration scheme}
\input{doc/2_3_.tex}

\subsection{Free-surface detection} \label{sec:fsd}
\input{doc/2_4_.tex}

\subsection{Particle relocation} \label{sec:ps}
\input{doc/2_5_.tex}

%% file: doc/2_1_.tex
The incompressible free-surface flows considered in this study are governed by the Navier--Stokes momentum equation and the continuity equation:
\begin{equation} \label{eq:ge_1}
	\frac{D\bm{v}}{Dt}
	= -\frac{1}{\rho}\nabla p
	+ \nu\nabla^2\bm{v}
	+ \bm{g},
\end{equation}
\begin{equation} \label{eq:ge_2}
	\nabla\cdot\bm{v}=0.
\end{equation}
Here, $D/Dt$ denotes the material derivative, and $\bm{v}$, $\rho$, $p$, $\nu$, and $\bm{g}$ denote the fluid velocity, density, pressure, kinematic viscosity, and gravitational acceleration, respectively.

Let $\Omega$ denote the fluid domain.
The free surface and wall portions of its boundary $\partial\Omega$ are denoted by $\partial\Omega_{\mathrm{fs}}$ and $\partial\Omega_{\mathrm{wall}}$, respectively.
The following boundary conditions are imposed:
\begin{equation} \label{eq:bc_fs}
	\left\{
	-p\bm{I}
	+\rho\nu
	\left[
	\nabla\bm{v}
	+\left(\nabla\bm{v}\right)^T
	\right]
	\right\}\bm{n}^\mathrm{fs}
	=\bm{0}
	\qquad
	\text{on } \partial\Omega_{\mathrm{fs}},
\end{equation}
\begin{equation} \label{eq:bc_wall}
	\bm{v}=\bm{v}^{\mathrm{wall}}
	\qquad
	\text{on } \partial\Omega_{\mathrm{wall}},
\end{equation}
where $\bm{n}^\mathrm{fs}$ is the unit normal vector on the free surface, and $\bm{I}$ is the second-order identity tensor.
The superscript $T$ denotes the transpose, and $\bm{v}^{\mathrm{wall}}$ is the velocity of the wall.

%% file: doc/2_2_.tex
In SPH, the value of a scalar field $\phi$ at position $\bm{x}$ is approximated by the following kernel-weighted integral:
\begin{equation} \label{eq:sph_ave}
	\phi(\bm{x})
	\approx
	\int_{\Omega}
	\phi(\bm{x}')
	W\left(\left|\bm{x}'-\bm{x}\right|,h\right)
	d\bm{x}',
\end{equation}
where $W$ is the smoothing kernel function and $h$ is the smoothing length.
In this study, the following quintic spline function is employed as the kernel:
\begin{equation} \label{eq:kernel}
	W(r,h)
	=
	\frac{7}{478\pi h^2}
	\begin{cases}
		\left(3-\dfrac{r}{h}\right)^5
		-6\left(2-\dfrac{r}{h}\right)^5
		+15\left(1-\dfrac{r}{h}\right)^5
		& (0\leq r<h), \\[2mm]
		\left(3-\dfrac{r}{h}\right)^5
		-6\left(2-\dfrac{r}{h}\right)^5
		& (h\leq r<2h), \\[2mm]
		\left(3-\dfrac{r}{h}\right)^5
		& (2h\leq r<3h), \\[2mm]
		0
		& (3h\leq r),
	\end{cases}
\end{equation}
where $r$ denotes the distance between the two points.

Upon discretizing the computational domain into a set of particles, the kernel integral in \eq{eq:sph_ave} is approximated by the following weighted summation:
\begin{equation}
	\phi(\bm{x}_i)
	\approx
	\left\langle\phi\right\rangle_i
	=
	\sum_{j=1}^{N_{\mathrm{p}}}
	\frac{m_j}{\rho_j}
	W\left(\left|\bm{x}_{ij}\right|,h\right)
	\phi_j.
\end{equation}
Throughout this paper, angle brackets $\langle\cdot\rangle$ denote an SPH approximation.
A subscript attached to a field variable denotes the particle index and indicates that the variable is evaluated at the corresponding particle; for example, $\phi_i=\phi(\bm{x}_i)$.
The vector $\bm{x}_{ij}=\bm{x}_j-\bm{x}_i$ is the relative position vector from particle $i$ to particle $j$.
The corresponding kernel value $W\left(\left|\bm{x}_{ij}\right|,h\right)$ is hereafter abbreviated as $W_{ij}$.
Additionally, $N_{\mathrm{p}}$ is the total number of particles, and $m_i$ denotes the mass of particle $i$.

The spatial differential operators appearing in the governing equations are discretized using the same SPH formulations as those adopted in the stabilized ISPH scheme proposed by Asai et al.~\cite{Asai2012}:
\begin{equation} \label{eq:sph_gp}
	\left\langle\nabla p\right\rangle_i
	=
	\rho_i
	\sum_{j=1}^{N_{\mathrm{p}}}
    \frac{m_j}{\rho_j}
	\left(
	\frac{p_i}{\rho_i}
	+
	\frac{p_j}{\rho_j}
	\right)
	\nabla W_{ij},
\end{equation}
\begin{equation} \label{eq:sph_lp}
	\left\langle\nabla^2p\right\rangle_i
	=
	2
	\sum_{j=1}^{N_{\mathrm{p}}}
    \frac{m_j}{\rho_j}
	\frac{p_j-p_i}{\left|\bm{x}_{ij}\right|}
	\frac{\bm{x}_{ij}}{\left|\bm{x}_{ij}\right|}
	\cdot\nabla W_{ij},
\end{equation}
\begin{equation} \label{eq:sph_lv}
	\left\langle\nabla^2\bm{v}\right\rangle_i
	=
	2
	\sum_{j=1}^{N_{\mathrm{p}}}
	\frac{m_j}{\rho_j}
	\frac{\bm{v}_j-\bm{v}_i}{\left|\bm{x}_{ij}\right|}
	\frac{\bm{x}_{ij}}{\left|\bm{x}_{ij}\right|}
	\cdot\nabla W_{ij},
\end{equation}
\begin{equation} \label{eq:sph_dv}
	\left\langle\nabla\cdot\bm{v}\right\rangle_i
	=
	\sum_{j=1}^{N_{\mathrm{p}}}
	\frac{m_j}{\rho_j}
	\left(\bm{v}_j-\bm{v}_i\right)
	\cdot\nabla W_{ij},
\end{equation}
where $\nabla W_{ij}$ denotes the gradient of the kernel function with respect to $\bm{x}_i$.

%% file: doc/2_3_.tex
The ISPH method advances the solution using a fractional-step procedure.
The momentum equation is split into a predictor step for an intermediate velocity field $\bm{v}^*$ and a corrector step.
Given the solution at the $n$-th time step, these two steps are expressed as
\begin{equation} \label{eq:fs1}
	\frac{\bm{v}_i^*-\bm{v}_i^n}{\Delta t}
	=
	\nu\left\langle\nabla^2\bm{v}\right\rangle_i^n
	+\bm{g},
\end{equation}
\begin{equation} \label{eq:fs2}
	\frac{\bm{v}_i^{n+1}-\bm{v}_i^*}{\Delta t}
	=
	-\frac{1}{\rho_i}
	\left\langle\nabla p\right\rangle_i^{n+1},
\end{equation}
where the asterisk denotes an intermediate value, and $\Delta t$ is the time step size.
The pressure field at the $(n+1)$-th time step, required in \eq{eq:fs2}, is obtained by solving the following pressure Poisson equation:
\begin{equation} \label{eq:ppe}
	\left\langle\nabla^2p\right\rangle_i^{n+1}
	=
	\frac{\rho_i}{\Delta t}
	\left\langle\nabla\cdot\bm{v}\right\rangle_i^*
	+
	\alpha
	\frac{\rho_i-\left\langle\rho\right\rangle_i^n}{\Delta t^2}.
\end{equation}
The second term on the right-hand side is a stabilization term that promotes a uniform particle distribution and helps preserve particle volume.
$\alpha$ controls the strength of this stabilization.
$\left\langle\rho\right\rangle_i^n$ is the numerical density computed from the particle configuration at the $n$-th time step:
\begin{equation} \label{eq:pos_rho}
	\left\langle\rho\right\rangle_i^n
	=
	\sum_{j=1}^{N_{\mathrm{p}}}
	m_jW_{ij}.
\end{equation}
{
Note that in this study, if the pressure obtained by solving \eq{eq:ppe} is negative, it is set to zero.
}

Finally, based on the symplectic Euler method, the particle positions at the $(n+1)$-th time step is given by:
\begin{equation}
	\bm{x}^{n+1} = \bm{x}^{n} + \bm{v}^{n+1}\Delta t.
\end{equation}

Neglecting the viscous contributions associated with velocity gradients, the boundary conditions in \eqq{eq:bc_fs}{eq:bc_wall} yield the following pressure conditions:
\begin{equation} \label{eq:pbc_fs}
	p=0
	\qquad
	\text{on } \partial\Omega_{\mathrm{fs}},
\end{equation}
\begin{equation} \label{eq:pbc_wall}
	\nabla p\cdot\bm{n}^{\mathrm{wall}}
	=
	\rho
	\left(
	\bm{g}
	-
	\frac{D\bm{v}^{\mathrm{wall}}}{Dt}
	\right)
	\cdot\bm{n}^{\mathrm{wall}}
	\qquad
	\text{on } \partial\Omega_{\mathrm{wall}},
\end{equation}
where $\bm{n}^{\mathrm{wall}}$ is the unit normal vector on the wall boundary pointing into the fluid domain.
\eq{eq:pbc_wall} represents the Neumann pressure condition associated with the pressure Poisson equation.
In the present implementation, however, this condition is not imposed directly in the linear system for \eq{eq:ppe}.
Instead,  following FGP models~\cite{marrone2011,Bouscasse2013,Morikawa2021}, it is incorporated into the evaluation of the pressure-gradient term in \eq{eq:fs2}.
This treatment provides the wall-normal pressure gradient required to suppress fluid-particle penetration into the wall.

%% file: doc/2_4_.tex
Application of the Dirichlet pressure condition in \eq{eq:pbc_fs} requires the identification of free-surface particles.
In this study, these particles are detected using the method proposed by Marrone et al.~\cite{marrone2010}.

For each particle $i$, a normal vector of the free surface, $\tilde{\bm{n}}^{\mathrm{fs}}_i$, and the associated moment matrix $\bm{M}_i$ are computed as
\begin{equation} \label{eq:fsd_n}
	\tilde{\bm{n}}^{\mathrm{fs}}_i
	=
	-\bm{M}_i^{-1}
	\sum_{j=1}^{N_{\mathrm{p}}}
	\nabla W_{ij}V_j,
\end{equation}
\begin{equation} \label{eq:fsd_m}
	\bm{M}_i
	=
	\sum_{j=1}^{N_{\mathrm{p}}}
	\nabla W_{ij}
	\otimes
	\bm{x}_{ij}
	V_j,
\end{equation}
where $\otimes$ denotes the dyadic product.
The smallest eigenvalue $\lambda_i$ of $\bm{M}_i$ approaches unity when particle $i$ is well inside the fluid and is surrounded by neighboring particles.
A particle is classified as an interior particle when $\lambda_i>0.8$ and as a free-surface particle when $\lambda_i<0.2$.

For particles satisfying $0.2\leq\lambda_i\leq0.8$, the classification is refined using an additional geometric detection in the free-surface normal direction.
Then, each neighboring particle $j$ is evaluated against the following two criteria:
\begin{equation}
	\left|\bm{x}_{ij}\right|
	\leq
	\sqrt{2}H
	\:\text{and}\:
	\cos^{-1}
	\left(
	\bm{n}^{\mathrm{fs}}_i
	\cdot
	\frac{\bm{x}_{ij}}
	{\left|\bm{x}_{ij}\right|}
	\right)
	\leq
	\frac{\pi}{4},
\end{equation}
\begin{equation}
	\left|\bm{x}_{ij}\right|
	>
	\sqrt{2}H
	\:\text{and}\:
	\left|
	\bm{x}_{ij}-H\bm{n}^{\mathrm{fs}}_i
	\right|
	\leq H.
\end{equation}
If at least one neighboring particle satisfies either criterion, particle $i$ is classified as an interior particle; otherwise, it is classified as a free-surface particle.
The unit free-surface normal vector is computed as
\begin{equation}
	\bm{n}^{\mathrm{fs}}_i
	=
	\frac{\tilde{\bm{n}}^{\mathrm{fs}}_i}
	{\left|\tilde{\bm{n}}^{\mathrm{fs}}_i\right|}.
\end{equation}
The parameter $H$ determines the size of the geometric search region.
In this study, $H=1.33 d_0$, where $d_0$ is the initial particle spacing.

%% file: doc/2_5_.tex
To mitigate the loss of accuracy in SPH differential operators caused by irregular particle distributions, particle-regularization corrections are introduced.
This study combines optimized particle shifting (OPS)~\cite{Khayyer2017}, which relocates particles according to the local gradient of particle concentration, with dynamic stabilization (DS)~\cite{tsuruta2013}, which applies a repulsive correction to particle pairs that become excessively close.
Near the free surface, OPS retains only the tangential component of the displacement, thereby suppressing unphysical particle motion in the free-surface normal direction.
DS is subsequently applied to separate particle pairs that remain closer than a prescribed threshold after OPS.

For particle $i$, the particle-concentration gradient estimated from the neighboring particle distribution is given by
\begin{equation}
	\left\langle\widehat{\nabla}C_i\right\rangle
	=
	\sum_{j=1}^{N_{\mathrm{p}}}
	\left\{
	1
	+
	R_{\mathrm{p}}
	\left(
	\frac{W_{ij}}{W(d_0, h)}
	\right)^n
	\right\}
	\nabla W_{ij}V_j,
\end{equation}
where $R_{\mathrm{p}}$ and $n$ are coefficients of the correction term.
The commonly adopted values $R_{\mathrm{p}}=0.2$ and $n=4$ are used in this study.
The displacement correction produced by OPS is
\begin{equation} \label{eq:ops_shift}
	\delta\bm{x}^{\mathrm{OPS}}_i
	=
	\begin{cases}
		\bm{0}
		& (\lambda_i<0.4), \\[1mm]
		-C_{\mathrm{shift}}d_0^2
		\left(
		\bm{I}
		-
		\bm{n}^{\mathrm{fs}}_i
		\otimes
		\bm{n}^{\mathrm{fs}}_i
		\right)
		\left\langle\widehat{\nabla}C_i\right\rangle
		& (\lambda_i \geq 0.4 \:\text{and}\: i\in\mathcal{N}_\mathrm{fs}), \\[1mm]
		-C_{\mathrm{shift}}d_0^2
		\left\langle\widehat{\nabla}C_i\right\rangle
		& (\text{otherwise}),
	\end{cases}
\end{equation}
where $C_{\mathrm{shift}}$ controls the magnitude of particle shifting and $\mathcal{N}_{\mathrm{fs}}$ denotes the set of particles located on or near the free surface.
{
Mathematically, it would be natural to unify the particle classification condition in \eq{eq:ops_shift} with the screening condition used in the free-surface detection.
However, prioritizing computational stability, the above condition derived through trial and error is employed here.
}
For particles on or near the free surface, the projection operator $\bm{I}-\bm{n}^{\mathrm{fs}}_i\otimes\bm{n}^{\mathrm{fs}}_i$ removes the normal component of the displacement.

Following the OPS correction, the resultant DS corrective force per unit volume acting on particle $i$ is evaluated as
\begin{equation} \label{eq:ds_total_force}
	\bm{F}^{\mathrm{DS}}_i
	=
	\sum_{j=1}^{N_{\mathrm{p}}}
	\bm{f}^{\mathrm{DS}}_{ij}
	W_{ij}V_j.
\end{equation}
The corresponding pairwise contribution is defined by
\begin{equation} \label{eq:ds_force}
	\bm{f}^{\mathrm{DS}}_{ij}
	=
	\begin{cases}
		\bm{0}
		& \left(
		\left|\bm{x}^{n+1}_{ij}\right|\geq d
		\right), \\[1mm]
		-\rho_i\Pi_{ij}
		\dfrac{\bm{x}^{n}_{ij}}
		{\left|\bm{x}^{n}_{ij}\right|}
		& \left(
		\left|\bm{x}^{n+1}_{ij}\right|<d
		\right),
	\end{cases}
\end{equation}
where
\begin{equation}
	\Pi_{ij}
	=
	\frac{\rho_j}
	{\Delta t^2(\rho_i+\rho_j)}
	\left(
	\sqrt{
	d^2
	-
	\left|\bm{x}^{n+1}_{ij\perp}\right|^2
	}
	-
	\left|\bm{x}^{n+1}_{ij\parallel}\right|
	\right).
\end{equation}
$\bm{x}^{n+1}_{ij\parallel}$ and $\bm{x}^{n+1}_{ij\perp}$ denote the components of $\bm{x}^{n+1}_{ij}$ parallel and perpendicular to $\bm{x}^{n}_{ij}$, respectively.
The threshold distance $d$ used to identify excessively close particle pairs is set to $d=0.8d_0$ in this study.
The velocity and position corrections produced by DS are
\begin{equation}
    \label{eq: ds_update_vel}
	\delta\bm{v}^{\mathrm{DS}}_i
	=
	\frac{\Delta t}{\rho_i}
	\bm{F}^{\mathrm{DS}}_i,
\end{equation}
\begin{equation}
    \label{eq: ds_update_posi}
	\delta\bm{x}^{\mathrm{DS}}_i
	=
	\delta\bm{v}^{\mathrm{DS}}_i\Delta t.
\end{equation}
At each time step, the OPS displacement in \eq{eq:ops_shift} is applied first.
The DS correction defined by \eqq{eq:ds_total_force}{eq:ds_force} is then applied.

%% file: doc/3_.tex
This study proposes a generalized GCB model that extends the original model formulated by Zheng et al.~\cite{zheng2020-1,zheng2020-2}, in which the wall domain is discretized into ghost cells and the associated wall contribution is evaluated by Gaussian quadrature within each cell.
The proposed model introduces the following three extensions:
\begin{enumerate}[label=(\Alph*)]
	\item The one-point quadrature formulation of the original model is generalized to higher-order tensor-product Gaussian quadrature with an arbitrary number of quadrature points in each coordinate direction of the reference domain.
	
	\item The original model, which was developed exclusively for EMPS, is reformulated for semi-implicit particle methods.
	
	\item In the original model, the pressure and velocity values assigned at the quadrature points were not fully consistent with the prescribed boundary conditions, and oscillations were observed in the numerical solution. The proposed model addresses this issue by introducing a consistent boundary-condition treatment based on the FGP models.
\end{enumerate}
Extension~(A) is detailed in Sections~\ref{sec:gg} and \ref{sec:gq}.
Extensions~(B) and (C) are presented in Sections~\ref{sec:ggcbisph} and \ref{sec:ggcbfgp}, respectively.

\subsection{Generalization to arbitrary-point Gaussian quadrature} \label{sec:gg}
\input{doc/3_1_.tex}

\subsection{Quadrature points and weights} \label{sec:gq}
\input{doc/3_2_.tex}

\subsection{Differential operators using the generalized GCB model} \label{sec:ggcbisph}
\input{doc/3_3_.tex}

\subsection{Boundary condition treatment based on the FGP model} \label{sec:ggcbfgp}
\input{doc/3_4_.tex}

\subsection{Computational procedure}
\input{doc/3_5_.tex}

%% file: doc/3_1_.tex
For a fluid particle $i$ located near a wall, the approximation of a field quantity $\phi(\bm{x}_i)$ is decomposed into fluid and wall contributions:
\begin{equation} \label{eq:gcb_split}
	\phi(\bm{x}_i)
	\approx
	\left\langle\phi\right\rangle^{\mathrm{fluid}}_i
	+
	\left\langle\phi\right\rangle^{\mathrm{wall}}_i.
\end{equation}
The fluid contribution is evaluated using the corresponding SPH summation introduced in Section~\ref{sec:isph}.

\begin{figure}[tbp]
	\centering
	\includegraphics[width=0.5\linewidth]{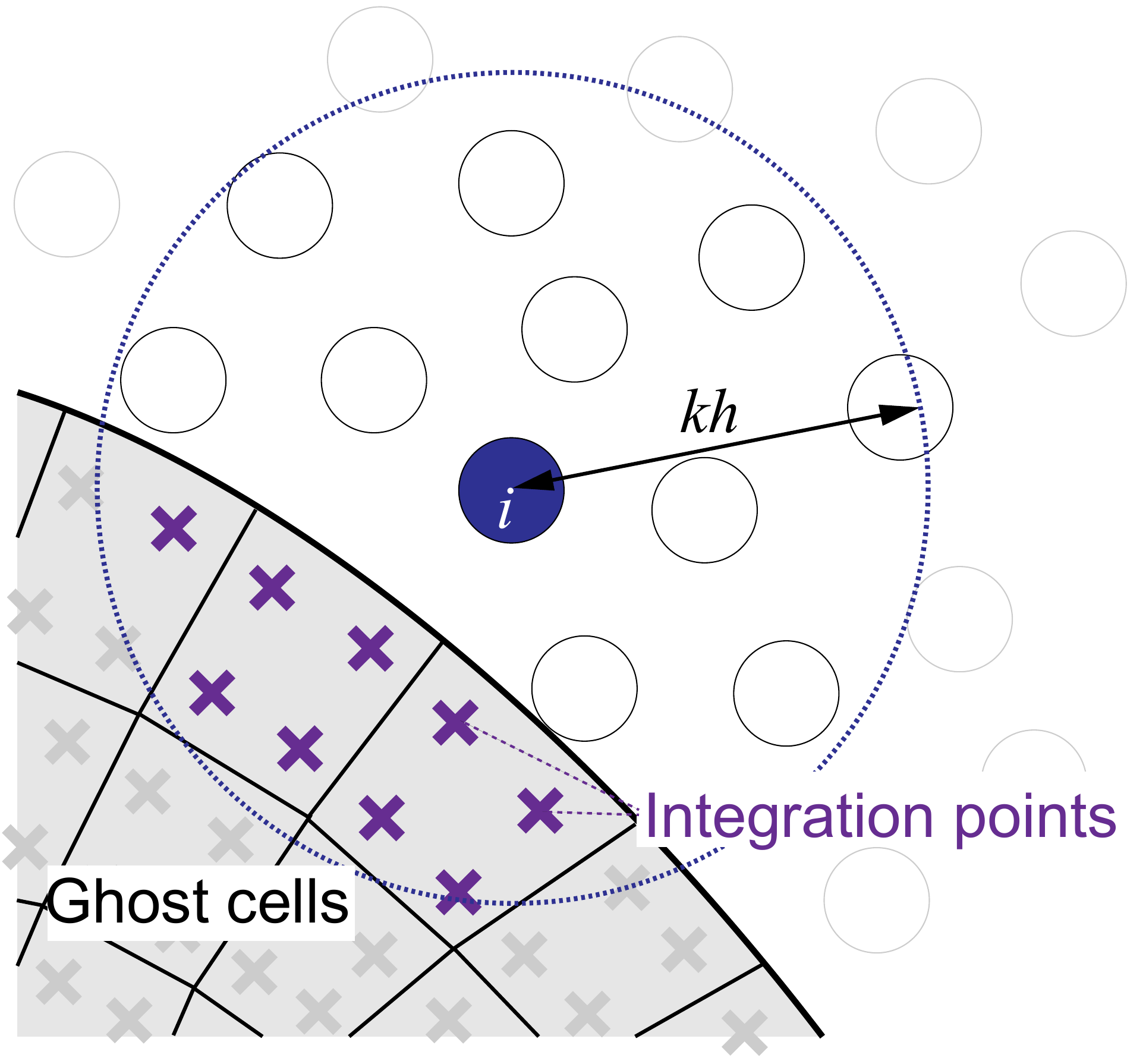}
	\caption{Schematic of ghost cells, quadrature points, and kernel support in the GCB model.}
	\label{fig:agcb}
\end{figure}

In the GCB model, as illustrated in \fig{fig:agcb}, the ghost cells are defined in the wall domain.
The wall contribution $\left\langle\phi\right\rangle^{\mathrm{wall}}_i$ is evaluated as a sum of volume integrals over these cells.
Applying Gaussian quadrature yields
\begin{equation} \label{eq:gcb_int}
	\left\langle\phi\right\rangle^{\mathrm{wall}}_i
	=
	\sum_{c=1}^{N_{\mathrm{cell}}}
	\sum_{m=1}^{N_{\mathrm{ip}}}
	\phi_{(c,m)}
	W_{i(c,m)}
	H^{\mathrm{ip}}_m
	\left|\bm{J}_{(c,m)}\right|,
\end{equation}
\begin{equation}
    \bm{x}_{(c,m)}
    =
    \bm{x}_c\left(\bm{\xi}^{\mathrm{ip}}_m\right),
\end{equation}
\begin{equation}
	\phi_{(c,m)}
	=
	\phi\left(
	\bm{x}_{(c,m)}
	\right),
\end{equation}
\begin{equation}
	W_{i(c,m)}
	=
	W\left(
	\left|
	\bm{x}_i
	-
	\bm{x}_{(c,m)}
	\right|,
	h
	\right),
\end{equation}
\begin{equation}
	\bm{J}_{(c,m)}
	=
	\bm{J}_c\left(\bm{\xi}^{\mathrm{ip}}_m\right).
\end{equation}
Here, $N_{\mathrm{cell}}$ is the number of ghost cells included in the support radius, and $N_{\mathrm{ip}}$ is the number of quadrature points in each cell.
Also, $\bm{x}_c$, $\bm{\xi}^{\mathrm{ip}}_m$, $H^{\mathrm{ip}}_m$, and $\bm{J}_c(\bm{\xi})$ are the coordinates of the $m$-th quadrature point in the physical and reference domains, the quadrature weight, and the Jacobian matrix, respectively.
The same formulation applies when $\phi$ represents a vector or a second-order tensor field.
The calculation of these quantities is described in detail in Section~\ref{sec:gq}.

%% file: doc/3_2_.tex
Consider an $l$-point Gauss--Legendre quadrature rule on the 1D reference interval $\xi\in[-1,1]$.
The $i$-th quadrature weight $w^{\mathrm{ip}}_i$ $(i=0,1,\ldots,l-1)$ is given by
\begin{equation}
	w^{\mathrm{ip}}_i
	=
	\frac{2}
	{
	\left[
	1-\left(\xi^{\mathrm{ip}}_i\right)^2
	\right]
	\left[
	\dfrac{dP_l}{d\xi}
	\left(\xi^{\mathrm{ip}}_i\right)
	\right]^2
	}.
\end{equation}
Here, $P_l(\xi)$ is the Legendre polynomial of degree $l$, defined by Rodrigues' formula as
\begin{equation}
	P_l(\xi)
	=
	\frac{1}{2^l l!}
	\frac{d^l}{d\xi^l}
	\left(\xi^2-1\right)^l.
\end{equation}
The quadrature points $\xi^{\mathrm{ip}}_i$ $(i=0,1,\ldots,l-1)$ are the roots of $P_l(\xi)$.

Taking the tensor product of the 1D rule in the two coordinate directions gives the $m$-th quadrature point and its associated weight in the 2D reference domain:
\begin{equation}
	H^{\mathrm{ip}}_m
	=
	w^{\mathrm{ip}}_i
	w^{\mathrm{ip}}_j,
\end{equation}
\begin{equation}
	\bm{\xi}^{\mathrm{ip}}_m
	=
	\begin{pmatrix}
		\xi^{\mathrm{ip}}_i \\
		\xi^{\mathrm{ip}}_j
	\end{pmatrix},
\end{equation}
\begin{equation}
	m=jl+i+1,
\end{equation}
where $i,j=0,1,\ldots,l-1$.
The numerical tests in this study employ the one-point $(l=1)$ and four-point $(l=2)$ tensor-product quadrature rules.
The corresponding quadrature-point coordinates $\bm{\xi}^{\mathrm{ip}}_m$ and weights $H^{\mathrm{ip}}_m$ are listed in \tab{tbl:ip}.

\begin{table}[tbp]
	\centering
	\caption{Quadrature-point coordinates and weights used in this study}
	\label{tbl:ip}
	\begin{tabular}{cccc}
		\toprule
		$l$ & $m$ & Quadrature points $\bm{\xi}^{\mathrm{ip}}_m$ & Weights $H^{\mathrm{ip}}_m$ \\
		\midrule
		$1$ & $1$ & $\left\{0 \quad 0\right\}^{T}$ & $4$ \\
		\midrule
		\multirow{4}{*}{$2$}
			& $1$ & $\left\{-\dfrac{1}{\sqrt{3}} \quad -\dfrac{1}{\sqrt{3}}\right\}^{T}$ & \multirow{4}{*}{$1$} \\
			& $2$ & $\left\{\dfrac{1}{\sqrt{3}} \quad -\dfrac{1}{\sqrt{3}}\right\}^{T}$ & \\
			& $3$ & $\left\{-\dfrac{1}{\sqrt{3}} \quad \dfrac{1}{\sqrt{3}}\right\}^{T}$ & \\
			& $4$ & $\left\{\dfrac{1}{\sqrt{3}} \quad \dfrac{1}{\sqrt{3}}\right\}^{T}$ & \\
		\bottomrule
	\end{tabular}
\end{table}

The geometry of each ghost cell in the physical domain is represented using Lagrange shape functions commonly employed in the FEM.
On the 1D reference interval $\xi\in[-1,1]$, the $i$-th Lagrange basis function of degree $q$ $(i=0,1,\ldots,q)$ is defined as
\begin{equation}
	L^q_i(\xi)
	=
	\prod_{\substack{j=0 \\ j\neq i}}^{q}
	\frac{\xi-\xi_j}{\xi_i-\xi_j},
\end{equation}
\begin{equation}
	\xi_i
	=
	\frac{2i}{q}-1.
\end{equation}
For a 2D square reference domain, with $\bm{\xi}=(\xi,\eta)^T$, the tensor-product shape functions are defined as
\begin{equation}
	\psi_k(\bm{\xi})
	=
	L^q_i(\xi)L^q_j(\eta),
\end{equation}
\begin{equation}
	k=(q+1)j+i+1.
\end{equation}
Although this study employs quadrilateral ghost cells, a triangular cell can also be represented as a degenerate quadrilateral by collapsing one edge to a point.
The isoparametric mapping from the reference domain to ghost cell $c$ in the physical domain and its Jacobian matrix are then evaluated as
\begin{equation}
	\bm{x}_c(\bm{\xi})
	=
	\sum_{k=1}^{(q+1)^2}
	\bm{x}^{\mathrm{cell}}_k
	\psi_k(\bm{\xi}),
\end{equation}
\begin{equation}
	\bm{J}_c(\bm{\xi})
	=
	\frac{\partial\bm{x}_c}{\partial\bm{\xi}}
	=
	\sum_{k=1}^{(q+1)^2}
	\bm{x}^{\mathrm{cell}}_k
	\frac{\partial\psi_k}{\partial\bm{\xi}}
	(\bm{\xi}),
\end{equation}
where $\bm{x}^{\mathrm{cell}}_k$ denotes the physical position of the $k$-th node of cell $c$.

%% file: doc/3_3_.tex
As in \eq{eq:gcb_split}, each differential operator is decomposed into fluid and wall contributions.
In the generalized GCB model, the wall contribution is evaluated by applying Gaussian quadrature to the volume integrals over the ghost cells.
The resulting operators are
\begin{equation} \label{eq:gcb_gp}
	\left\langle\nabla p\right\rangle^{\mathrm{wall}}_i
	=
	\sum_{c=1}^{N_{\mathrm{cell}}}
	\sum_{m=1}^{N_{\mathrm{ip}}}
	\left(
	p_i+p_{(c,m)}
	\right)
	\nabla W_{i(c,m)}
	H^{\mathrm{ip}}_m
	\left|\bm{J}_{(c,m)}\right|,
\end{equation}
\begin{equation} \label{eq:gcb_lp}
	\left\langle\nabla^2p\right\rangle^{\mathrm{wall}}_i
	=
	2
	\sum_{c=1}^{N_{\mathrm{cell}}}
	\sum_{m=1}^{N_{\mathrm{ip}}}
    \left(p_{(c,m)}-p_i\right)
	\frac{\bm{x}_{i(c,m)}}
	{\left|\bm{x}_{i(c,m)}\right|^2}
	\cdot
	\nabla W_{i(c,m)}
	H^{\mathrm{ip}}_m
	\left|\bm{J}_{(c,m)}\right|,
\end{equation}
\begin{equation} \label{eq:gcb_lv}
	\left\langle\nabla^2\bm{v}\right\rangle^{\mathrm{wall}}_i
	=
	2
	\sum_{c=1}^{N_{\mathrm{cell}}}
	\sum_{m=1}^{N_{\mathrm{ip}}}
    \left(\bm{v}_{(c,m)}-\bm{v}_i\right)
	\frac{\bm{x}_{i(c,m)}}
	{\left|\bm{x}_{i(c,m)}\right|^2}
	\cdot
	\nabla W_{i(c,m)}
	H^{\mathrm{ip}}_m
	\left|\bm{J}_{(c,m)}\right|,
\end{equation}
\begin{equation} \label{eq:gcb_dv}
	\left\langle\nabla\cdot\bm{v}\right\rangle^{\mathrm{wall}}_i
	=
	\sum_{c=1}^{N_{\mathrm{cell}}}
	\sum_{m=1}^{N_{\mathrm{ip}}}
	\left(
	\bm{v}_{(c,m)}-\bm{v}_i
	\right)
	\cdot
	\nabla W_{i(c,m)}
	H^{\mathrm{ip}}_m
	\left|\bm{J}_{(c,m)}\right|,
\end{equation}
where $\bm{x}_{i(c,m)} = \bm{x}_{(c,m)}-\bm{x}_i$ and $\nabla W_{i(c,m)}$ denotes the gradient of the kernel with respect to $\bm{x}_i$ in the direction of $\bm{x}_{i(c,m)}$.

%

%% file: doc/3_4_.tex
In the original GCB model~\cite{zheng2020-1,zheng2020-2}, the field value at every quadrature point within the kernel support of a target fluid particle is set equal to the value carried by that fluid particle. 
As noted by Mitsume et al.~\cite{mitsume2015} for a polygonal wall-boundary model, this treatment does not generally satisfy the wall boundary conditions, inducing velocity and pressure oscillations.
To address this issue, the present study adapts and extends the FGP models~\cite{marrone2011,Bouscasse2013,Morikawa2021} to the GCB model, providing a more accurate and consistent treatment of the boundary conditions.

\begin{figure}[tbp]
	\centering
	\includegraphics[width=0.9\linewidth]{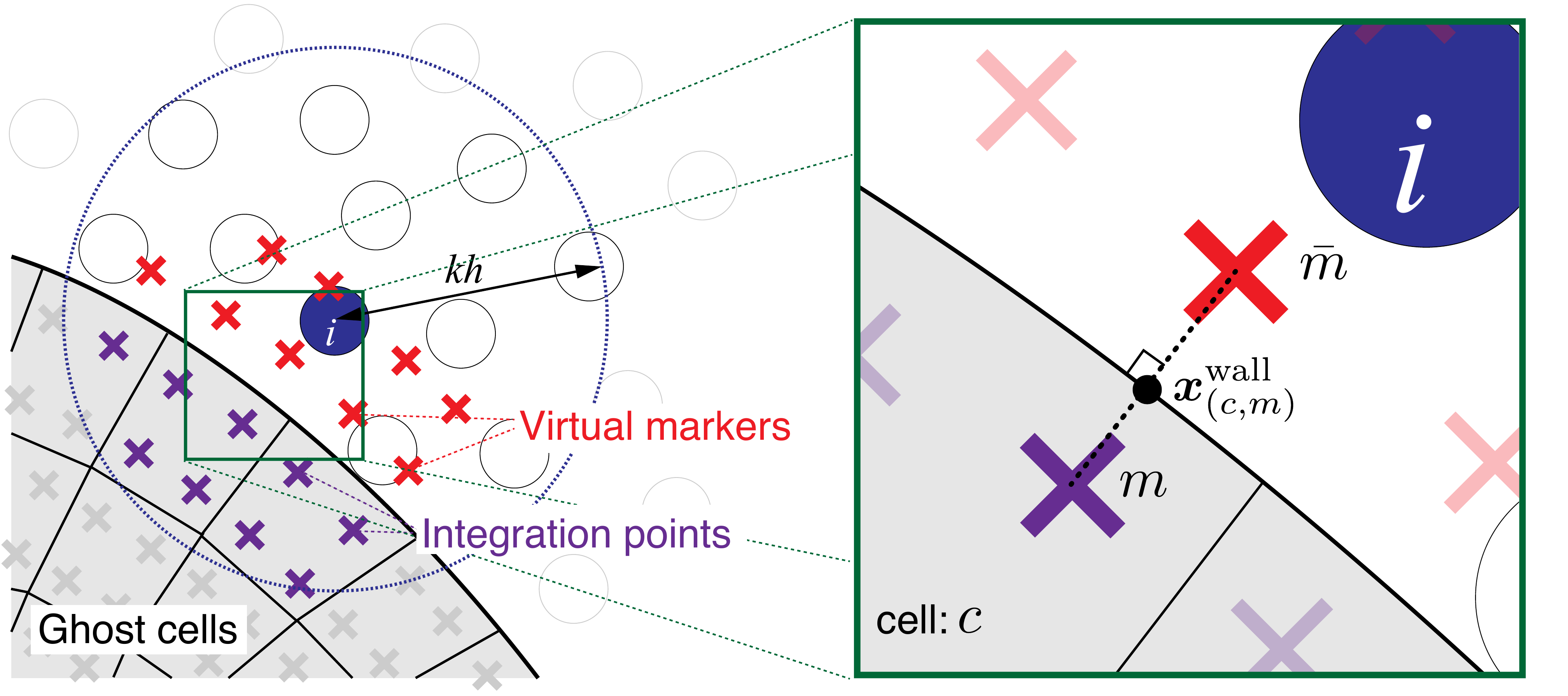}
	\caption{Quadrature points and their corresponding virtual markers in the generalized GCB model.}
	\label{fig:agcb_gp}
\end{figure}

In the generalized GCB model, a virtual marker $\bar{m}$ is associated with each quadrature point $m$, as illustrated in \fig{fig:agcb_gp}.
The virtual marker is placed at the reflection of the quadrature point with respect to the closest point on the wall surface.
Its position $\bm{x}_{(c,\bar{m})}$ is given by
\begin{equation}
    \label{eq: mirror}
	\bm{x}_{(c,\bar{m})}
	=
	\bm{x}_{(c,m)}
	+
	2
	\left(
	\bm{x}^{\mathrm{wall}}_{(c,m)}
	-
	\bm{x}_{(c,m)}
	\right).
\end{equation}
{
$\bm{x}^\mathrm{wall}_{(c,m)}$ is the closest point on the wall surface to the quadrature point $(c,m)$, including corners.
That is, the generalized GCB model consistently applies \eq{eq: mirror} to walls of arbitrary geometry without introducing any exceptional treatment for corners.
However, this approach may introduce errors in the calculation of wall boundary conditions for walls with non-orthogonal corners.
Therefore, for more practical flow simulations, the present model needs to be enhanced by referring to high-accuracy FGP models (e.g., \cite{Antuono2023}); this development is left for future work.
}

The field value at each quadrature point is determined from the corresponding virtual-marker value so that the wall boundary condition is satisfied.
Following the FGP models~\cite{marrone2011,Bouscasse2013,Morikawa2021}, the field value at a virtual marker is evaluated from the neighboring fluid particles using normalized SPH interpolation:
\begin{equation}
    \label{eq: vm_int}
	\phi\left(\bm{x}_{(c,\bar{m})}\right)
	\approx
	\left\langle\phi\right\rangle_{(c,\bar{m})}
	=
	\sum_{j=1}^{N_{\mathrm{p}}}
	\frac{m_j}{\rho_j}
	\tilde{W}
	\left(
	\left|
	\bm{x}_{(c,\bar{m})j}
	\right|,
	h
	\right)
	\phi_j,
\end{equation}
\begin{equation}
    \label{eq: vm_kernel}
	\tilde{W}
	\left(
	\left|
	\bm{x}_{(c,\bar{m})j}
	\right|,
	h
	\right)
	=
	\frac{
	W\left(
	\left|
	\bm{x}_{(c,\bar{m})j}
	\right|,
	h
	\right)
	}{
	\displaystyle
	\sum_{k=1}^{N_{\mathrm{p}}}
	\frac{m_k}{\rho_k}
	W\left(
	\left|
	\bm{x}_{(c,\bar{m})k}
	\right|,
	h
	\right)
	}.
\end{equation}
{
When the number of neighboring fluid particles around a virtual marker is extremely small, the interpolation accuracy of \eqq{eq: vm_int}{eq: vm_kernel} decreases.
However, such a situation occurs only when isolated splash particles approach the wall, meaning that these fluid particles also have few neighbors and do not affect the behavior of the main fluid body.
Therefore, the decrease in interpolation accuracy in such sparse regions has a negligible impact on the overall accuracy and stability of the fluid simulation.
}

The velocity $\bm{v}_{(c,m)}$ at a quadrature point is constructed from the interpolated velocity $\left\langle\bm{v}\right\rangle_{(c,\bar{m})}$ at the corresponding virtual marker and the wall velocity $\bm{v}^{\mathrm{wall}}_{(c,m)}$:
\begin{equation} \label{eq:vm_bc_v}
	\bm{v}_{(c,m)}
	-
	\bm{v}^{\mathrm{wall}}_{(c,m)}
	=
	\bm{M}
	\left(
	\left\langle\bm{v}\right\rangle_{(c,\bar{m})}
	-
	\bm{v}^{\mathrm{wall}}_{(c,m)}
	\right).
\end{equation}
The transformation matrix $\bm{M}$ is selected according to the prescribed velocity boundary condition:
\begin{equation}
	\bm{M}
	=
	\begin{cases}
		-\bm{I}
		& \text{(no-slip)}, \\[1mm]
		\bm{I}
		-
		2
		\bm{n}^{\mathrm{wall}}_{(c,m)}
		\otimes
		\bm{n}^{\mathrm{wall}}_{(c,m)}
		& \text{(free-slip)}.
	\end{cases}
\end{equation}
Here, $\bm{n}^{\mathrm{wall}}_{(c,m)}$ is the unit normal vector pointing from the wall into the fluid domain at the closest wall point $\bm{x}^{\mathrm{wall}}_{(c,m)}$.
The no-slip velocity condition in \eq{eq:bc_wall} is adopted in this study.

The nonhomogeneous Neumann pressure condition in \eq{eq:pbc_wall} is imposed by adapting the pressure treatment in the FGP models~\cite{marrone2011,Bouscasse2013,Morikawa2021} to the GCB model.
The pressure $p_{(c,m)}$ at a quadrature point is extrapolated from the pressure $\left\langle p\right\rangle_{(c,\bar{m})}$ at the corresponding virtual marker as
\begin{equation} \label{eq:vm_bc_p}
	p_{(c,m)}
	=
	\left\langle p\right\rangle_{(c,\bar{m})}
	-
	\left|
    \bm{x}_{(c,m)}
	-
	\bm{x}_{(c,\bar{m})}
    \right|
	\rho
	\left(
	\bm{g}
	-
	\bm{a}^{\mathrm{wall}}_{(c,m)}
	\right)
	\cdot
	\bm{n}^{\mathrm{wall}}_{(c,m)},
\end{equation}
where $\bm{a}^{\mathrm{wall}}_{(c,m)}$ is the acceleration of the wall at $\bm{x}^{\mathrm{wall}}_{(c,m)}$.
To avoid destabilizing particle motion near the free surface, the free-surface detection procedure is also applied at each virtual-marker position.
If a virtual marker is classified as being on the free surface, the pressure at the corresponding quadrature point is set to zero.

%% file: doc/3_5_.tex
\begin{figure}[tbp]
	\centering
	\includegraphics[width=0.7\linewidth]{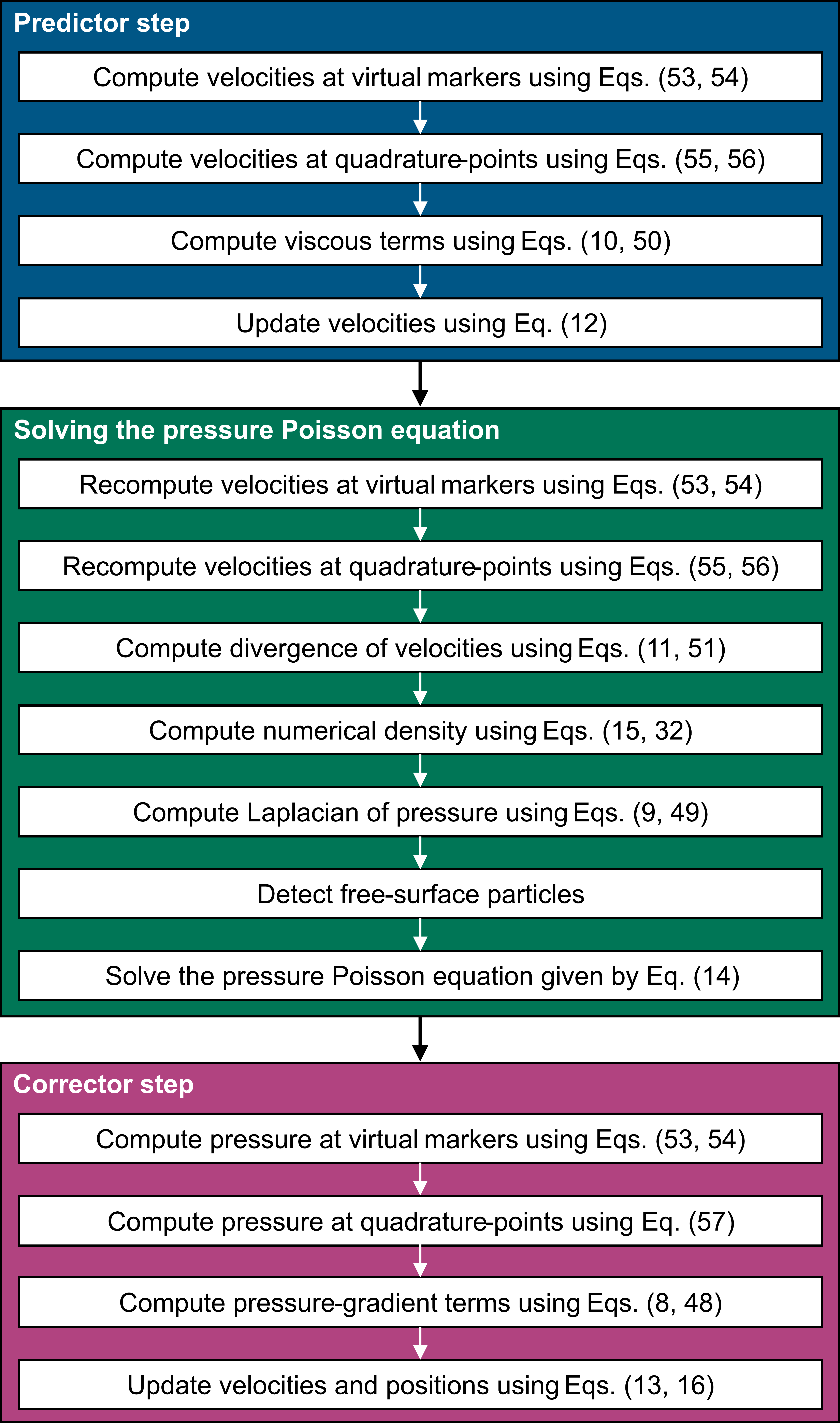}
	\caption{Flowchart of the computational procedure for a time step in the fluid simulation based on the stabilized ISPH method with the generalized GCB model.}
	\label{fig:flow_fluid}
\end{figure}

The computational procedure for advancing the solution is summarized in \fig{fig:flow_fluid}.
{
Note that, for splash particles isolated from the main fluid body, the following artificial repulsive force is applied to prevent wall penetration:
\begin{equation}
  \label{eq: rep_force}
  \bm{f}_i^{\mathrm{rep}} = 
  \begin{cases} 
    \dfrac{\rho_i}{\Delta t^2} \left(\frac{d}{2} - \lvert \bm{x}^\mathrm{wall}_i - \bm{x}_i\rvert \right) \bm{n}^{\mathrm{wall}} & (0 \le \lvert \bm{x}^\mathrm{wall}_i - \bm{x}_i\rvert < d/2) \\
    \bm{0} & (d/2 \le \lvert \bm{x}^\mathrm{wall}_i - \bm{x}_i \rvert) 
  \end{cases},
\end{equation}
where $\bm{x}^\mathrm{wall}_i$ is the closest point on the wall surface to particle $i$.
The repulsive force $\bm{f}^\mathrm{rep}_i$ is incorporated into the right-hand side of \eq{eq: ds_update_vel} to be reflected in the velocity and position updates.
}

%% file: doc/4_.tex
This section extends the generalized GCB model formulated in Section~\ref{sec:ggcb} to fluid--rigid-body coupling.
The rigid body is discretized by ghost cells, and the reaction to the wall contribution computed in the fluid analysis is applied to the rigid body as the hydrodynamic force.
This treatment is consistent with coupling approaches that represent fluid--solid interaction through momentum exchange with particles or virtual boundary points~\cite{Bouscasse2013}, and is closely related to the MPS--FEM coupling framework based on the GCB model~\cite{zheng2020-2}.
A feature of the present formulation is that the hydrodynamic force and moment are evaluated from the wall contribution associated with each quadrature point in the ghost cells.
Consequently, the force exerted on the fluid particles and its reaction on the rigid body are obtained from the same discrete operation, while the rigid-body geometry remains represented explicitly by the cells.

\subsection{Governing equations of rigid-body motion}
\input{doc/4_1_.tex}

\subsection{Hydrodynamic forces and moments}
\input{doc/4_2_.tex}

\subsection{Time-marching algorithm}
\input{doc/4_3_.tex}

\subsection{Coupling procedure}
\input{doc/4_4_.tex}

%% file: doc/4_1_.tex
A 2D rigid body of uniform density is considered.
Assuming unit depth, let $A$ and $\rho_{\mathrm{r}}$ denote the area and density of the rigid body, respectively.
When only the hydrodynamic force $\bm{F}^{\mathrm{fluid}}$ and gravity act on the body, its translational motion and rotation about the center of mass are governed by
\begin{equation}
    \rho_{\mathrm{r}}A\frac{D\bm{v}_{\mathrm{g}}}{Dt}
    =
    \bm{F}^{\mathrm{fluid}}
    +
    \rho_{\mathrm{r}}A\bm{g},
\end{equation}
\begin{equation} \label{eq:fsi_angular}
    I_{\mathrm{g}}\frac{D\omega_{\mathrm{g}}}{Dt}
    =
    M^{\mathrm{fluid}},
\end{equation}
where $\bm{v}_{\mathrm{g}}$ is the velocity of the center of mass, $I_{\mathrm{g}}$ is the mass moment of inertia about the center of mass, $\omega_{\mathrm{g}}$ is the angular velocity, and $M^{\mathrm{fluid}}$ is the hydrodynamic moment.

When the rigid-body domain $\Omega_{\mathrm{r}}$ is discretized by $N_{\mathrm{cell}}$ ghost cells, its area, center-of-mass position, and mass moment of inertia are evaluated by Gaussian quadrature over the ghost cells as
\begin{equation}
    A
    =
    \int_{\Omega_{\mathrm{r}}}d\bm{x}
    \approx
    \sum_{c=1}^{N_{\mathrm{cell}}}
    \sum_{m=1}^{N_{\mathrm{ip}}}
    H_m^{\mathrm{ip}}
    \left|\bm{J}_{(c,m)}\right|,
\end{equation}
\begin{equation}
    \bm{x}_{\mathrm{g}}
    =
    \frac{1}{A}
    \int_{\Omega_{\mathrm{r}}}\bm{x}\,d\bm{x}
    \approx
    \frac{1}{A}
    \sum_{c=1}^{N_{\mathrm{cell}}}
    \sum_{m=1}^{N_{\mathrm{ip}}}
    \bm{x}_{(c,m)}
    H_m^{\mathrm{ip}}
    \left|\bm{J}_{(c,m)}\right|,
\end{equation}
\begin{equation}
    I_{\mathrm{g}}
    =
    \rho_{\mathrm{r}}
    \int_{\Omega_{\mathrm{r}}}
    \left|\bm{x}-\bm{x}_{\mathrm{g}}\right|^2
    d\bm{x}
    \approx
    \rho_{\mathrm{r}}
    \sum_{c=1}^{N_{\mathrm{cell}}}
    \sum_{m=1}^{N_{\mathrm{ip}}}
    \left|\bm{x}_{(c,m)}-\bm{x}_{\mathrm{g}}\right|^2
    H_m^{\mathrm{ip}}
    \left|\bm{J}_{(c,m)}\right|.
\end{equation}
These properties are determined by the rigid-body geometry and density and are evaluated using the same ghost cells and quadrature points as those employed for the boundary treatment.
Thus, the geometric representation used for the boundary treatment and the mass properties used in the rigid-body equations are obtained from a common discretization scheme.

%% file: doc/4_2_.tex
In the generalized GCB model, the force exerted by the wall domain on a fluid particle $i$ is evaluated from the wall components of the pressure-gradient and viscous terms.
For fluid--rigid-body coupling, only the ghost cells belonging to the moving rigid body are included in the evaluation of the hydrodynamic force; ghost cells representing fixed walls are excluded.
This set of moving-body ghost cells is denoted by $\mathcal{C}_{\mathrm{r}}$.
Unless otherwise stated, the wall contributions in this subsection refer only to contributions from $\mathcal{C}_{\mathrm{r}}$.

Multiplying the Navier--Stokes momentum equation by the particle mass $m_i$ and isolating the wall contribution associated with the moving rigid body gives
\begin{equation}
    m_i\frac{D\bm{v}_i}{Dt}
    =
    m_i
    \left\{
    -\frac{1}{\rho}
    \left\langle\nabla p\right\rangle^{\mathrm{fluid}}_i
    +
    \nu
    \left\langle\nabla^2\bm{v}\right\rangle^{\mathrm{fluid}}_i
    +
    \bm{g}
    \right\}
    +
    \bm{f}^{\mathrm{wall}}_i,
\end{equation}
\begin{equation}
    \bm{f}^{\mathrm{wall}}_i
    =
    m_i
    \left(
    -\frac{1}{\rho}
    \left\langle\nabla p\right\rangle^{\mathrm{wall}}_i
    +
    \nu
    \left\langle\nabla^2\bm{v}\right\rangle^{\mathrm{wall}}_i
    \right),
\end{equation}
where $\bm{f}^{\mathrm{wall}}_i$ is the force exerted by the moving rigid-body boundary on fluid particle $i$.
As shown in \eqq{eq:gcb_gp}{eq:gcb_lv}, each wall contribution is expressed as a sum over the quadrature points in the ghost cells.
Accordingly, $\bm{f}^{\mathrm{wall}}_i$ can be decomposed into the forces $\bm{f}^{\mathrm{wall}}_{i(c,m)}$ exerted on particle $i$ by individual quadrature points $(c,m)$:
\begin{equation}
    \bm{f}^{\mathrm{wall}}_i
    =
    \sum_{c\in\mathcal{C}_{\mathrm{r}}}
    \sum_{m=1}^{N_{\mathrm{ip}}}
    \bm{f}^{\mathrm{wall}}_{i(c,m)},
\end{equation}
\begin{equation} \label{eq:fsi_wall_force_cm}
    \bm{f}^{\mathrm{wall}}_{i(c,m)}
    =
    m_i
    \left(
    -\frac{1}{\rho}\bm{G}^{p}_{i(c,m)}
    +
    \nu\bm{L}^{v}_{i(c,m)}
    \right).
\end{equation}
Here, $\bm{G}^{p}_{i(c,m)}$ and $\bm{L}^{v}_{i(c,m)}$ are the individual quadrature-point contributions to the pressure gradient and velocity Laplacian, respectively, and are defined by
\begin{equation}
    \bm{G}^{p}_{i(c,m)}
    =
    \left(
    p_i+p_{(c,m)}
    \right)
    \nabla W_{i(c,m)}
    H_m^{\mathrm{ip}}
    \left|\bm{J}_{(c,m)}\right|,
\end{equation}
\begin{equation}
    \bm{L}^{v}_{i(c,m)}
    =
    2
    \left(
    \bm{v}_{(c,m)}-\bm{v}_i
    \right)
    \frac{\bm{x}_{i(c,m)}}
    {\left|\bm{x}_{i(c,m)}\right|^2}
    \cdot
    \nabla W_{i(c,m)}
    H_m^{\mathrm{ip}}
    \left|\bm{J}_{(c,m)}\right|.
\end{equation}

\begin{figure}[tbp]
    \centering
    \includegraphics[width=1.0\linewidth]{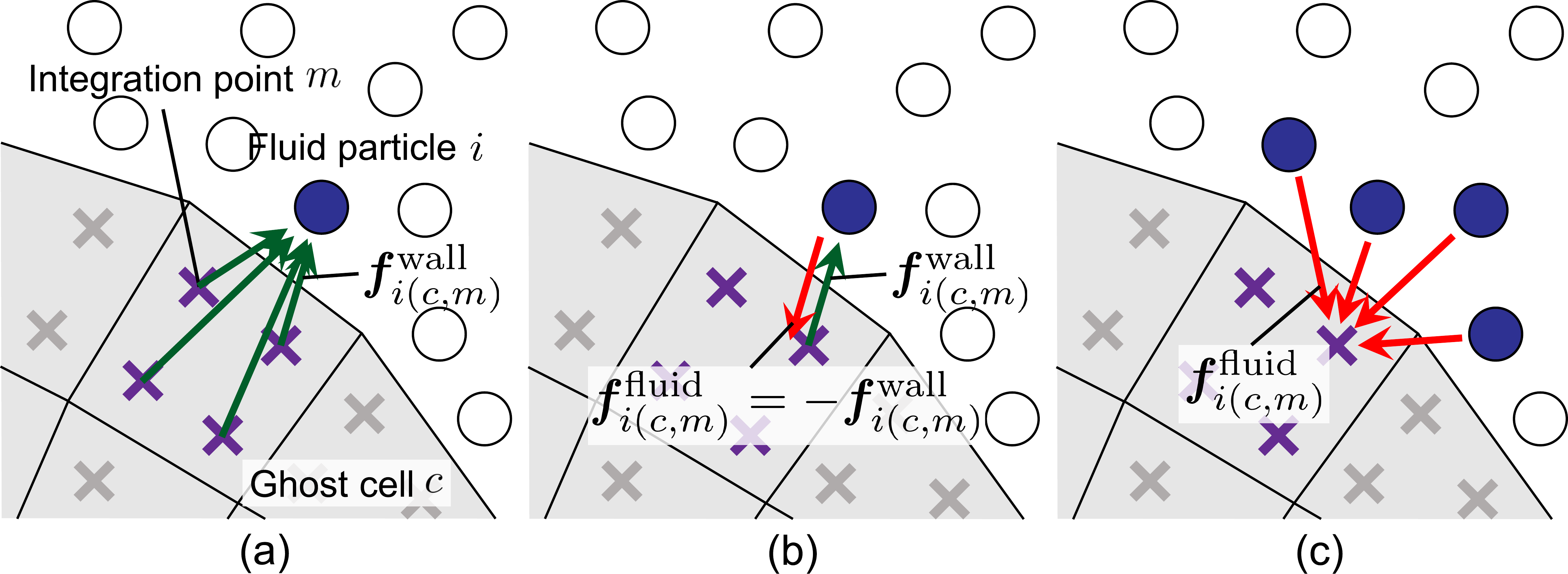}
    \caption{Schematic of hydrodynamic-force evaluation from the GCB wall-side contribution: (a) wall force acting on a fluid particle, (b) reaction force acting at a quadrature point, and (c) resultant force at the quadrature point obtained by summing the reactions from surrounding fluid particles.}
    \label{fig:fsi_force_concept}
\end{figure}

As illustrated in \fig{fig:fsi_force_concept}, the reaction to the force in \eq{eq:fsi_wall_force_cm} is interpreted as the force $\bm{f}^{\mathrm{fluid}}_{i(c,m)}$ exerted by fluid particle $i$ on quadrature point $(c,m)$:
\begin{equation} \label{eq:fsi_reaction_cm}
    \bm{f}^{\mathrm{fluid}}_{i(c,m)}
    =
    -\bm{f}^{\mathrm{wall}}_{i(c,m)}.
\end{equation}
The resultant force exerted by the fluid on quadrature point $(c,m)$ is obtained by summing the reactions from all surrounding fluid particles:
\begin{equation}
    \bm{f}^{\mathrm{fluid}}_{(c,m)}
    =
    \sum_{i=1}^{N_{\mathrm{p}}}
    \bm{f}^{\mathrm{fluid}}_{i(c,m)}.
\end{equation}
The total hydrodynamic force and moment exerted on the rigid body are then evaluated by summing over the ghost cells and quadrature points belonging to the body:
\begin{equation} \label{eq:fsi_force}
    \bm{F}^{\mathrm{fluid}}
    =
    \sum_{c\in\mathcal{C}_{\mathrm{r}}}
    \sum_{m=1}^{N_{\mathrm{ip}}}
    \bm{f}^{\mathrm{fluid}}_{(c,m)},
\end{equation}
\begin{equation} \label{eq:fsi_moment}
    M^{\mathrm{fluid}}
    =
    \sum_{c\in\mathcal{C}_{\mathrm{r}}}
    \sum_{m=1}^{N_{\mathrm{ip}}}
    \left(
    \bm{x}_{(c,m)}-\bm{x}_{\mathrm{g}}
    \right)
    \times
    \bm{f}^{\mathrm{fluid}}_{(c,m)}.
\end{equation}

This formulation pairs, at each quadrature point, the force exerted by the moving rigid-body boundary on the fluid with the equal and opposite reaction exerted by the fluid on the body.
To show this discrete balance, let the total force exerted on the fluid by the moving rigid-body boundary be
\begin{equation}
    \bm{F}^{\mathrm{wall}}
    =
    \sum_{i=1}^{N_{\mathrm{p}}}
    \bm{f}^{\mathrm{wall}}_i
    =
    \sum_{i=1}^{N_{\mathrm{p}}}
    \sum_{c\in\mathcal{C}_{\mathrm{r}}}
    \sum_{m=1}^{N_{\mathrm{ip}}}
    \bm{f}^{\mathrm{wall}}_{i(c,m)}.
\end{equation}
\eq{eq:fsi_reaction_cm} then gives
\begin{equation}
    \bm{F}^{\mathrm{fluid}}
    =
    -
    \sum_{i=1}^{N_{\mathrm{p}}}
    \sum_{c\in\mathcal{C}_{\mathrm{r}}}
    \sum_{m=1}^{N_{\mathrm{ip}}}
    \bm{f}^{\mathrm{wall}}_{i(c,m)},
\end{equation}
and hence
\begin{equation}
    \bm{F}^{\mathrm{wall}}
    +
    \bm{F}^{\mathrm{fluid}}
    =
    \bm{0}.
\end{equation}
Thus, in the generalized GCB model, the interaction forces associated with the moving rigid body satisfy an exact discrete action--reaction balance and, correspondingly, a discrete balance of linear momentum between the fluid and the body.
{
This fundamental principle, while well recognized in earlier FSI studies~\cite{Bouscasse2013}, has been further emphasized and refined in recent high-fidelity coupling algorithms~\cite{Sun2019, Gotoh2026b, Chen2026}.
}
However, this statement applies strictly to the balance of the interaction forces.
It should be noted that in actual computations, various factors affect the conservation of momentum.

%% file: doc/4_3_.tex
The rigid-body equations are advanced using a first-order explicit scheme.
Let $\Delta t_{\mathrm{r}}$ denote the time-step size for the rigid-body analysis.
{
Because the present study focuses on problems without rigid-body contact or collision, $\Delta t_\mathrm{r}$ is set equal to the fluid time-step size $\Delta t$.
}
The center-of-mass velocity and angular velocity at the $(n+1)$-th time step are updated as
\begin{equation} \label{eq:fsi_vg_update}
    \bm{v}_{\mathrm{g}}^{n+1}
    =
    \bm{v}_{\mathrm{g}}^{n}
    +
    \frac{1}{\rho_{\mathrm{r}}A}
    \left(
    \bm{F}^{\mathrm{fluid}}
    +
    \rho_{\mathrm{r}}A\bm{g}
    \right)
    \Delta t_{\mathrm{r}},
\end{equation}
\begin{equation} \label{eq:fsi_omega_update}
    \omega_{\mathrm{g}}^{n+1}
    =
    \omega_{\mathrm{g}}^{n}
    +
    \frac{M^{\mathrm{fluid}}}{I_{\mathrm{g}}}
    \Delta t_{\mathrm{r}}.
\end{equation}

The nodes of the ghost cells, the Gaussian quadrature points, and the virtual markers associated with those points are moved with the rigid body.
Using $k$ as a collective index for these points, the velocity of point $k$ at the $(n+1)$-th time step is given by
\begin{equation} \label{eq:fsi_point_velocity}
    \bm{v}^{n+1}_{k}
    =
    \bm{v}^{n+1}_{\mathrm{g}}
    +
    \omega^{n+1}_{\mathrm{g}}
    \bm{R}_{\perp}
    \left(
    \bm{x}^{n}_{k}-\bm{x}^{n}_{\mathrm{g}}
    \right),
\end{equation}
\begin{equation} \label{eq:fsi_inverse}
    \bm{R}_{\perp}
    =
    \begin{pmatrix}
        0 & -1 \\
        1 & 0
    \end{pmatrix},
\end{equation}
where $\bm{R}_{\perp}$ rotates an in-plane vector counterclockwise by $90^{\circ}$.
The position of point $k$ and the center-of-mass position are then updated as
\begin{equation} \label{eq:fsi_ck_update}
    \bm{x}^{n+1}_{k}
    =
    \bm{x}^{n}_{k}
    +
    \bm{v}^{n+1}_{k}
    \Delta t_{\mathrm{r}},
\end{equation}
\begin{equation} \label{eq:fsi_cg_update}
    \bm{x}^{n+1}_{\mathrm{g}}
    =
    \bm{x}^{n}_{\mathrm{g}}
    +
    \bm{v}^{n+1}_{\mathrm{g}}
    \Delta t_{\mathrm{r}}.
\end{equation}

Because the pressure boundary condition in \eq{eq:vm_bc_p} requires the wall acceleration, the acceleration at point $k$ is approximated by
\begin{equation} \label{eq:fsi_ip_acceleration}
    \bm{a}^{n+1}_{k}
    =
    \frac{
    \bm{v}^{n+1}_{k}
    -
    \bm{v}^{n}_{k}
    }
    {\Delta t_{\mathrm{r}}}.
\end{equation}

%% file: doc/4_4_.tex
The fluid and rigid-body solvers are coupled using a serial staggered partitioned procedure corresponding to the conventional serial staggered (CSS) scheme~\cite{Farhat2000}.
At the $n$-th time step, the fluid problem is solved using the rigid-body position, velocity, and acceleration at that time step.
After the fluid solution has been advanced to the $(n+1)$-th time step, the hydrodynamic force and moment are evaluated using the pressure and velocity at the time steps specified below.
Subsequently, the rigid-body motion is updated.

In the fluid analysis, the viscous term is evaluated using the velocity at the $n$-th time step, whereas the pressure-gradient term is evaluated using the pressure at the $(n+1)$-th time step.
The same time steps are therefore used in the evaluation of the hydrodynamic force:
\begin{equation}
    \bm{F}^{\mathrm{fluid},n+1} =
    - \sum_{i=1}^{N_{\mathrm{p}}}
    \sum_{c\in\mathcal{C}_{\mathrm{r}}}
    \sum_{m=1}^{N_{\mathrm{ip}}}
    m_i
    \left(
    -\frac{1}{\rho}
    \bm{G}^{p,n+1}_{i(c,m)}
    +
    \nu
    \bm{L}^{v,n}_{i(c,m)}
    \right).
\end{equation}
The hydrodynamic moment is evaluated in the same manner using \eq{eq:fsi_moment}.
Using the same temporal levels preserves consistency between the wall contribution applied in the fluid-particle momentum equation and the reaction force transferred to the rigid-body solver.

\begin{figure}[tbp]
	\centering
	\includegraphics[width=1.0\linewidth]{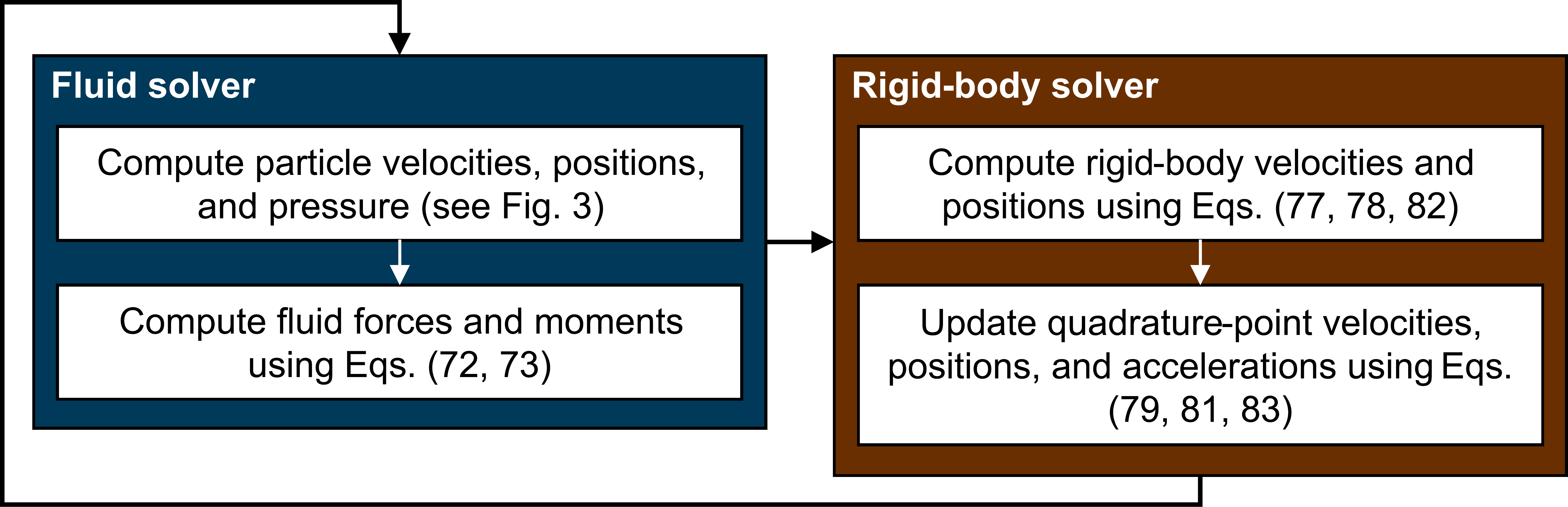}
	\caption{Flowchart of the computational procedure for a time step in the fluid--rigid-body coupling analysis.}
	\label{fig:flow_coupling}
\end{figure}

The computational procedure for advancing the coupled solution is summarized in \fig{fig:flow_coupling}.
This partitioned scheme provides a straightforward explicit coupling procedure and directly extends the moving-wall boundary treatment developed in Section~\ref{sec:ggcb} to fluid--rigid-body interaction.
Its coupling stability, however, requires particular attention for problems with low solid-to-fluid density ratios or strong added-mass effects.
This issue is assessed separately in the validation problems involving rigid-body motion.

%% file: doc/5_.tex
\subsection{Hydrostatic pressure problem}
\input{doc/5_1_.tex} \label{sec:hsp}

\subsection{Dam-break problem} \label{sec:dambreak}
\input{doc/5_2_.tex}

\subsection{Sloshing problem} \label{sec:sloshing}
\input{doc/5_3_.tex}

\subsection{Falling circular cylinder problem} \label{sec:rigid_circle}
\input{doc/5_4_.tex}

\subsection{Wedge water-entry problem} \label{sec:rigid_wedge}
\input{doc/5_5_.tex}

%% file: doc/5_1_.tex
\begin{figure}[tbp]
	\centering
	\includegraphics[width=0.5\linewidth]{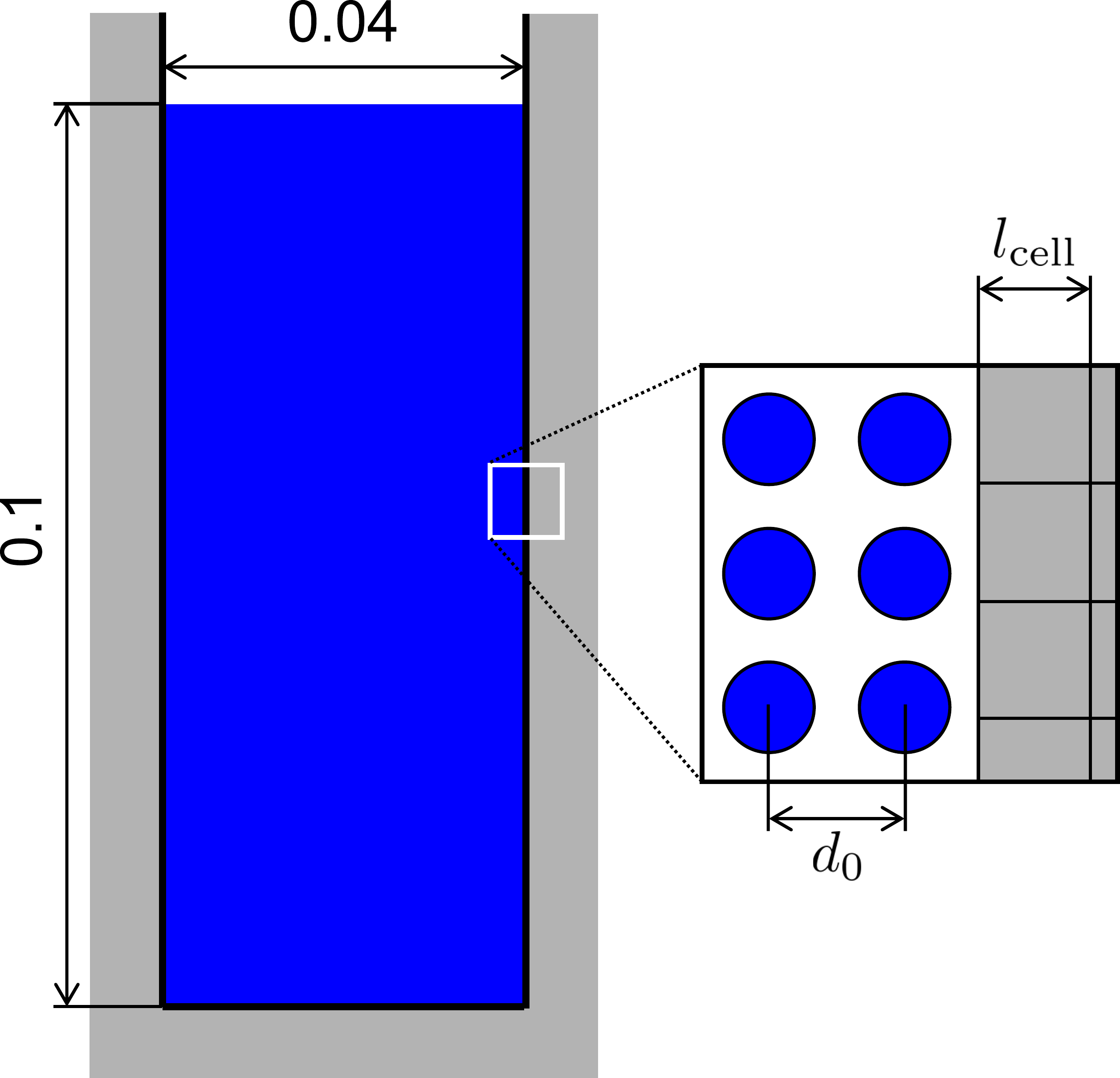} 
	\caption{Hydrostatic pressure problem: initial configuration with initial particle spacing $d_0$ and length of ghost cells $l_{\mathrm{cell}}$. All dimensions are in $\mathrm{(m)}$.}
	\label{fig:hsp}
\end{figure}

To verify the accuracy and stability of the proposed generalized GCB model, a hydrostatic pressure problem is computed and the numerical results are compared with the analytical solution.
For an incompressible viscous fluid in hydrostatic equilibrium, the analytical velocity and pressure at a depth $d$ below the free surface are given by
\begin{equation}
	\bm{v} = \bm{0},
\end{equation}
\begin{equation}
	p = \rho |\bm{g}| d.
\end{equation}
An overview of the hydrostatic pressure problem is shown in \fig{fig:hsp}, and the analysis conditions are listed in \tab{tbl:hsp}.
To investigate the relationship between the ratio of the element length $l_{\mathrm{cell}}$ to the initial particle spacing $d_0$ and the resulting accuracy and computational stability, numerical tests are performed with $d_0 = 0.5$, $1.0$, $2.0\times10^{-3}\:\mathrm{m}$ and $l_\mathrm{cell} = 0.25d_0$, $0.5d_0$, $1.0d_0$, $2.0d_0$, $4.0d_0$.
In addition, the following verification calculations vary the number of quadrature points (one-point and four-point quadrature) and the element arrangement (structured and unstructured meshes).
{
\tab{tbl:hsp_all_cases} lists the cases for the hydrostatic pressure analysis.
Hereafter, each case is identified by the ID listed in the first column.
}
All numerical results in this subsection are shown at $t = 10.0\,\mathrm{s}$, at which the solution is judged to have reached a sufficiently steady state.

\begin{table}[tbp]
	\centering
	\caption{Hydrostatic pressure problem: analysis conditions.}
	\label{tbl:hsp}
	\begin{tabular}{lc}
        \toprule
        Parameter & Value \\
		\midrule
		Time step size & $5.0 \times 10^{-4} \: \mathrm{s}$ \\
		Initial particle spacing $d_0$ & $0.5,\,1.0,\,2.0 \times 10^{-3} \: \mathrm{m}$ \\
		Cell length $l_{\mathrm{cell}}$ & $0.25d_0,\,0.5d_0,\,1.0d_0,\,2.0d_0,\,4.0d_0$ \\
		Smoothing length $h$ & $1.1 d_0$ \\
		Density $\rho$ & $1.0 \times 10^3 \: \mathrm{kg/m^3}$ \\
		Kinematic viscosity $\nu$ & $1.0 \times 10^{-4} \: \mathrm{m^2/s}$ \\
		Gravitational acceleration $|\bm{g}|$ & $9.8 \: \mathrm{m/s^2}$ \\
		Stabilization parameter $\alpha$ & $2.0 \times 10^{-2}$ \\
		\bottomrule
	\end{tabular}
\end{table}

\begin{table}[tbp]
	\centering
	\caption{Hydrostatic pressure problem: Summary of analysis cases.}
	\label{tbl:hsp_all_cases}
	\begin{tabular}{lcccc}
        \toprule
        Case ID & $d_0\:\mathrm{(m)}$ & $l_\mathrm{cell}\:\mathrm{(m)}$ & $N_\mathrm{ip}$ & Mesh \\
		\midrule
        1  & \multirow{5}{*}{$1.0\times10^{-3}$} & $0.25\times10^{-3}$ & \multirow{5}{*}{1} & \multirow{5}{*}{Structured} \\
        2  & & $0.5\times10^{-3}$ & & \\
        3  & & $1.0\times10^{-3}$ & & \\
        4  & & $2.0\times10^{-3}$ & & \\
        5  & & $4.0\times10^{-3}$ & & \\
        \midrule
        6  & \multirow{5}{*}{$1.0\times10^{-3}$} & $0.25\times10^{-3}$ & \multirow{5}{*}{4} & \multirow{5}{*}{Structured} \\
        7  & & $0.5\times10^{-3}$ & & \\
        8  & & $1.0\times10^{-3}$ & & \\
        9  & & $2.0\times10^{-3}$ & & \\
        10 & & $4.0\times10^{-3}$ & & \\
        \midrule
        11 & $2.0\times10^{-3}$ & $2.0\times10^{-3}$ & \multirow{2}{*}{1} & \multirow{2}{*}{Structured} \\
        12 & $0.5\times10^{-3}$ & $0.5\times10^{-3}$ & & \\
        \midrule
        13 & $2.0\times10^{-3}$ & $4.0\times10^{-3}$ & \multirow{2}{*}{4} & \multirow{2}{*}{Structured} \\
        14 & $0.5\times10^{-3}$ & $1.0\times10^{-3}$ & & \\
        \midrule
        15 & \multirow{2}{*}{$1.0\times10^{-3}$} & $1.0\times10^{-3}$  & 1 & \multirow{2}{*}{Unstructured} \\
        16 & & $2.0\times10^{-3}$ & 4 & \\
		\bottomrule
	\end{tabular}
\end{table}

\subsubsection{Comparison between the conventional and generalized GCB models}
\begin{figure}[tbp]
	\centering
	\includegraphics[width=\linewidth]{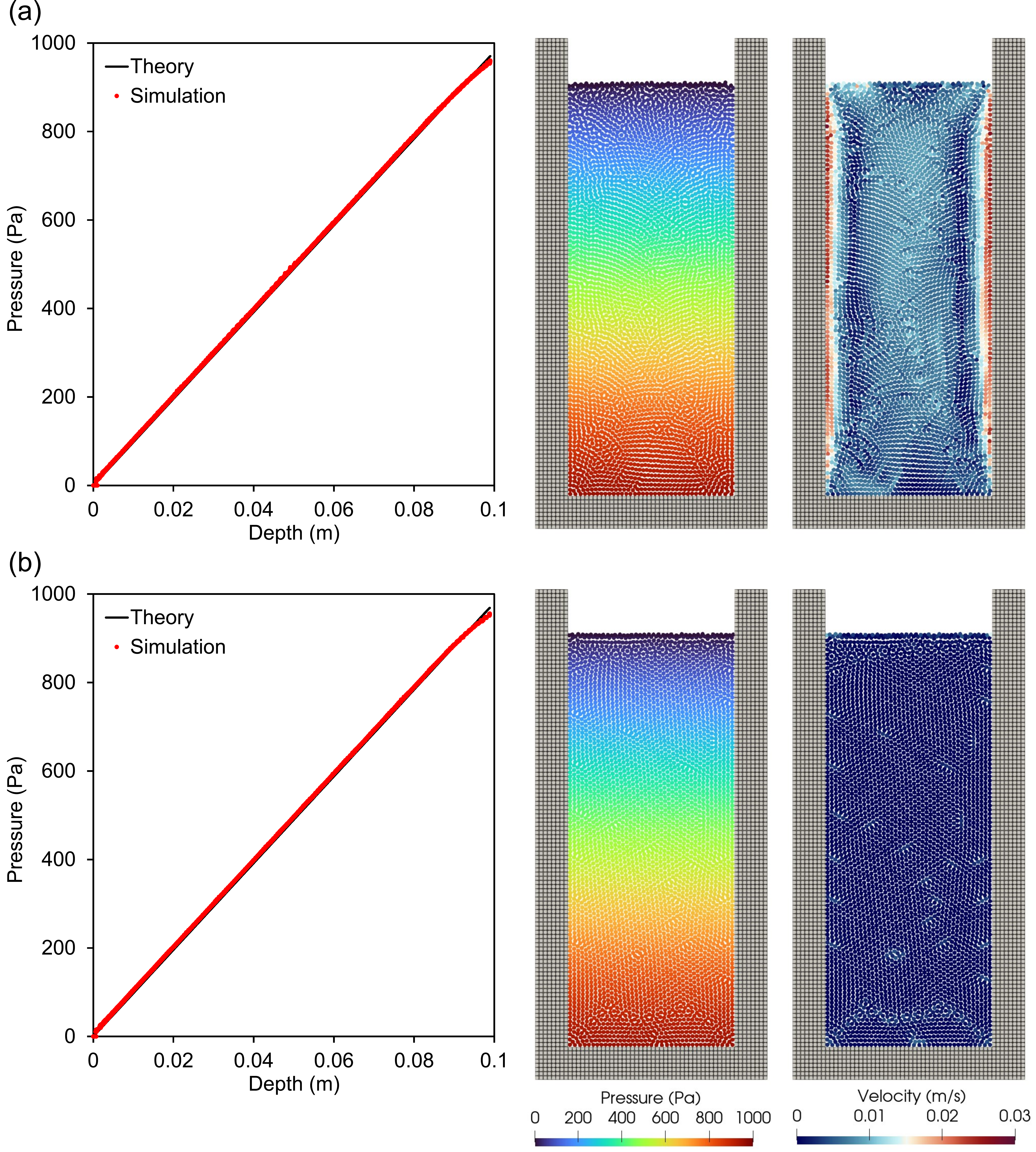} 
	\caption{Hydrostatic pressure problem: comparison between (a) the conventional GCB model and (b) the generalized GCB model for Case 3. The left, center, and right panels show the relationship between pressure and water depth for all particles, the pressure contours, and the velocity contours, respectively.}
	\label{fig:hsp_comp_scheme}
\end{figure}

First, to compare the conventional GCB model with the generalized GCB model proposed in this study, Case 3 is computed.
The results obtained with the conventional and proposed models are shown in \fig{fig:hsp_comp_scheme}.
Although the conventional model is formulated for explicit schemes, it is incorporated into the ISPH method in the present comparison so that both models can be evaluated under identical conditions.
Thus, by excluding the effects of the relative resolutions of the particles and elements, this test focuses solely on the boundary-condition treatment described in Section~\ref{sec:ggcbfgp} to compare the accuracy of the two models.
In the conventional GCB model, the velocity at each quadrature point is set to $\bm{0}$ in this boundary treatment, and the pressure at every quadrature point used in the pressure-gradient calculation is assigned the same value as that of the target fluid particle.

\begin{figure}[tbp]
	\centering
	\includegraphics[width=0.9\linewidth]{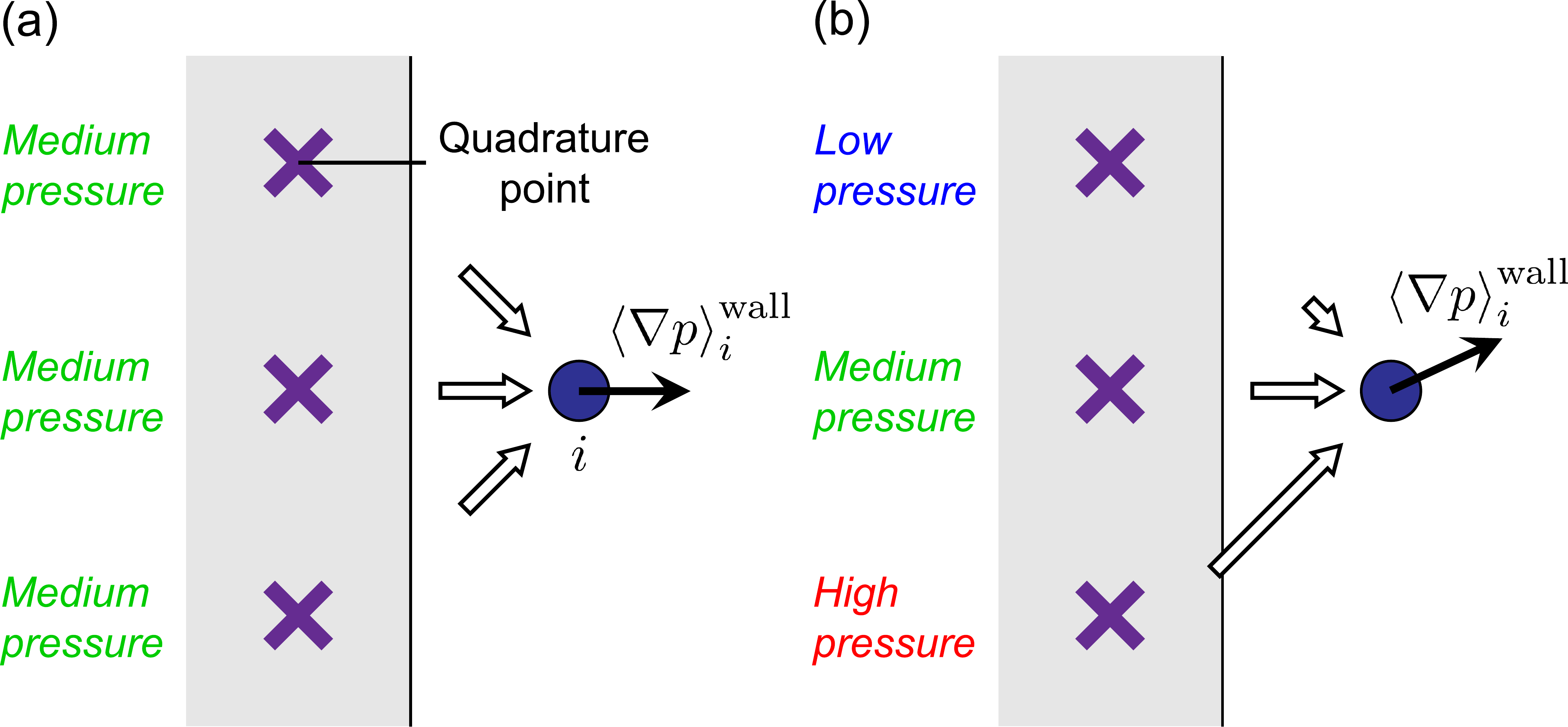} 
	\caption{Hydrostatic pressure problem: wall contribution to the pressure gradient in (a) the conventional and (b) generalized GCB models.}
	\label{fig:hsp_dis}
\end{figure}

Regarding the velocity field, the conventional GCB model produces nonphysical flow near the wall, whereas no such flow occurs with the proposed model, which appropriately reproduces the steady state.
In the conventional model, as illustrated in \fig{fig:hsp_dis}(a), the pressure of the target fluid particle $i$ is assigned to the quadrature points within its kernel support; consequently, the wall contribution to the pressure gradient has only a horizontal component and no vertical component.
Therefore, the pressure-gradient force acting on fluid particles near the wall does not balance gravity, and the downward motion of the fluid particles generates nonphysical flow.
By contrast, in the proposed method, the FGP model determines the pressure at the quadrature points by reflecting the fluid-particle pressure field, as shown in \fig{fig:hsp_dis}(b); consequently, the pressure-gradient force acting on fluid particles near the wall balances gravity, and no nonphysical flow occurs.
This verification demonstrates that the proposed boundary-condition treatment is effective, at least for static problems, and successfully resolves the nonphysical flow issue associated with the conventional model.

\subsubsection{Effects of relative spatial resolution and quadrature-point density} \label{sec:iptest}
Next, the effects of the initial particle spacing and the element length on the accuracy and computational stability of the generalized GCB model are examined.
Additionally, the influence of increasing the number of quadrature points, and hence their spatial density, is also investigated.

{
\tab{tbl:hsp_all_results} summarizes the computational results for Cases 1 through 10.
As observed from the table, when the quadrature points are more sparsely distributed than the fluid particles (Cases 4, 5, and 10), the computation immediately becomes unstable.
This behavior is considered to arise because the quadrature points responsible for the wall calculations are more sparsely distributed than the fluid particles within the kernel supports of the SPH differential operators.
Each quadrature point represents a portion of the volume and field-variable distribution of its element; when the quadrature points are sparse, a situation can occur in which an element lies within the kernel support while its quadrature point lies outside it, and the portion of the integration domain within the kernel support that should be included in the calculation is consequently omitted.
Conversely, even when $l_\mathrm{cell}$ is larger than $d_0$, as in Case 9, the computation remains stable if four-point quadrature is employed, because the spatial density of the quadrature points becomes comparable to that of the particles.
These results suggest that it is crucial to select an appropriate number of quadrature points depending on the ratio of the cell length to the particle spacing.
}

\begin{table}[tbp]
	\centering
	\caption{Hydrostatic pressure problem: Summary of $L_2$ errors of pressure and velocity for Cases 1--10.}
	\label{tbl:hsp_all_results}
	\begin{tabular}{lcc}
        \toprule
        Case ID & $L_2$ error of $p$ (\%) & $L_2$ error of $\lvert \bm{v} \rvert$ (\%) \\
		\midrule
        1  & 0.9990 & 0.0478 \\
        2  & 1.0287 & 0.0436 \\
        3  & 0.9944 & 0.0238 \\
        4  & \multicolumn{2}{c}{Diverged} \\
        5  & \multicolumn{2}{c}{Diverged} \\
        6  & \multicolumn{2}{c}{Diverged} \\
        7  & \multicolumn{2}{c}{Diverged} \\
        8  & 1.2454 & 0.0542 \\
        9  & 1.1013 & 0.0473 \\
        10 & \multicolumn{2}{c}{Diverged} \\
		\bottomrule
	\end{tabular}
\end{table}

\begin{figure}[tbp]
	\centering
	\includegraphics[width=\linewidth]{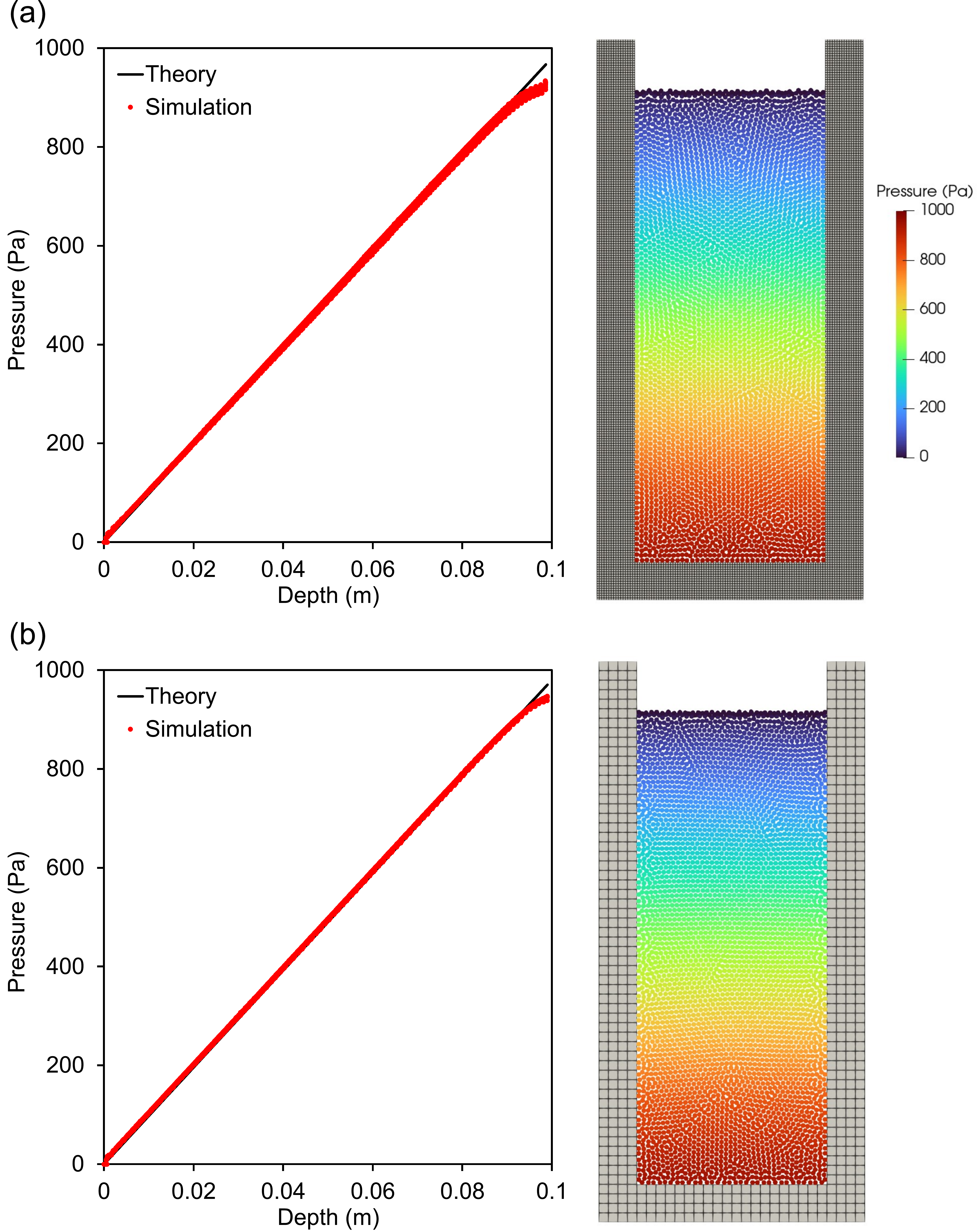} 
		\caption{Hydrostatic pressure problem: visualization of pressure distributions for (a) a fine cell with one-point quadrature (Case 2) and (b) a coarse cell with four-point quadrature (Case 9).}
	\label{fig:hsp_comp_length_nip}
\end{figure}

{
As shown in \fig{fig:hsp_comp_length_nip}, when the distribution density of the quadrature points is higher than that of the particles, or when four-point quadrature is used, a local degradation in the accuracy of the pressure solution is observed near the bottom.
If the quadrature points are more densely distributed than the particles or are arranged non-uniformly, a slight error arises in the spatial integration values between the wall and fluid domains.
This error is considered to adversely affect the evaluation of the pressure gradient, inducing nonphysical flow velocities for particles near the wall.
Specifically, to satisfy the conservation of momentum, this study employs a pressure gradient model based on the sum of pressures (\eq{eq:sph_gp}).
Because this model evaluates the absolute magnitude of the pressure gradient to be large, it reacts sensitively to minute integration errors, generating unbalanced forces.
Therefore, this problem can be resolved by applying a gradient model based on pressure differences, such as those presented by Randles and Libersky~\cite{Randles1996} and Bonet and Lok~\cite{Bonet1999}.
The computational results obtained using the difference-based pressure gradient model are presented in Appendix A.1.
}

\subsubsection{Convergence verification with respect to particle resolution}
\begin{figure}[tbp]
	\centering
	\includegraphics[width=\linewidth]{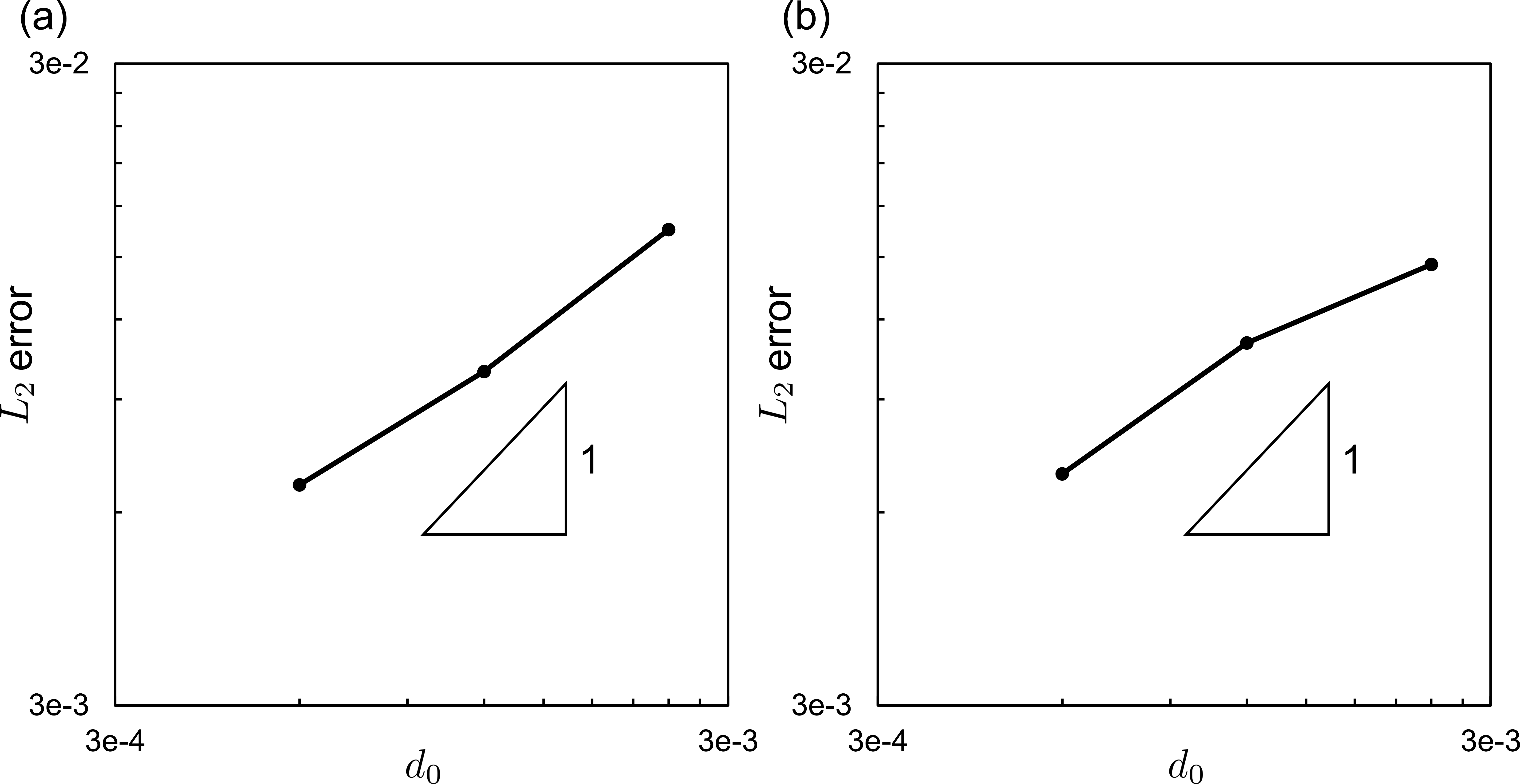} 
		\caption{Hydrostatic pressure problem: $L_2$ errors of pressure with respect to the fluid spatial resolution for (a) $l_\mathrm{cell} = d_0$ with one-point quadrature (Cases 3, 11, and 12) and (b) $l_\mathrm{cell} = 2d_0$ with four-point quadrature (Cases 9, 13, and 14).}
	\label{fig:hsp_converge}
\end{figure}

{
\fig{fig:hsp_converge} shows the $L_2$ error of pressure with respect to the spatial resolution of the particles when the ratio of the spatial densities between the fluid particles and quadrature points is fixed. 
The left panel shows the results for the case with $l_\mathrm{cell} = d_0$ and one-point quadrature (Cases 3, 11, and 12), and the right panel shows the results for the case with $l_\mathrm{cell} = 2d_0$ and four-point quadrature (Cases 9, 13, and 14).
In both the one-point and four-point quadrature cases, the error converges with respect to the spatial resolution.
Furthermore, at least within the range examined in this study, the convergence of the pressure solution was approximately first-order.
}

\subsubsection{Verification for unstructured meshes}
\begin{figure}[tbp]
	\centering
	\includegraphics[width=\linewidth]{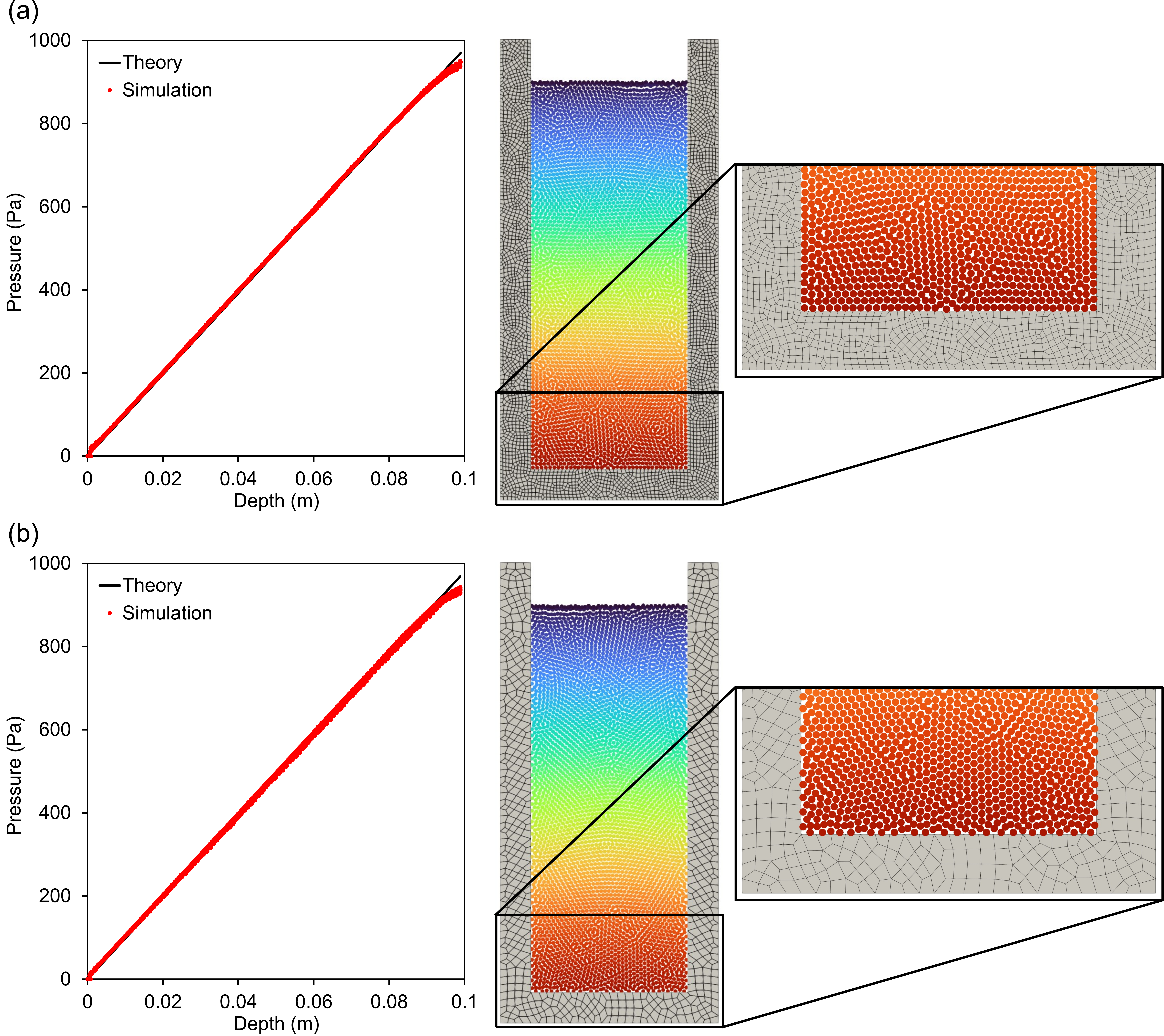} 
	\caption{Hydrostatic pressure problem: visualization of pressure distributions for (a) $l_{\mathrm{cell}}=d_0$ with one-point quadrature (Case 15) and (b) $l_{\mathrm{cell}}=2d_0$ with four-point quadrature (Case 16) using unstructured meshes.}
	\label{fig:hsp_comp_st}
\end{figure}

The preceding investigations assumed structured element arrangements; here, the applicability of the proposed method is examined for an unstructured mesh generated using Gmsh~\cite{geuzaine2009}.
\fig{fig:hsp_comp_st} shows the pressure distributions for Cases 15 and 16.
The results show that the numerical solutions obtained with the proposed method agree well with the analytical solution even for unstructured meshes.
Together with the discussion in Section~\ref{sec:iptest}, these results suggest that elements with poor aspect ratios may require different numbers of quadrature points along each coordinate direction of the reference domain; nevertheless, the present verification demonstrates that the proposed method can be applied with appropriate accuracy to unstructured meshes whose elements have comparatively favorable aspect ratios, as considered here.
A comprehensive examination of the applicability of the proposed method to meshes with nonuniform element lengths and to elements with various aspect ratios, as well as the development of a dynamic procedure for determining the number of quadrature points, falls outside the scope of this study and is left for future work.

%% file: doc/5_2_.tex
\begin{figure}[tbp]
	\centering
	\includegraphics[width=0.9\linewidth]{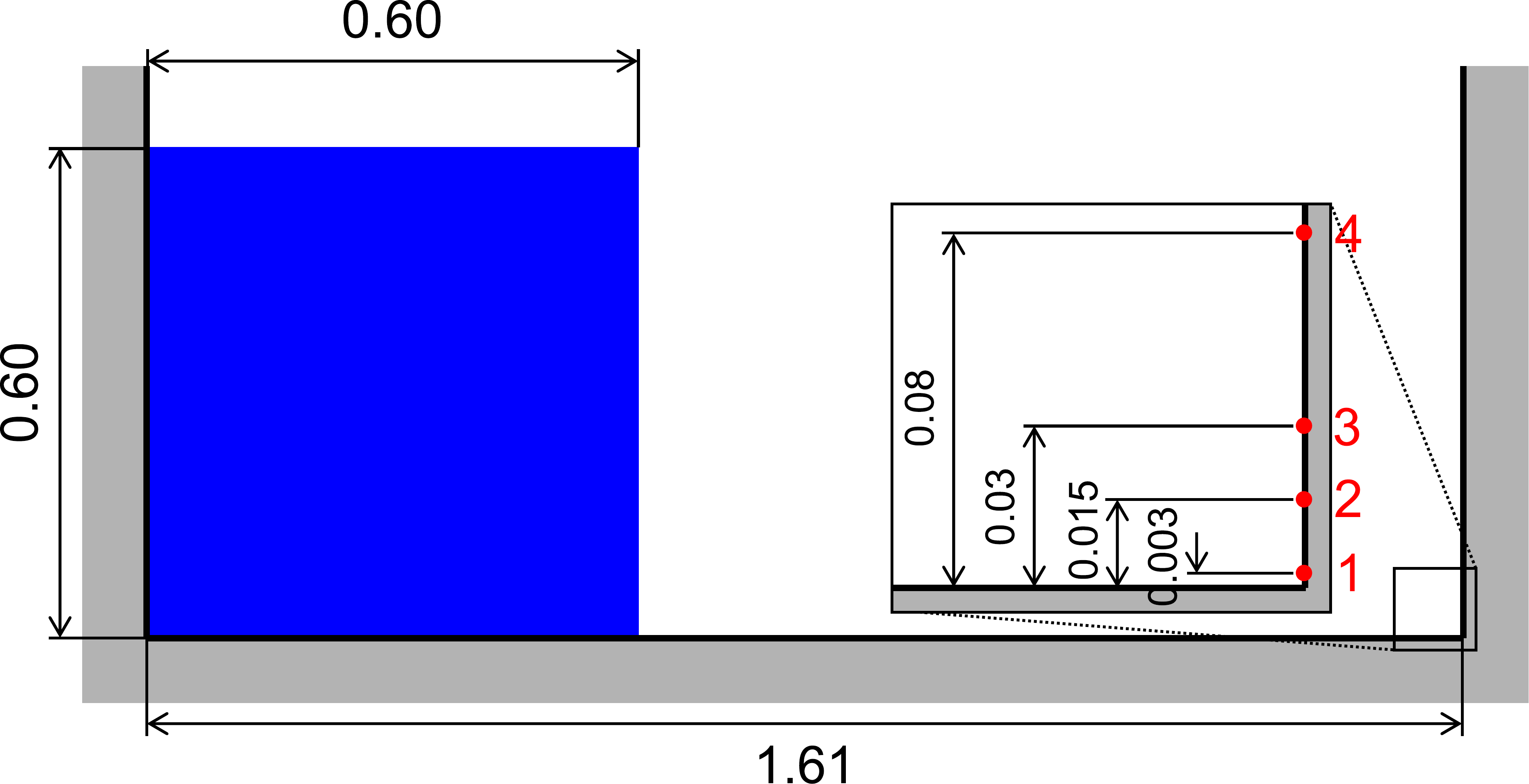} 
	\caption{Dam-break problem: initial configuration and positions of pressure sensors. All dimensions are in $\mathrm{(m)}$.}
	\label{fig:dam}
\end{figure}

To examine the validity of the generalized GCB model for a dynamic problem, the dam-break problem is computed using the experimental tank configuration of Lobovsk\'{y} et al.~\cite{Lobovsky2014}.
An overview of this verification problem is shown in \fig{fig:dam}.
The initial dimensions of the water column are $0.6\:\mathrm{m} \times 0.6\:\mathrm{m}$, and a total of four pressure measurement points are located on the wall opposite the water column.
The gate-opening motion is neglected in the simulation.
The analysis conditions are listed in \tab{tbl:dam}.
In addition, eight layers of cells are arranged in a structured manner outside the boundary, and one-point quadrature is used.
{
As discussed in Section 5.1, numerical instability occurs when the sum-based pressure gradient model is used and $l_\mathrm{cell}$ is set smaller than $d_0$.
The computational results obtained by applying the difference-based model with various cell lengths are presented in Appendix A.2.
}

\begin{table}[tbp]
	\centering
	\caption{Dam-break problem: analysis conditions.}
	\label{tbl:dam}
	\begin{tabular}{lc}
		\toprule
        Parameter & Value \\
        \midrule
		Time step size & $1.0 \times 10^{-4} \: \mathrm{s}$ \\
		Initial particle spacing $d_0$ & $2.0 \times 10^{-3} \: \mathrm{m}$ \\
		Cell length $l_{\mathrm{cell}}$ & $1.0 d_0$ \\
		Smoothing length $h$ & $1.1 d_0$ \\
		Density $\rho$ & $9.97 \times 10^2 \: \mathrm{kg/m^3}$ \\
		Kinematic viscosity $\nu$ & $8.9 \times 10^{-7} \: \mathrm{m^2/s}$ \\
		Gravitational acceleration $|\bm{g}|$ & $9.8 \: \mathrm{m/s^2}$ \\
		Stabilization parameter $\alpha$ & $3.0 \times 10^{-2}$ \\
		\bottomrule
	\end{tabular}
\end{table}

In the numerical analysis, the pressure at a measurement point is evaluated, with $\bm{x}_{\mathrm{e}}$ denoting the evaluation-point position, as the sum of the contributions from the fluid particles and the Gaussian quadrature points in the ghost cells:
\begin{equation} \label{eq:pressure_eval}
	p_{\mathrm{e}} = \sum_{j=1}^{N_{\mathrm{p}}} p_j W_{ej} V_j
	+ \sum_{c=1}^{N_{\mathrm{cell}}} \sum_{m=1}^{N_{\mathrm{ip}}} p_{(c,m)} W_{e(c,m)} H_m^{\mathrm{ip}} |\bm{J}_{(c,m)}|.
\end{equation}

\begin{figure}[tbp]
	\centering
	\includegraphics[width=1.0\linewidth]{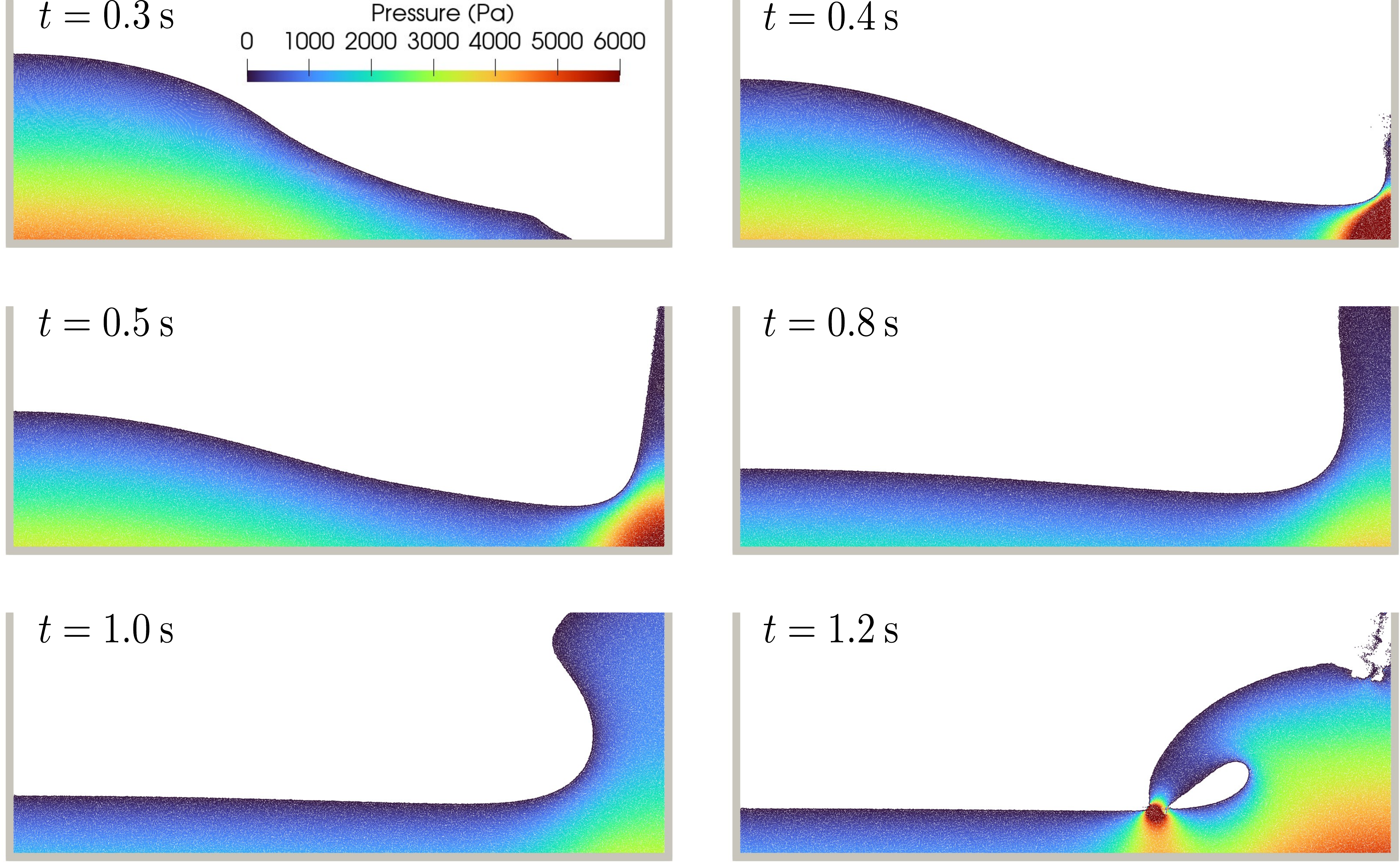} 
	\caption{Dam-break problem: visualization results.}
	\label{fig:dam_result_vis}
\end{figure}

\begin{figure}[tbp]
	\centering
	\includegraphics[width=1.0\linewidth]{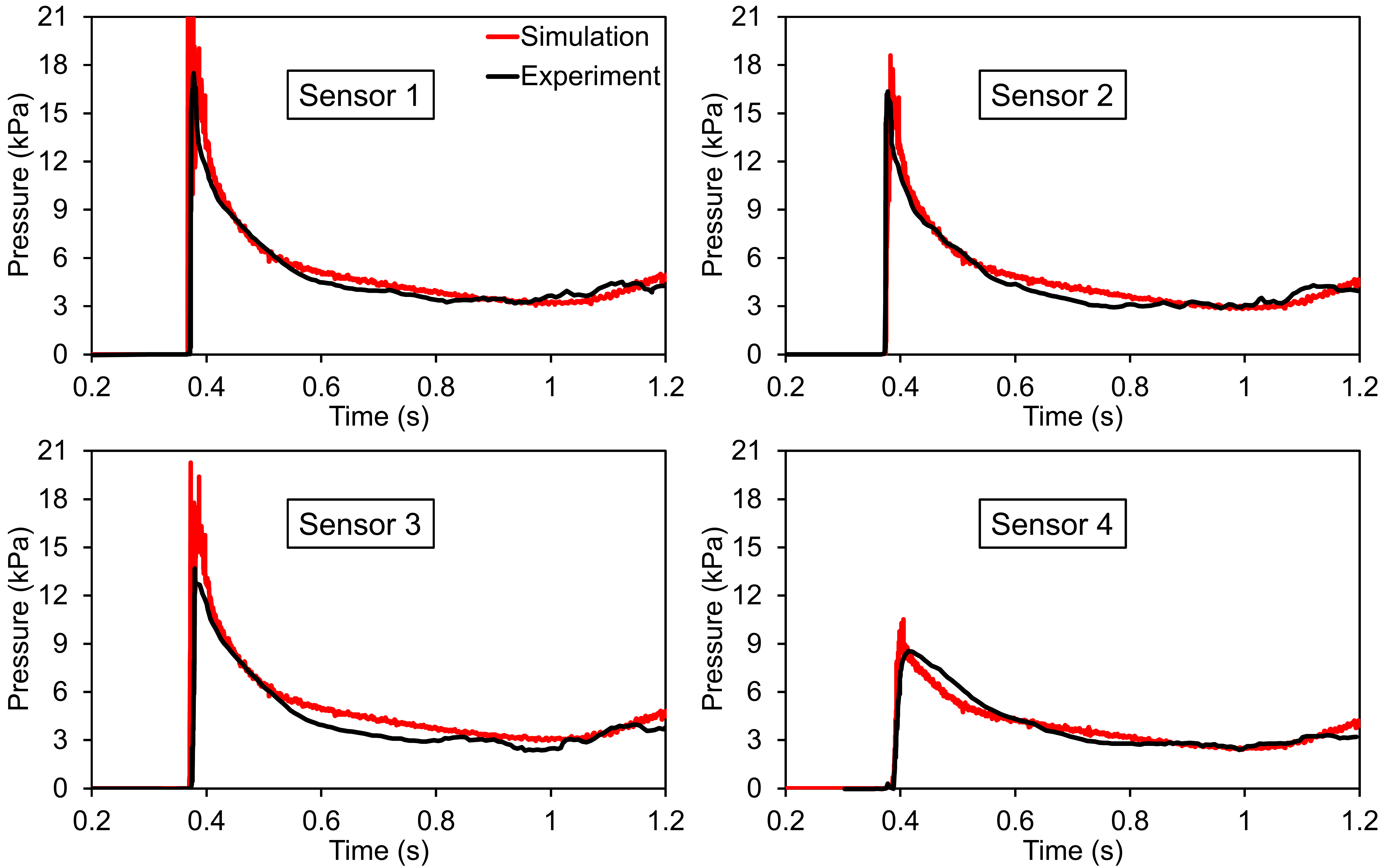} 
	\caption{Dam-break problem: time histories of pressure.}
	\label{fig:dam_result_pressure}
\end{figure}

\fig{fig:dam_result_vis} and \fig{fig:dam_result_pressure} show snapshots of the numerical results at selected times and the pressure histories at the respective pressure measurement points.
The collapsed water column flows toward the wall on which the sensors are installed and eventually impacts it.
Upon impact, a pressure peak is observed at Sensor 1, followed by pressure peaks at the other measurement points.
Subsequently, the upward flow along the wall reverses downward under gravity, and the pressure gradually decreases as the wave height and dynamic pressure decline.
{
Furthermore, as seen in the snapshot at $t = 1.2\:\mathrm{s}$ in \fig{fig:dam_result_vis}, even isolated splash particles generated by the plunging wave and subsequent impact are computed without penetrating the wall, owing to the artificial repulsive force introduced in \eq{eq: rep_force} (Section 3.5).
}
At all sensors, the numerical results obtained with the proposed method exhibit trends similar to the experimental results, indicating that the experimental behavior is reproduced appropriately.
Pressure oscillations occur in the numerical analysis during the impact of the water column; however, similar behavior has been observed in comparable analyses using ISPH, for example in~\cite{Asai2012}, and is therefore not considered to originate from the wall-boundary model and is not discussed further in this paper.

%% file: doc/5_3_.tex
\begin{figure}[tbp]
	\centering
	\includegraphics[width=0.9\linewidth]{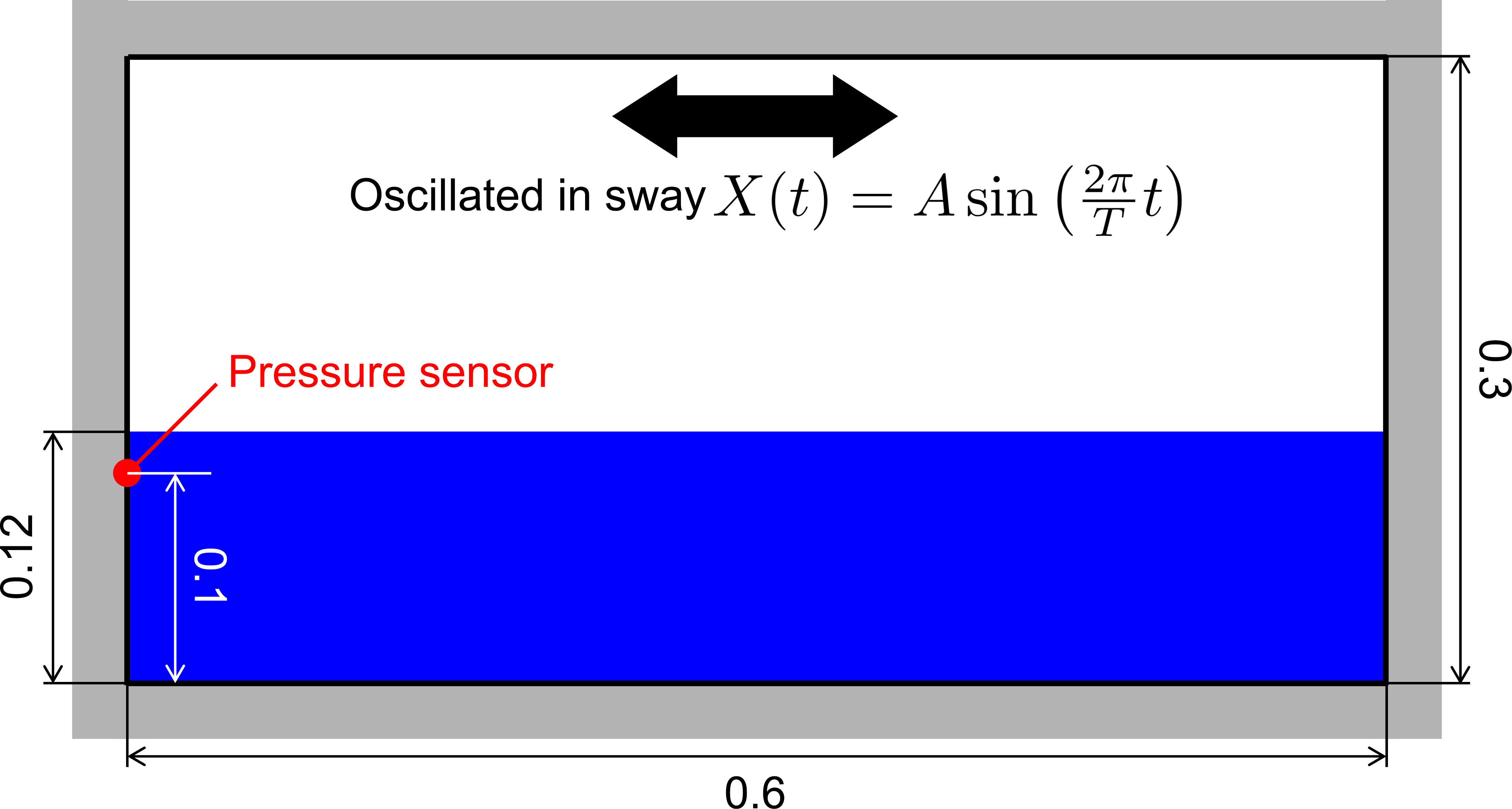} 
	\caption{Sloshing problem: initial configuration and position of the pressure sensor. All dimensions are in $\mathrm{(m)}$.}
	\label{fig:sloshing}
\end{figure}

{
To verify the stability of the ISPH method incorporating the generalized GCB model in long-duration dynamic simulations, the sloshing experiment conducted by Kishev et al.~\cite{Kishev2006} is reproduced.
A schematic of the target system is shown in \fig{fig:sloshing}.
The initial water depth is set to $0.12\;\mathrm{m}$, and a pressure measurement point is located on the left wall at a height of $0.1\:\mathrm{m}$.
The tank is oscillated horizontally according to $A\sin{\left(2\pi t / T\right)}$.
The amplitude $A$ and period $T$ of the oscillatory displacement are $0.05\:\mathrm{m}$ and $1.5\:\mathrm{s}$, respectively.
The analysis conditions are listed in \tab{tbl:sloshing}.
Square ghost cells are employed with one-point Gaussian quadrature. 
}

\begin{table}[tbp]
	\centering
	\caption{Sloshing problem: analysis conditions.}
	\label{tbl:sloshing}
	\begin{tabular}{lc}
		\toprule
        Parameter & Value \\
        \midrule
		Time step size & $2.0 \times 10^{-4} \: \mathrm{s}$ \\
		Initial particle spacing $d_0$ & $4.0 \times 10^{-3} \: \mathrm{m}$ \\
		Cell length $l_{\mathrm{cell}}$ & $1.0 d_0$ \\
		Smoothing length $h$ & $1.1 d_0$ \\
		Density $\rho$ & $1.0 \times 10^3 \: \mathrm{kg/m^3}$ \\
		Kinematic viscosity $\nu$ & $1.0 \times 10^{-6} \: \mathrm{m^2/s}$ \\
		Gravitational acceleration $|\bm{g}|$ & $9.8 \: \mathrm{m/s^2}$ \\
		Stabilization parameter $\alpha$ & $5.0 \times 10^{-3}$ \\
		\bottomrule
	\end{tabular}
\end{table}

\begin{figure}[tbp]
	\centering
	\includegraphics[width=1.0\linewidth]{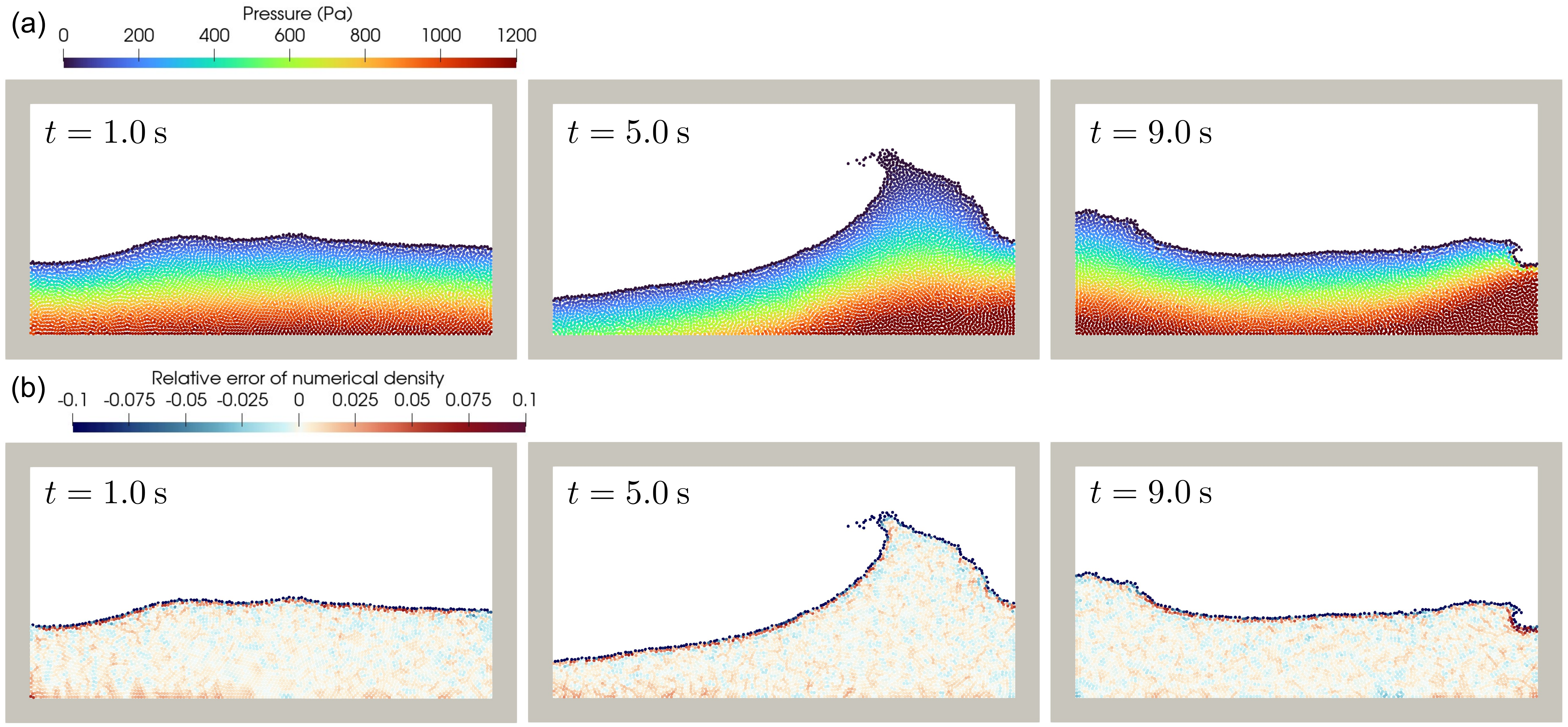} 
	\caption{Sloshing problem: visualization results of (a) pressure and (b) relative error of numerical density at $t = 1.0$, $5.0$, and $9.0\:\mathrm{s}$.}
	\label{fig:sloshing_vis}
\end{figure}

\begin{figure}[tbp]
	\centering
	\includegraphics[width=0.9\linewidth]{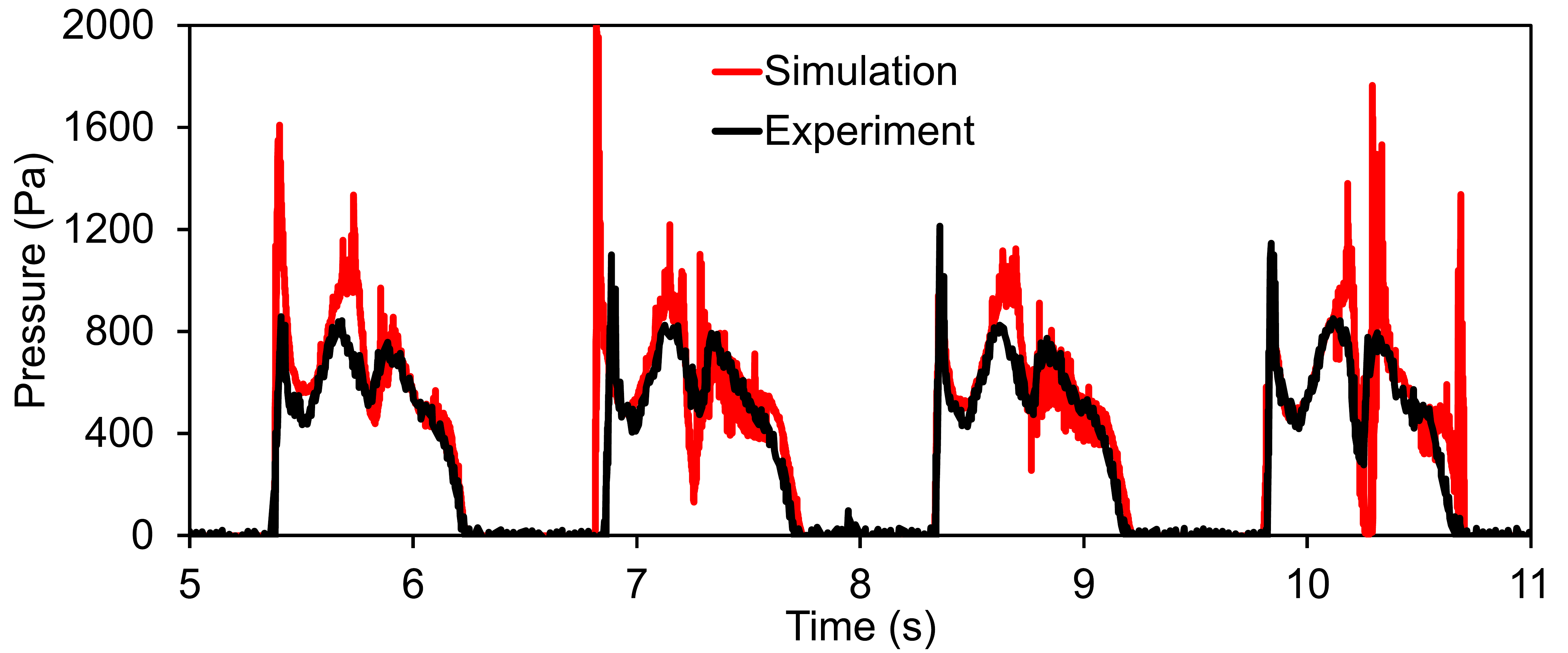} 
	\caption{Sloshing problem: time history of pressure at the sensor.}
	\label{fig:sloshing_pressure}
\end{figure}

{
\fig{fig:sloshing_vis} shows the contours of the pressure and the relative error of the numerical density $\left<\rho\right>$ obtained from the analysis.
The error of the numerical density is evaluated by $\left(\left<\rho\right>-\rho\right)/\rho$.
Although minor density errors occur locally, it is confirmed that the volume conservation law is well satisfied even in a long-duration dynamic problem.
\fig{fig:sloshing_pressure} shows the time history of the pressure at the sensor.
As in Section 5.2, the numerical value of pressure is evaluated using \eq{eq:pressure_eval}.
Similar to the analysis of the dam-break problem, although the reproducibility of the peak pressure generated when the water impacts the wall is relatively low, the phase and waveform characteristics of the periodically acting pressure are generally well captured.  
}

%% file: doc/5_4_.tex
In this subsection, a 2D falling circular cylinder problem is computed to examine the fundamental validity of the fluid--rigid-body coupling method described in Section~\ref{sec:fsi}.
In this problem, the cylinder is released from rest with its lower end in contact with the free surface, and the equilibrium position ultimately reached under the hydrodynamic force is compared with the theoretical solution.
Accordingly, this test provides a fundamental verification of the formulation used to evaluate the hydrodynamic force acting on the rigid body as the reaction to the wall contribution, as well as the consistency between the updates of the rigid-body position and the ghost cells.

\begin{figure}[tbp]
	\centering
	\includegraphics[width=1.0\linewidth]{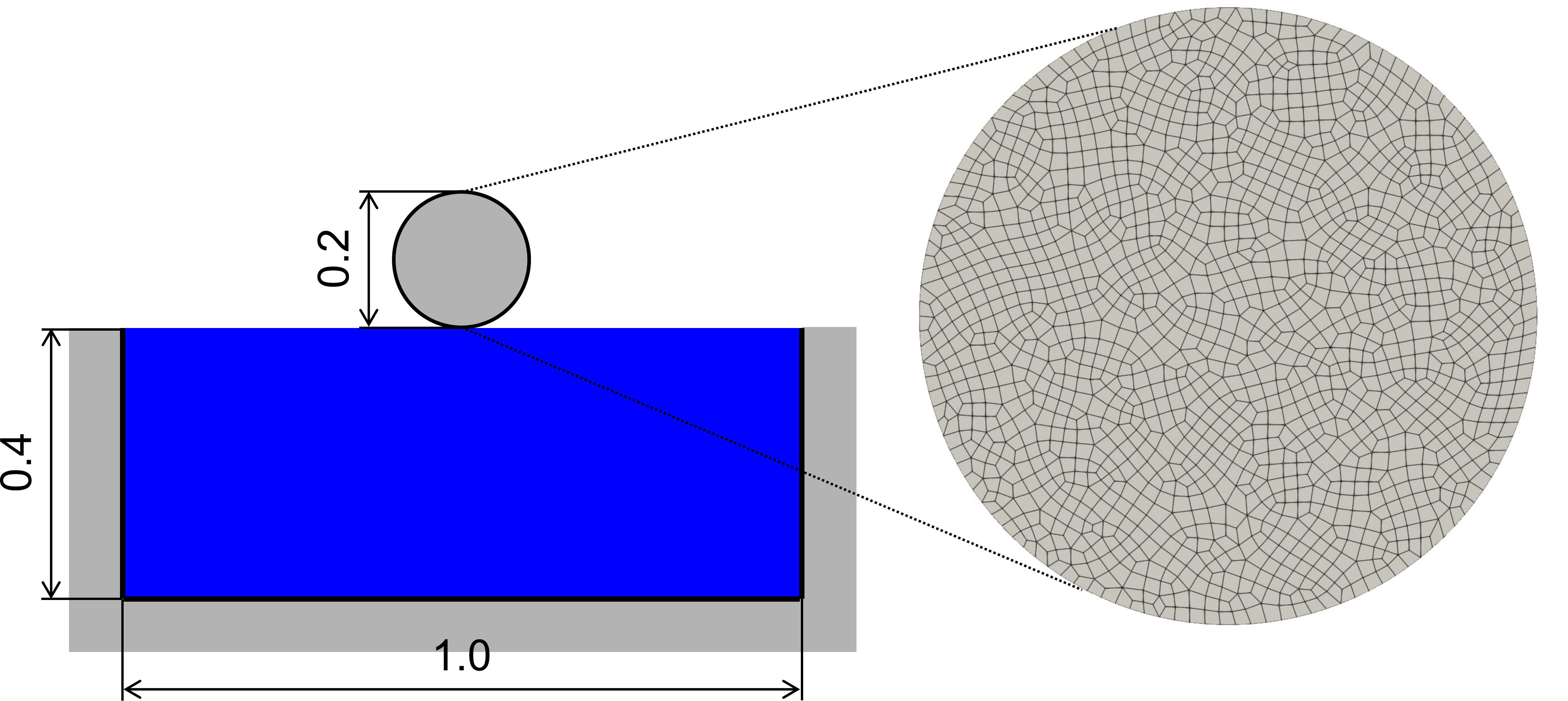}
	\caption{Falling circular cylinder problem: computational domain and quadrilateral ghost-cell mesh of the cylinder. All dimensions are in $\mathrm{(m)}$.}
	\label{fig:circle_configuration}
\end{figure}

\fig{fig:circle_configuration} shows an overview of the computational system and the quadrilateral mesh of the cylinder generated using Gmsh~\cite{geuzaine2009}.
Fluid particles are arranged on a regular lattice in the blue region of the tank, while the fixed walls surrounding the fluid and the interior of the cylinder are represented by ghost cells.
The size of the ghost cells in the cylinder is set to be comparable to $d_0$, and one-point quadrature is used.
The analysis conditions are listed in \tab{tbl:circle_condition}.
Because the rigid body is discretized using an unstructured mesh, $l_\mathrm{cell}$ represents the average cell length.
The kinematic viscosity is set higher than that used in the practical verification of free-surface flow so that the cylinder reaches equilibrium comparatively quickly.
The rigid-body density $\rho_{\mathrm{r}}$ is varied from $100$ to $900\,\mathrm{kg/m^3}$, and a total of nine cases are considered.

\begin{table}[tbp]
	\centering
	\caption{Falling circular cylinder problem: analysis conditions.}
	\label{tbl:circle_condition}
	\begin{tabular}{lc}
		\toprule
        Parameter & Value \\
        \midrule
		Time step size & $5.0 \times 10^{-4} \: \mathrm{s}$ \\
		Number of particles & $16{,}000$ \\
		Initial particle spacing $d_0$ & $5.0 \times 10^{-3} \: \mathrm{m}$ \\
        Cell length $l_{\mathrm{cell}}$ & $\approx 1.0 d_0$ \\
		Smoothing length $h$ & $1.1d_0$ \\
		Fluid density $\rho$ & $1.0 \times 10^{3} \: \mathrm{kg/m^3}$ \\
		Kinematic viscosity $\nu$ & $1.0 \times 10^{-3} \: \mathrm{m^2/s}$ \\
		Gravitational acceleration $|\bm{g}|$ & $9.8 \: \mathrm{m/s^2}$ \\
		Stabilization parameter $\alpha$ & $5.0 \times 10^{-3}$ \\
		\bottomrule
	\end{tabular}
\end{table}

\begin{figure}[tbp]
	\centering
	\includegraphics[width=1.0\linewidth]{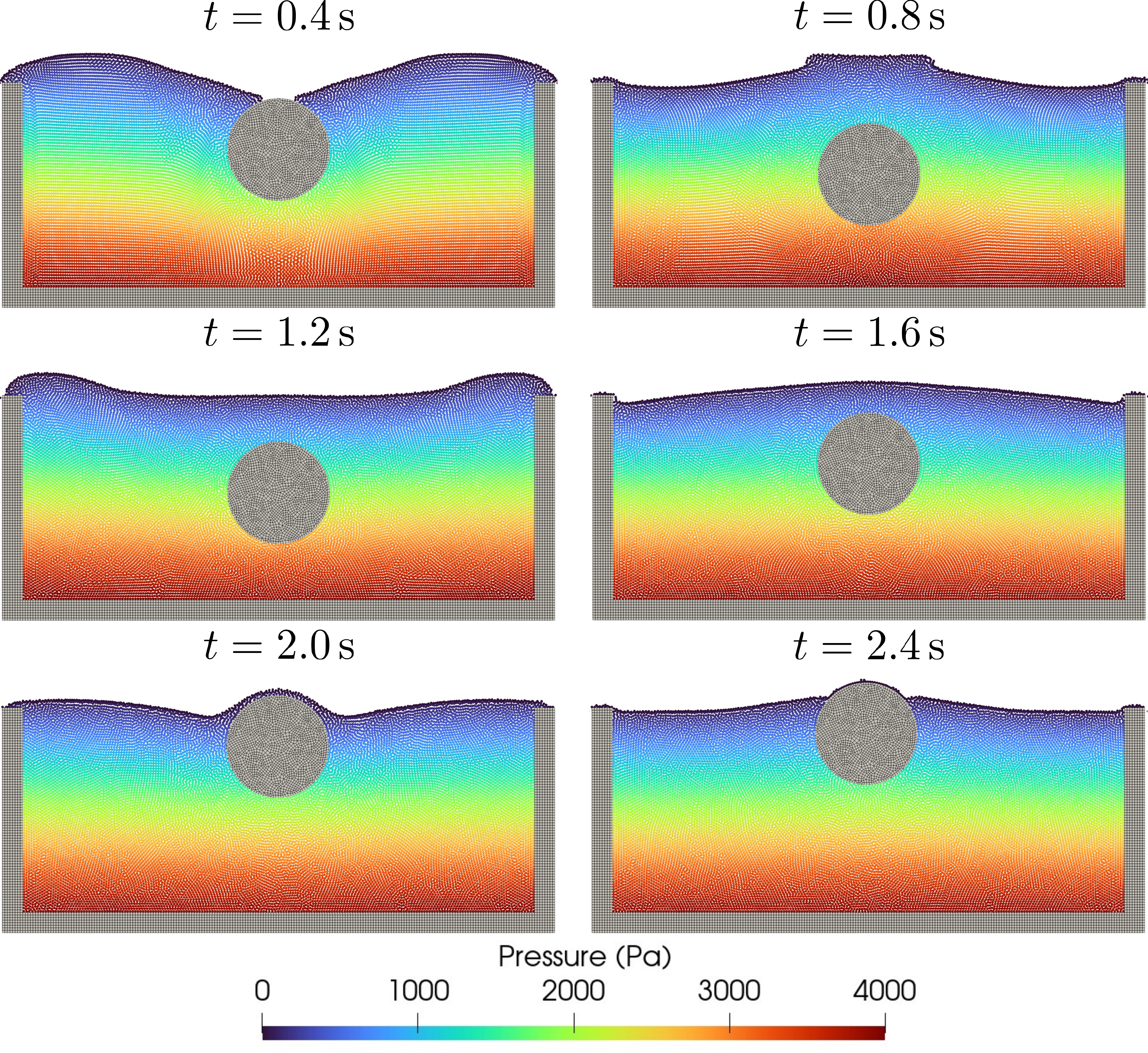}
	\caption{Falling circular cylinder problem: pressure-field snapshots for $\rho_{\mathrm{r}}=900\:\mathrm{kg/m^3}$.}
	\label{fig:circle_rho900_snapshots}
\end{figure}

\begin{figure}[tbp]
	\centering
	\includegraphics[width=1.0\linewidth]{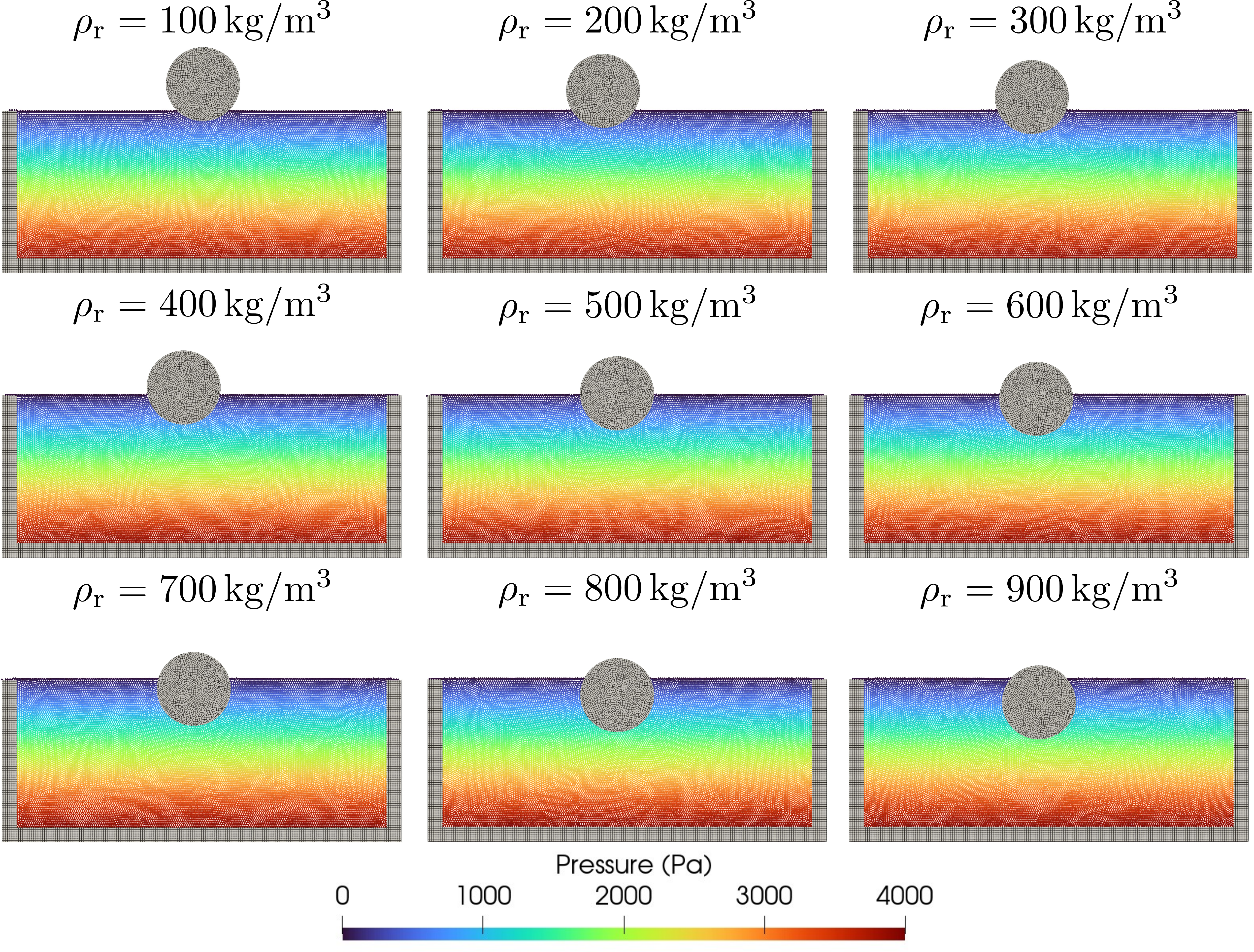}
	\caption{Falling circular cylinder problem: numerical results at $t=20\:\mathrm{s}$ for the various rigid-body densities.}
	\label{fig:circle_density_snapshots}
\end{figure}

\fig{fig:circle_rho900_snapshots} shows the pressure distribution and cylinder position at selected times for the case with $\rho_{\mathrm{r}}=900\:\mathrm{kg/m^3}$.
After initially descending into the water, the cylinder reverses to an upward motion because of buoyancy and approaches an equilibrium position near the free surface while undergoing damped oscillations.
This behavior is governed by the balance between the gravity acting on the rigid body and the buoyancy and hydrodynamic forces exerted by the fluid, demonstrating that the proposed method can update the rigid-body motion in conjunction with the fluid pressure field.
\fig{fig:circle_density_snapshots} shows visualization results at $t=20\:\mathrm{s}$ for the different rigid-body densities.
The immersion depth of the cylinder increases with increasing rigid-body density, and a qualitatively reasonable density dependence is obtained.

\begin{figure}[tbp]
	\centering
	\includegraphics[width=0.6\linewidth]{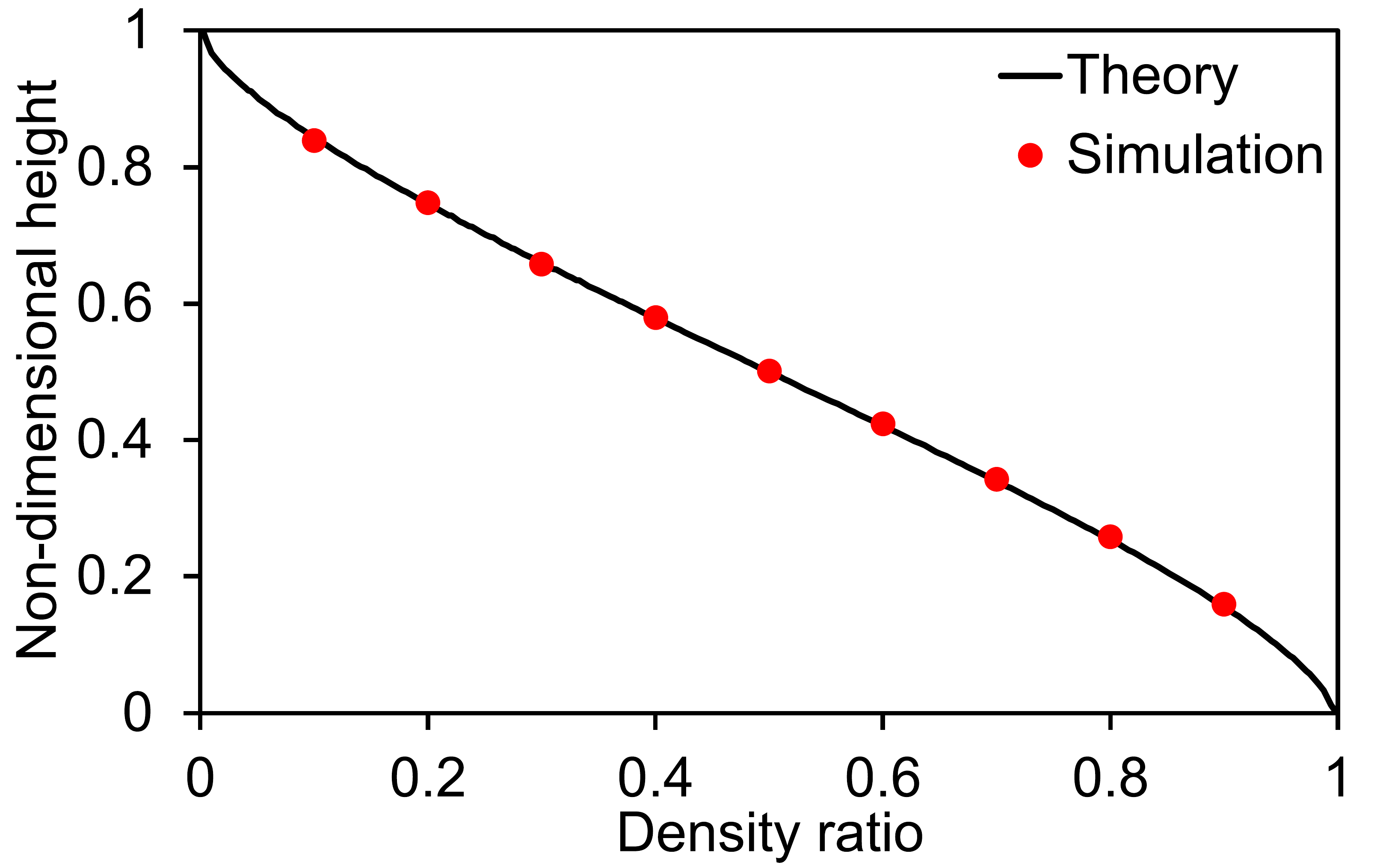}
	\caption{Falling circular cylinder problem: equilibrium position compared with the theoretical solution.}
	\label{fig:circle_equilibrium}
\end{figure}

Next, the equilibrium position of the cylinder is compared with the theoretical solution.
Let $D$ denote the cylinder diameter, and let $l$ denote the protruding height of the cylinder, defined as the vertical distance from the free surface to its upper end.
Considering the balance between the gravity acting on the cylinder and the buoyancy determined by the area of its submerged portion, the following equation is obtained for $0 \leq l \leq D$:
\begin{equation}
\label{eq:circle_theoretical}
    \frac{\rho_{\mathrm{r}}}{\rho} = \frac{1}{\pi} \left\{ \frac{\pi}{2} + \left(1 - \frac{l}{D}\right) \sqrt{\frac{l}{D} \left( 2 - \frac{l}{D} \right)} + \sin^{-1} \left(1 - \frac{l}{D} \right) \right\}.
\end{equation}
The above equation is nonlinear with respect to $l/D$, and the theoretical solution is obtained using Newton's method for each prescribed value of $\rho_{\mathrm{r}}/\rho$.
For the numerical result, the free-surface elevation is defined as the position $d_0/2$ vertically above the highest fluid particle, and $l$ is evaluated as the height by which the highest node among the ghost cells constituting the cylinder lies above the free surface.
\fig{fig:circle_equilibrium} shows the relationship between the cylinder density ratio $\rho_{\mathrm{r}}/\rho$ and the nondimensional equilibrium position $l/D$.
In addition, \tab{tbl:circle_equilibrium} presents the theoretical solutions, numerical results, and relative errors.
For every density, the numerical result agrees well with the theoretical solution, and the maximum relative error is limited to $2.23\:\%$.
These results confirm that the proposed reaction-based hydrodynamic force evaluation and rigid-body position update reproduce the balance between buoyancy and gravity in still water with good accuracy.

\begin{table}[tbp]
    \centering
	\caption{Falling circular cylinder problem: theoretical and numerical equilibrium positions.}
	\label{tbl:circle_equilibrium}
	\begin{tabular}{cccc}
		\toprule
		$\frac{\rho_\mathrm{r}}{\rho}$ & $\frac{l}{D}$ (theory) & $\frac{l}{D}$ (simulation) & Relative error $(\%)$ \\
		\midrule
        0.1 & 0.83963 & 0.84353 & 0.462 \\
        0.2 & 0.74837 & 0.74593 & 0.326 \\
        0.3 & 0.65855 & 0.65985 & 0.196 \\
        0.4 & 0.57984 & 0.57887 & 0.168 \\
        0.5 & 0.50170 & 0.50000 & 0.340 \\
        0.6 & 0.42393 & 0.42113 & 0.665 \\
        0.7 & 0.34254 & 0.34016 & 0.701 \\
        0.8 & 0.25837 & 0.25407 & 1.690 \\
        0.9 & 0.15997 & 0.15648 & 2.230 \\
		\bottomrule
	\end{tabular}
\end{table}

%% file: doc/5_5_.tex
Finally, the wedge water-entry experiment conducted by Barjasteh et al.~\cite{Barjasteh2016} is reproduced as a fluid--rigid-body coupling problem involving more severe dynamic pressure loading.
In this problem, a vertically downward initial velocity is imposed on a wedge initially in contact with the free surface, and the pressure histories acting on the wedge surface during water entry are compared with the experimental results.
Whereas the falling circular cylinder problem primarily verifies static buoyancy equilibrium, the present problem examines a highly transient pressure distribution.

\begin{figure}[tbp]
	\centering
	\includegraphics[width=1.0\linewidth]{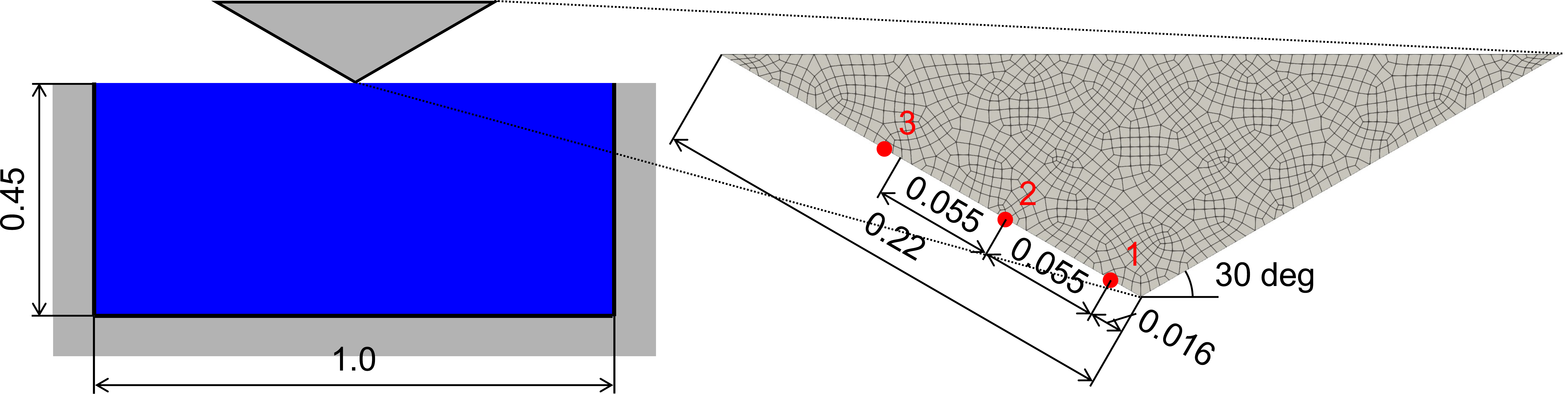}
	\caption{Wedge water-entry problem: computational domain, wedge geometry, and ghost-cell mesh. All length dimensions are in ($\mathrm{m}$).}
	\label{fig:wedge_configuration}
\end{figure}

\fig{fig:wedge_configuration} shows the computational configuration and the quadrilateral mesh generated using Gmsh~\cite{geuzaine2009}.
Fluid particles are arranged on a regular lattice in the fluid domain, while the fixed walls surrounding the fluid and the interior of the wedge are represented by ghost cells.
Three pressure sensors are placed on the left surface of the wedge for comparison with the experimental results.
The analysis conditions are listed in \tab{tbl:wedge_condition}.
Note that $l_\mathrm{cell}$ indicates the average target cell length for the unstructured mesh, and one-point quadrature is used.
Because the wedge was dropped from a height of $0.5\:\mathrm{m}$ in the experiment, an initial velocity of $\sqrt{2 \times 9.8 \times 0.5} \simeq 3.13\:\mathrm{m/s}$ is imposed vertically downward in the analysis based on conservation of energy.
In addition, $\rho_{\mathrm{r}}=4562.138\:\mathrm{kg/m^3}$ is specified as the rigid-body density corresponding to the experimental mass of $44\:\mathrm{kg}$.

\begin{table}[tbp]
	\centering
	\caption{Wedge water-entry problem: analysis conditions.}
	\label{tbl:wedge_condition}
	\begin{tabular}{lc}
		\toprule
        Parameter & Value \\
        \midrule
		Time step size & $1.0 \times 10^{-4} \: \mathrm{s}$ \\
		Number of particles & $10{,}000$ \\
		Initial particle spacing $d_0$ & $5.0 \times 10^{-3} \: \mathrm{m}$ \\
        Cell length $l_{\mathrm{cell}}$ & $\approx 1.0 d_0$ \\
		Smoothing length $h$ & $1.1d_0$ \\
		Fluid density $\rho$ & $1.0 \times 10^3 \: \mathrm{kg/m^3}$ \\
		Rigid-body density $\rho_{\mathrm{r}}$ & $4562.138 \: \mathrm{kg/m^3}$ \\
		Kinematic viscosity $\nu$ & $1.0 \times 10^{-6} \: \mathrm{m^2/s}$ \\
		Gravitational acceleration $|\bm{g}|$ & $9.8 \: \mathrm{m/s^2}$ \\
		Stabilization parameter $\alpha$ & $5.0 \times 10^{-3}$ \\
		Initial downward velocity & $3.13 \: \mathrm{m/s}$ \\
		\bottomrule
	\end{tabular}
\end{table}

\begin{figure}[tbp]
	\centering
	\includegraphics[width=1.0\linewidth]{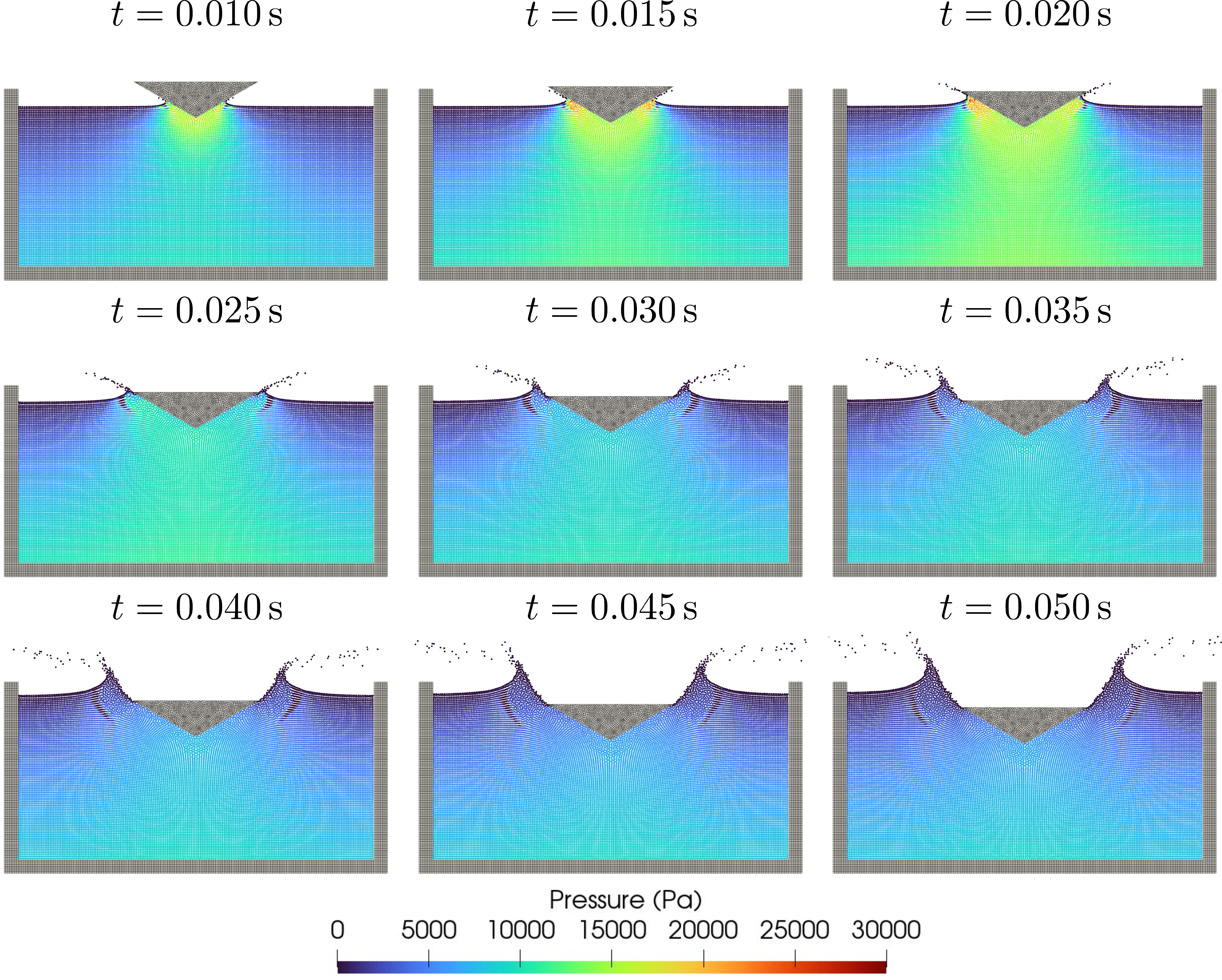}
	\caption{Wedge water-entry problem: pressure-field snapshots.}
	\label{fig:wedge_snapshots}
\end{figure}

\begin{figure}[tbp]
	\centering
	\includegraphics[width=0.6\linewidth]{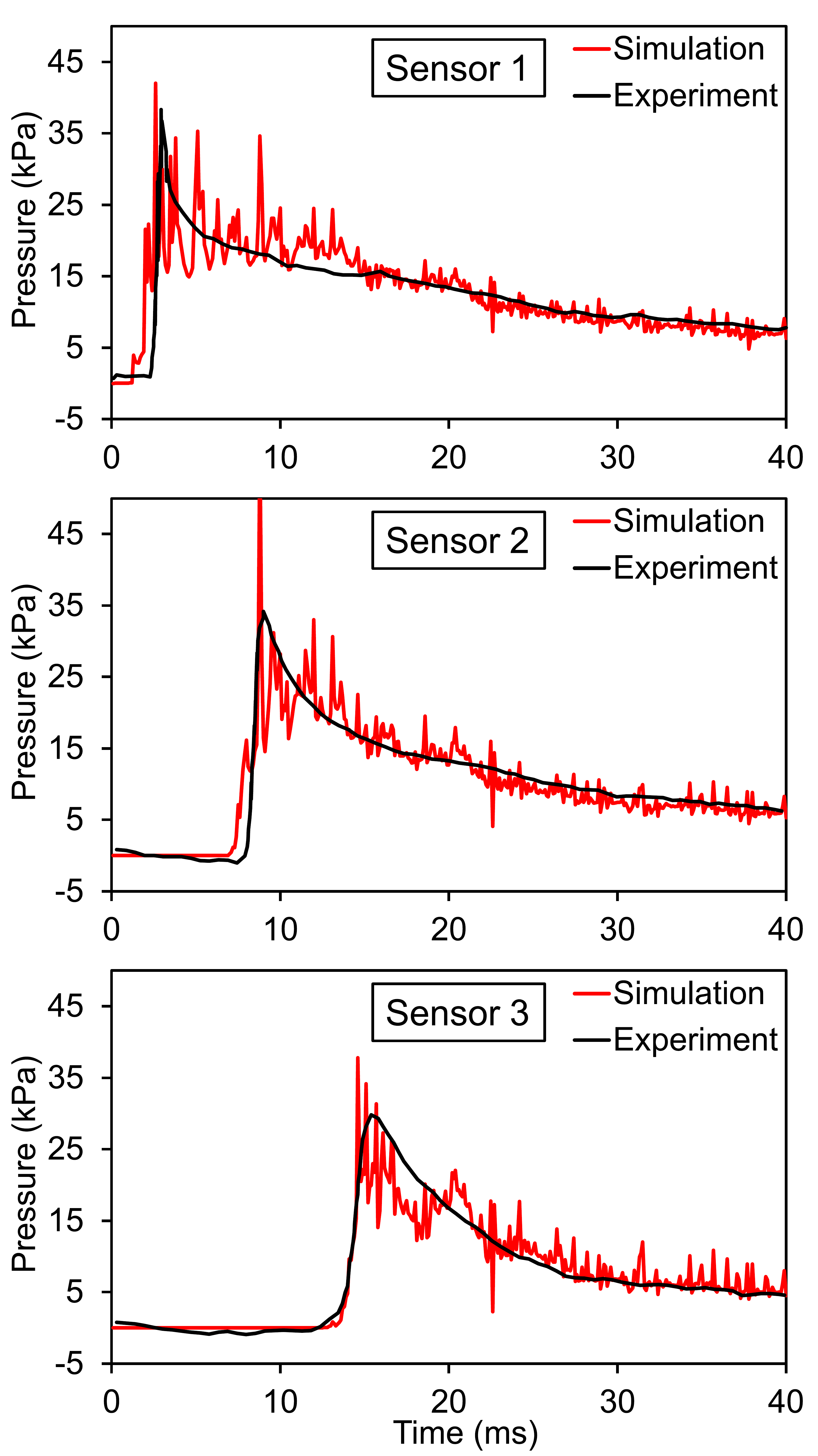}
	\caption{Wedge water-entry problem: time histories of pressure at the sensors compared with the experimental results~\cite{Barjasteh2016}.}
	\label{fig:wedge_pressure_histories}
\end{figure}

\fig{fig:wedge_snapshots} shows snapshots of the pressure distribution and wedge position.
High pressures develop near the wedge surfaces, and the entry of the wedge into the water is reproduced.
In addition, \fig{fig:wedge_pressure_histories} compares the pressure histories at the three sensors on the wedge surface with the experimental results~\cite{Barjasteh2016}.
As in \eq{eq:pressure_eval}, the pressure at each sensor is evaluated as the sum of the contributions from the fluid particles and the Gaussian quadrature points in the ghost cells.
Because the definition of the reference time is not specified in the referenced experiment, the experimental and numerical results are compared after aligning the times at which the pressure peak occurs at Sensor 1.
Although oscillations occur in the numerical solution at each sensor, the timing of the pressure rise, the post-peak decay trend, and the overall waveform agree well with the experimental results.
These results demonstrate that the generalized GCB model successfully evaluates the unsteady pressure field during water entry in conjunction with the rigid-body motion.
{
In \fig{fig:wedge_snapshots}, the propagation of a zero-pressure field to the left and right is observed as the wedge enters the water.
This nonphysical phenomenon, in which the zero-pressure region propagates, is considered to have occurred because the process of replacing negative pressures with zero altered the original phenomenon, which should have been the propagation of vortices generated by the water entry of the wedge.
}
Possible causes of the pressure oscillations include the particle distribution near the free surface, differences between the experimental system and its 2D representation, and local kernel-interpolation errors around the pressure sensors; a detailed investigation of these factors is left for future work.

%% file: doc/6_.tex
This study introduced the following improvements and generalizations to the GCB model, in which the wall contribution is evaluated by discretizing the wall domain into ghost cells and applying Gaussian quadrature:
\begin{enumerate}[label=(\Alph*)]
    \item Whereas the original GCB model is limited to one-point quadrature, the proposed model generalizes it to higher-order Gaussian quadrature with an arbitrary number of quadrature points in each coordinate direction.
    \item Whereas the pressure and velocity extrapolated to the quadrature points in the original model do not satisfy the wall boundary conditions, the proposed model establishes a physically consistent boundary-condition treatment based on the FGP model.
    \item Finally, a fluid--rigid-body coupling method is formulated based on the reaction force associated with the wall contribution evaluated by the generalized GCB model.
\end{enumerate}

Verification using the hydrostatic pressure problem demonstrated that the FGP-based extrapolation of pressure and velocity to the quadrature points suppresses the nonphysical velocity field near the wall observed with the original GCB model.
It was also confirmed that, when the element length exceeds the particle spacing, one-point quadrature becomes unstable due to an insufficient density of quadrature points, whereas increasing the quadrature-point density through four-point quadrature yields a stable and physically reasonable pressure distribution.
The proposed model also produces results in good agreement with the analytical solution even for unstructured ghost-cell meshes.

Verification using the dam-break and sloshing problems confirmed that the proposed model reproduces the overall pressure response observed in the experiments.
Pressure oscillations occurred during the impact of the water; however, similar behavior has been reported in analyses using ISPH and is therefore considered to arise from inherent limitations in the pressure-evaluation accuracy of particle methods.

To validate the fluid--rigid-body coupling method, the falling circular cylinder problem was first considered, and the equilibrium position determined by the balance between buoyancy and gravity was compared with the analytical solution.
For the nine cases in which the rigid-body density was varied over the range $100$--$900\:\mathrm{kg/m^3}$, the numerical results agreed well with the analytical solution, with a maximum relative error of $2.23\:\%$.
In addition, the pressure histories on the wedge surface were compared with the experimental measurements in the wedge water-entry problem.
Although pressure oscillations occurred, the timing of the pressure peak and the post-peak decay trend agreed well with the experimental results.
These results confirmed that the fluid--rigid-body coupling method based on the generalized GCB model can be used to analyze both static and dynamic problems with good accuracy.

Although the present study focused on 2D problems, the proposed generalized GCB model is inherently extensible to three dimensions.
Such an extension would involve the use of hexahedral ghost cells and corresponding 3D Gaussian quadrature rules. 
Future work will also focus on the automatic selection of the quadrature-point density and the extension of the proposed model to fluid--structure interactions.

%% file: doc/app_.tex
{
In the appendix, as mentioned in Section 5.1.2, we present the computational results obtained using the difference-based pressure gradient model~\cite{Randles1996, Bonet1999} instead of \eq{eq:sph_gp}.
This gradient model, derived based on a first-order Taylor expansion, is given by the following equation:
\begin{equation}
    \label{eq:isph_cgp}
    \left< \nabla p \right>_i = \bm{M}_i^{-1}
    \sum_{j=1}^{N_\mathrm{p}} \frac{m_j}{\rho_j}
    \left(p_j - p_i\right) \nabla W_{ij},
\end{equation}
where $\bm{M}_i$ is the correction matrix defined in \eq{eq:fsd_m}.
Conventionally, this model has been considered inferior to the model in \eq{eq:sph_gp} in terms of stability; however, recent studies have shown that combining it with particle shifting techniques can achieve both high stability and accuracy.
}

\subsection{Hydrostatic pressure problem} \label{app1}
\input{doc/app_1_.tex}

\subsection{Dam-break problem} \label{app2}
\input{doc/app_2_.tex}

%% file: doc/app_1_.tex
\begin{figure}[tbp]
	\centering
	\includegraphics[width=\linewidth]{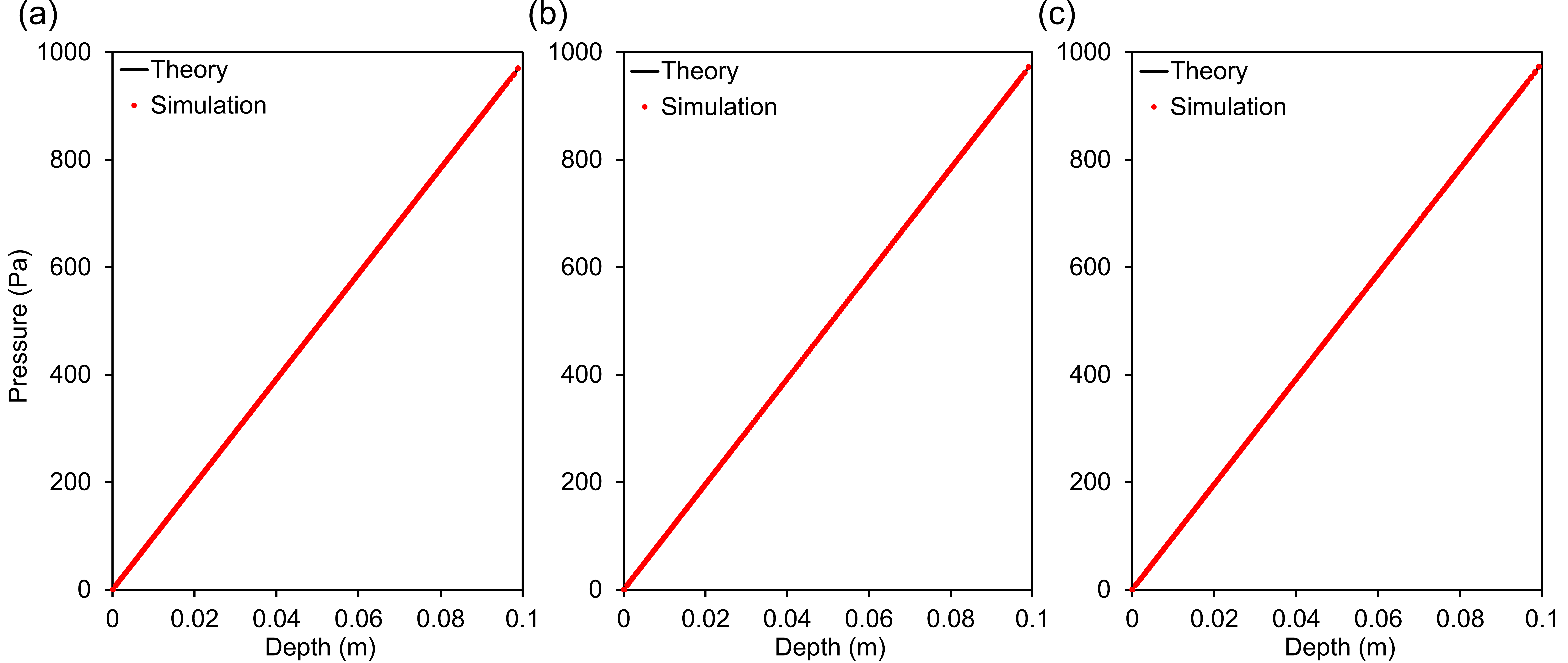} 
		\caption{Hydrostatic pressure problem: pressure distributions for (a) a standard cell with one-point quadrature (Case 3), (b) a fine cell with one-point quadrature (Case 2), and (c) a coarse cell with four-point quadrature (Case 9) obtained by the difference-based pressure gradient model.}
	\label{fig:hsp_comp_length_nip_dif}
\end{figure}

{
\tab{tbl:hsp_all_results_dif} summarizes the computational results obtained using the difference-based pressure gradient model.
The computational conditions and corresponding case IDs are identical to those in \tab{tbl:hsp_all_cases} in Section 5.1.
For Cases 1, 2, 3, 8, and 9, the \(L_2\) errors in pressure and velocity are lower than the corresponding errors for the sum-based model in \tab{tbl:hsp_all_results}.
Furthermore, Cases 6 and 7, which diverged under the sum-based model, remained stable when the difference-based model was applied.
\fig{fig:hsp_comp_length_nip_dif} shows the pressure distributions for Cases 3, 2, and 9 computed using the difference-based model.
As shown in the figure, the local errors near the bottom observed with the sum-based model have been successfully resolved.
}

\begin{table}[tbp]
	\centering
	\caption{Hydrostatic pressure problem: Summary of $L_2$ errors of pressure and velocity for Cases 1--10 with the difference-based pressure gradient model.}
	\label{tbl:hsp_all_results_dif}
	\begin{tabular}{lcc}
        \toprule
        Case ID & $L_2$ error of $p$ (\%) & $L_2$ error of $\lvert \bm{v} \rvert$ (\%) \\
		\midrule
        1  & 0.0259 & 0.0141 \\
        2  & 0.0503 & 0.0109 \\
        3  & 0.0214 & 0.0070 \\
        4  & \multicolumn{2}{c}{Diverged} \\
        5  & \multicolumn{2}{c}{Diverged} \\
        6  & 0.0259 & 0.0116 \\
        7  & 0.0259 & 0.0114 \\
        8  & 0.0393 & 0.0119 \\
        9  & 0.0288 & 0.0104 \\
        10 & \multicolumn{2}{c}{Diverged} \\
		\bottomrule
	\end{tabular}
\end{table}

%% file: doc/app_2_.tex
\begin{figure}[tbp]
	\centering
	\includegraphics[width=1.0\linewidth]{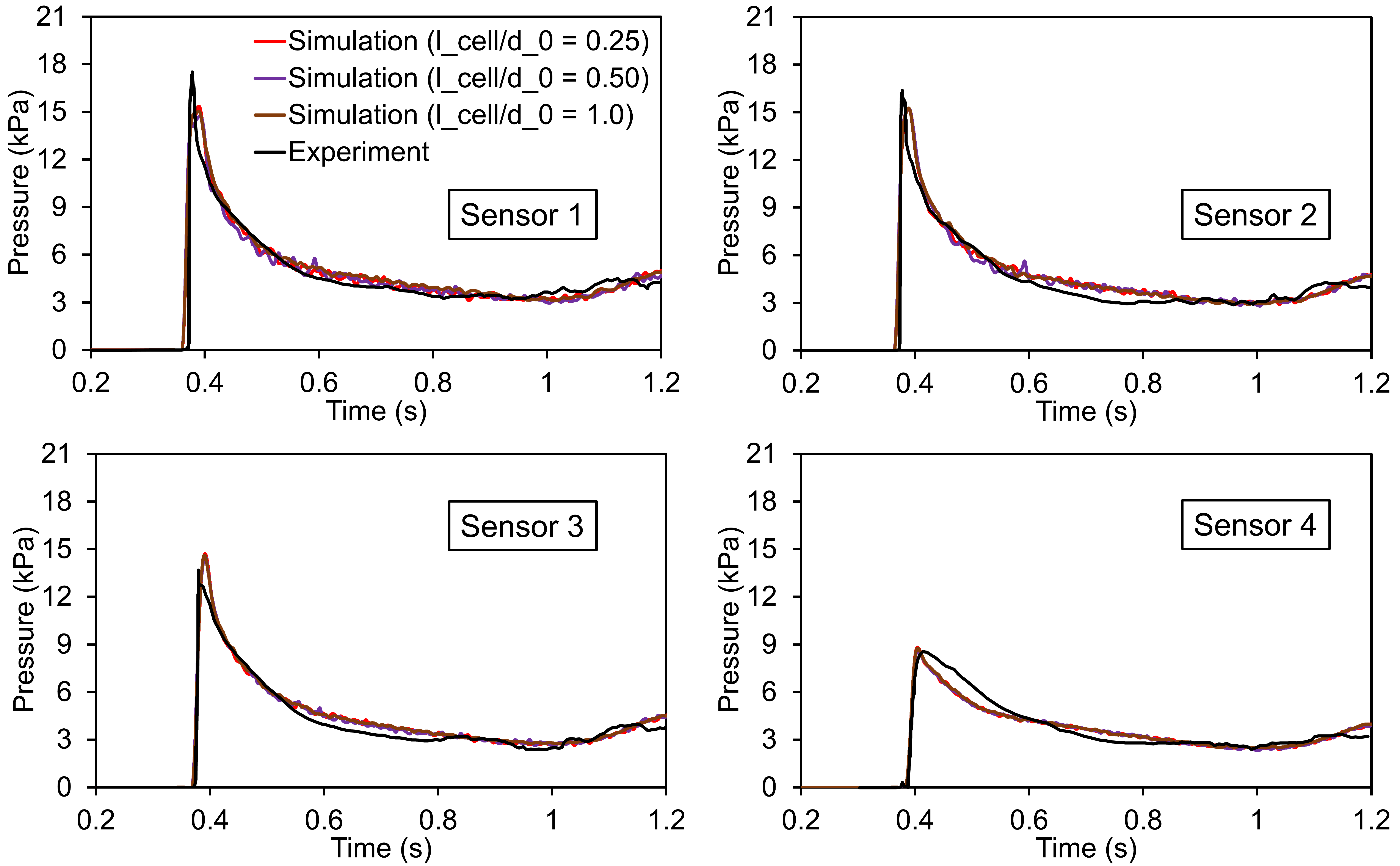} 
	\caption{Dam-break problem: time histories of pressure obtained using the difference-based pressure gradient model.}
	\label{fig:dam_result_pressure_dif}
\end{figure}

{
\fig{fig:dam_result_pressure_dif} shows the pressure histories obtained from the dam-break simulation using the difference-based pressure gradient model.
All sensor positions and computational conditions are identical to those in Section 5.2, except that three cell lengths are considered: $l_\mathrm{cell} = 0.25d_0$, $0.5d_0$, and $1.0d_0$.
The overestimation of the peak pressure upon flow impact observed in \fig{fig:dam_result_pressure} is resolved. Furthermore, it is confirmed that stable computations are achieved regardless of the cell resolution.
}

%% file: cas-refs.bib
@article{Gingold1977,
  author  = {Gingold, A. R. and Monaghan, J. J.},
  title   = {Smoothed particle hydrodynamics: theory and application to non-spherical stars},
  journal = {Monthly Notices of the Royal Astronomical Society},
  volume  = {181},
  number  = {3},
  pages   = {375--389},
  year    = {1977}
}

@article{Lucy1977,
  author  = {Lucy, L. B.},
  title   = {A numerical approach to the testing of the fission hypothesis},
  journal = {The Astronomical Journal},
  volume  = {82},
  pages   = {1013--1024},
  year    = {1977}
}

@article{Koshizuka1996,
  author  = {Koshizuka, S. and Oka, Y.},
  title   = {Moving-particle semi-implicit method for fragmentation of incompressible fluid},
  journal = {Nuclear Science and Engineering},
  volume  = {123},
  number  = {3},
  pages   = {421--434},
  year    = {1996}
}

@article{Koshizuka1998,
  author  = {Koshizuka, S. and Nobe, A. and Oka, Y.},
  title   = {Numerical analysis of breaking waves using the moving particle semi-implicit method},
  journal = {International Journal for Numerical Methods in Fluids},
  volume  = {26},
  number  = {7},
  pages   = {751--769},
  year    = {1998}
}

@article{Monaghan1994,
  author  = {Monaghan, J. J.},
  title   = {Simulating free surface flows with {SPH}},
  journal = {Journal of Computational Physics},
  volume  = {110},
  number  = {2},
  pages   = {399--406},
  year    = {1994}
}

@article{Monaghan2009,
  author  = {Monaghan, J. J. and Kajtar, J. B.},
  title   = {{SPH} particle boundary forces for arbitrary boundaries},
  journal = {Computer Physics Communications},
  volume  = {180},
  number  = {10},
  pages   = {1811--1820},
  year    = {2009}
}

@article{Colagrossi2003,
  author  = {Colagrossi, A. and Landrini, M.},
  title   = {Numerical simulation of interfacial flows by smoothed particle hydrodynamics},
  journal = {Journal of Computational Physics},
  volume  = {191},
  number  = {2},
  pages   = {448--475},
  year    = {2003}
}

@article{Morris1997,
  author  = {Morris, J. P. and Fox, P. J. and Zhu, Y.},
  title   = {Modeling low {R}eynolds number incompressible flows using {SPH}},
  journal = {Journal of Computational Physics},
  volume  = {136},
  number  = {1},
  pages   = {214--226},
  year    = {1997}
}

@article{mitsume2015,
  author  = {Mitsume, N. and Yoshimura, S. and Murotani, K. and Yamada, T.},
  title   = {Explicitly represented polygon wall boundary model for the {E}xplicit {MPS} method},
  journal = {Computational Particle Mechanics},
  volume  = {2},
  number  = {1},
  pages   = {73--89},
  year    = {2015}
}

@article{zheng2020-1,
  author  = {Zheng, Z. and Duan, G. and Mitsume, N. and Chen, S. and Yoshimura, S.},
  title   = {A novel ghost cell boundary model for the explicit moving particle simulation method in two dimensions},
  journal = {Computational Mechanics},
  volume  = {66},
  pages   = {87--102},
  year    = {2020}
}

@article{zheng2020-2,
  author  = {Zheng, Z. and Duan, G. and Mitsume, N. and Chen, S. and Yoshimura, S.},
  title   = {An explicit {MPS/FEM} coupling algorithm for three-dimensional fluid--structure interaction analysis},
  journal = {Engineering Analysis with Boundary Elements},
  volume  = {121},
  pages   = {192--206},
  year    = {2020}
}

@article{Bouscasse2013,
  author  = {Bouscasse, B. and Colagrossi, A. and Marrone, S. and Antuono, M.},
  title   = {Nonlinear water wave interaction with floating bodies in {SPH}},
  journal = {Journal of Fluids and Structures},
  volume  = {42},
  pages   = {112--129},
  year    = {2013}
}

@article{Morikawa2021,
  author  = {Morikawa, D. and Senadheera, H. and Asai, M.},
  title   = {Explicit incompressible smoothed particle hydrodynamics in a multi-{GPU} environment for large-scale simulations},
  journal = {Computational Particle Mechanics},
  volume  = {8},
  number  = {3},
  pages   = {493--510},
  year    = {2021}
}

@article{Farhat2000,
  author  = {Farhat, C. and Lesoinne, M.},
  title   = {Two efficient staggered algorithms for the serial and parallel solution of three-dimensional nonlinear transient aeroelastic problems},
  journal = {Computer Methods in Applied Mechanics and Engineering},
  volume  = {182},
  number  = {3--4},
  pages   = {499--515},
  year    = {2000}
}

@article{Asai2012,
  author  = {Asai, M. and Aly, A. M. and Sonoda, Y. and Sakai, Y.},
  title   = {A stabilized incompressible {SPH} method by relaxing the density invariance condition},
  journal = {Journal of Applied Mathematics},
  volume  = {2012},
  pages   = {139583},
  year    = {2012}
}

@article{marrone2010,
  author  = {Marrone, S. and Colagrossi, A. and Le Touz\'{e}, D. and Graziani, G.},
  title   = {Fast free-surface detection and level-set function definition in {SPH} solvers},
  journal = {Journal of Computational Physics},
  volume  = {229},
  number  = {10},
  pages   = {3652--3663},
  year    = {2010}
}

@article{Khayyer2017,
  author  = {Khayyer, A. and Gotoh, H. and Shimizu, Y.},
  title   = {Comparative study on accuracy and conservation properties of two particle regularization schemes and proposal of an optimized particle shifting scheme in {ISPH} context},
  journal = {Journal of Computational Physics},
  volume  = {332},
  pages   = {236--256},
  year    = {2017}
}

@article{tsuruta2013,
  author  = {Tsuruta, N. and Khayyer, A. and Gotoh, H.},
  title   = {A short note on dynamic stabilization of moving particle semi-implicit method},
  journal = {Computers \& Fluids},
  volume  = {82},
  pages   = {158--164},
  year    = {2013}
}

@article{marrone2011,
  author  = {Marrone, S. and Antuono, M. and Colagrossi, A. and Colicchio, G. and Le Touz\'{e}, D. and Graziani, G.},
  title   = {$\delta$-{SPH} model for simulating violent impact flows},
  journal = {Computer Methods in Applied Mechanics and Engineering},
  volume  = {200},
  pages   = {1526--1542},
  year    = {2011}
}

@article{geuzaine2009,
  author  = {Geuzaine, C. and Remacle, J. F.},
  title   = {Gmsh: A 3-{D} finite element mesh generator with built-in pre- and post-processing facilities},
  journal = {International Journal for Numerical Methods in Engineering},
  volume  = {79},
  number  = {11},
  pages   = {1309--1331},
  year    = {2009}
}

@article{Lobovsky2014,
  author  = {Lobovsk\'{y}, L. and Botia-Vera, E. and Castellana, F. and Mas-Soler, J. and Souto-Iglesias, A.},
  title   = {Experimental investigation of dynamic pressure loads during dam break},
  journal = {Journal of Fluids and Structures},
  volume  = {48},
  pages   = {407--434},
  year    = {2014}
}

@article{Kishev2006,
  author  = {Kishev, Z. R. and Hu, C. and Kashiwagi, M.},
  title   = {Numerical simulation of violent sloshing by a {CIP}-based method},
  journal = {Journal of Marine Science and Technology},
  volume  = {11},
  number  = {2},
  pages   = {111--122},
  year    = {2006}
}

@article{Liu2010,
  author  = {Liu, M. B. and Liu, G. R.},
  title   = {Smoothed {P}article {H}ydrodynamics ({SPH}): an overview and recent developments},
  journal = {Archives of Computational Methods in Engineering},
  volume  = {17},
  pages   = {25--76},
  year    = {2010}
}

@article{Vacondio2021,
  author  = {Vacondio, R. and Altomare, C. and De Leffe, M. and Hu, X. and Le Touz\'{e}, D. and Lind, S. and Marongiu, J. C. and Marrone, S. and Rogers, B. D. and Souto-Iglesias, A.},
  title   = {Grand challenges for smoothed particle hydrodynamics numerical schemes},
  journal = {Computational Particle Mechanics},
  volume  = {8},
  pages   = {575--588},
  year    = {2021}
}

@article{Cummins1999,
  author  = {Cummins, S. J. and Rudman, M.},
  title   = {An {SPH} projection method},
  journal = {Journal of Computational Physics},
  volume  = {152},
  number  = {2},
  pages   = {584--607},
  year    = {1999}
}

@article{Shao2003,
  author  = {Shao, S. and Lo, E. Y. M.},
  title   = {Incompressible {SPH} method for simulating {N}ewtonian and non-{N}ewtonian flows with a free surface},
  journal = {Advances in Water Resources},
  volume  = {26},
  number  = {7},
  pages   = {787--800},
  year    = {2003}
}

@article{Adami2012,
  author  = {Adami, S. and Hu, X. Y. and Adams, N. A.},
  title   = {A generalized wall boundary condition for smoothed particle hydrodynamics},
  journal = {Journal of Computational Physics},
  volume  = {231},
  number  = {21},
  pages   = {7057--7075},
  year    = {2012}
}

@article{Violeau2016,
  author  = {Violeau, D. and Rogers, B. D.},
  title   = {Smoothed {P}article {H}ydrodynamics ({SPH}) for free-surface flows: past, present and future},
  journal = {Journal of Hydraulic Research},
  volume  = {54},
  number  = {1},
  pages   = {1--26},
  year    = {2016}
}

@article{Bierbrauer2009,
  author  = {Bierbrauer, F. and Bollada, P. C. and Phillips, T. N.},
  title   = {A consistent reflected image particle approach to the treatment of boundary conditions in smoothed particle hydrodynamics},
  journal = {Computer Methods in Applied Mechanics and Engineering},
  volume  = {198},
  number  = {41--44},
  pages   = {3400--3410},
  year    = {2009}
}

@article{Antuono2023,
  author  = {Antuono, M. and Pilloton, C. and Colagrossi, A. and Durante, D.},
  title   = {Clone particles: A simplified technique to enforce solid boundary conditions in {SPH}},
  journal = {Computer Methods in Applied Mechanics and Engineering},
  volume  = {409},
  pages   = {115973},
  year    = {2023}
}

@article{Rezavand2022,
  author  = {Rezavand, M. and Zhang, C. and Hu, X.},
  title   = {Generalized and efficient wall boundary condition treatment in {GPU}-accelerated smoothed particle hydrodynamics},
  journal = {Computer Physics Communications},
  volume  = {281},
  pages   = {108507},
  year    = {2022}
}

@article{Harlow1965,
  author  = {Harlow, F. H. and Welch, J. E.},
  title   = {Numerical calculation of time-dependent viscous incompressible flow of fluid with free surface},
  journal = {Physics of Fluids},
  volume  = {8},
  number  = {12},
  pages   = {2182--2189},
  year    = {1965}
}

@article{Fedkiw1999,
  author  = {Fedkiw, R. P. and Aslam, T. and Merriman, B. and Osher, S.},
  title   = {A non-oscillatory {E}ulerian approach to interfaces in multimaterial flows (the ghost fluid method)},
  journal = {Journal of Computational Physics},
  volume  = {152},
  number  = {2},
  pages   = {457--492},
  year    = {1999}
}

@article{Johansen1998,
  author  = {Johansen, H. and Colella, P.},
  title   = {A {C}artesian grid embedded boundary method for {P}oisson's equation on irregular domains},
  journal = {Journal of Computational Physics},
  volume  = {147},
  number  = {1},
  pages   = {60--85},
  year    = {1998}
}

@article{Gibou2002,
  author  = {Gibou, F. and Fedkiw, R. P. and Cheng, L. T. and Kang, M.},
  title   = {A second-order-accurate symmetric discretization of the {P}oisson equation on irregular domains},
  journal = {Journal of Computational Physics},
  volume  = {176},
  number  = {1},
  pages   = {205--227},
  year    = {2002}
}

@article{Ye1999,
  author  = {Ye, T. and Mittal, R. and Udaykumar, H. S. and Shyy, W.},
  title   = {An accurate {C}artesian grid method for viscous incompressible flows with complex immersed boundaries},
  journal = {Journal of Computational Physics},
  volume  = {156},
  number  = {2},
  pages   = {209--240},
  year    = {1999}
}

@article{Udaykumar2001,
  author  = {Udaykumar, H. S. and Mittal, R. and Rampunggoon, P. and Khanna, A.},
  title   = {A sharp interface {C}artesian grid method for simulating flows with complex moving boundaries},
  journal = {Journal of Computational Physics},
  volume  = {174},
  number  = {1},
  pages   = {345--380},
  year    = {2001}
}

@article{Tseng2003,
  author  = {Tseng, Y. H. and Ferziger, J. H.},
  title   = {A ghost-cell immersed boundary method for flow in complex geometry},
  journal = {Journal of Computational Physics},
  volume  = {192},
  number  = {2},
  pages   = {593--623},
  year    = {2003}
}

@article{Mittal2005,
  author  = {Mittal, R. and Iaccarino, G.},
  title   = {Immersed boundary methods},
  journal = {Annual Review of Fluid Mechanics},
  volume  = {37},
  pages   = {239--261},
  year    = {2005}
}

@article{Ghias2007,
  author  = {Ghias, R. and Mittal, R. and Dong, H.},
  title   = {A sharp interface immersed boundary method for compressible viscous flows},
  journal = {Journal of Computational Physics},
  volume  = {225},
  number  = {1},
  pages   = {528--553},
  year    = {2007}
}

@article{Berthelsen2008,
  author  = {Berthelsen, P. A. and Faltinsen, O. M.},
  title   = {A local directional ghost cell approach for incompressible viscous flow problems with irregular boundaries},
  journal = {Journal of Computational Physics},
  volume  = {227},
  number  = {9},
  pages   = {4354--4397},
  year    = {2008}
}

@article{Crespo2007,
  author  = {Crespo, A. J. C. and G\'{o}mez-Gesteira, M. and Dalrymple, R. A.},
  title   = {Boundary conditions generated by dynamic particles in {SPH} methods},
  journal = {Computers, Materials \& Continua},
  volume  = {5},
  number  = {3},
  pages   = {173--184},
  year    = {2007}
}

@article{Vela2019,
  author  = {Vela, L. and Reynolds-Barredo, J. M. and S\'{a}nchez, R.},
  title   = {A positioning algorithm for {SPH} ghost particles in smoothly curved geometries},
  journal = {Journal of Computational and Applied Mathematics},
  volume  = {353},
  pages   = {140--153},
  year    = {2019}
}

@article{Fourtakas2019,
  author  = {Fourtakas, G. and Dom\'{i}nguez, J. M. and Vacondio, R. and Rogers, B. D.},
  title   = {Local uniform stencil ({LUST}) boundary condition for arbitrary {3-D} boundaries in parallel smoothed particle hydrodynamics ({SPH}) models},
  journal = {Computers \& Fluids},
  volume  = {190},
  pages   = {346--361},
  year    = {2019}
}

@article{English2022,
  author  = {English, A. and Dom\'{i}nguez, J. M. and Vacondio, R. and Crespo, A. J. C. and Stansby, P. K. and Lind, S. J. and Chiapponi, L. and G\'{o}mez-Gesteira, M.},
  title   = {Modified dynamic boundary conditions ({mDBC}) for general-purpose smoothed particle hydrodynamics ({SPH}): application to tank sloshing, dam break and fish pass problems},
  journal = {Computational Particle Mechanics},
  volume  = {9},
  pages   = {1--15},
  year    = {2022}
}

@article{Barjasteh2016,
  author  = {Barjasteh, M. and Zeraatgar, H. and Javaherian, M. J.},
  title   = {An experimental study on water entry of asymmetric wedges},
  journal = {Applied Ocean Research},
  volume  = {58},
  pages   = {292--304},
  year    = {2016}
}

@article{Xie2026,
  title     = {A two-way partitioned {MPS-CZM} coupling method for dynamic fluid-structure-crack interaction simulations},
  volume    = {347},
  journal   = {Ocean Engineering},
  publisher = {Elsevier BV},
  author    = {Xie, Si and Wang, Di and Zheng, Zumei and Fan, Ningge and Zeng, Haiyang and Qin, Yifang and Mitsume, Naoto and Chen, Shunhua},
  year      = {2026},
  pages     = {123902}
}

@article{Shakibaeinia2010,
  title     = {A weakly compressible {MPS} method for modeling of open-boundary free-surface flow},
  volume    = {63},
  number    = {10},
  journal   = {International Journal for Numerical Methods in Fluids},
  publisher = {Wiley},
  author    = {Shakibaeinia, Ahmad and Jin, Yee-Chung},
  year      = {2010},
  month     = aug,
  pages     = {1208--1232}
}

@article{Khayyer2018,
  title     = {An enhanced {ISPH--SPH} coupled method for simulation of incompressible fluid--elastic structure interactions},
  volume    = {232},
  journal   = {Computer Physics Communications},
  publisher = {Elsevier BV},
  author    = {Khayyer, Abbas and Gotoh, Hitoshi and Falahaty, Hosein and Shimizu, Yuma},
  year      = {2018},
  month     = nov,
  pages     = {139--164}
}

@article{Randles1996,
  title     = {Smoothed {P}article {H}ydrodynamics: {S}ome recent improvements and applications},
  volume    = {139},
  number    = {1--4},
  journal   = {Computer Methods in Applied Mechanics and Engineering},
  publisher = {Elsevier BV},
  author    = {Randles, P. W. and Libersky, L. D.},
  year      = {1996},
  month     = dec,
  pages     = {375--408}
}

@article{Fang2006,
  title     = {A numerical study of the {SPH} method for simulating transient viscoelastic free surface flows},
  volume    = {139},
  number    = {1--2},
  journal   = {Journal of Non-Newtonian Fluid Mechanics},
  publisher = {Elsevier BV},
  author    = {Fang, Jiannong and Owens, Robert G. and Tacher, Laurent and Parriaux, Aur\`{e}le},
  year      = {2006},
  month     = nov,
  pages     = {68--84}
}

@article{Bonet1999,
  title   = {Variational and momentum preservation aspects of Smooth Particle Hydrodynamic formulations},
  volume  = {180},
  number  = {1},
  journal = {Computer Methods in Applied Mechanics and Engineering},
  author  = {Bonet, J. and Lok, T-S L.},
  year    = {1999},
  month   = nov,
  pages   = {97--115}
}

@article{Khan2025,
  title     = {A motif based wall boundary model for smoothed particle hydrodynamics},
  volume    = {37},
  number    = {5},
  journal   = {Physics of Fluids},
  publisher = {AIP Publishing},
  author    = {Khan, Piru Mohan and Nayak, Subrat Kumar and Nair, Prapanch},
  year      = {2025},
  pages     = {053317}
}

@article{Gotoh2026a,
  title     = {Wall Boundary Potential model for incompressible smoothed particle hydrodynamics with arbitrary kernel functions},
  volume    = {188},
  ISSN      = {0955-7997},
  journal   = {Engineering Analysis with Boundary Elements},
  publisher = {Elsevier BV},
  author    = {Gotoh, Takafumi and Tsuruta, Naoki and Khayyer, Abbas and Liang, Dongfang and Gotoh, Hitoshi},
  year      = {2026},
  pages     = {106749}
}

@article{Tsuruta2015,
  title     = {Space potential particles to enhance the stability of projection-based particle methods},
  volume    = {29},
  ISSN      = {1061-8562},
  number    = {1},
  journal   = {International Journal of Computational Fluid Dynamics},
  publisher = {Informa UK Limited},
  author    = {Tsuruta, Naoki and Khayyer, Abbas and Gotoh, Hitoshi},
  year      = {2015},
  month     = jan,
  pages     = {100--119}
}

@article{Ruan2024,
  title     = {An improved incompressible {SPH} with unified semi-analytical wall boundary conditions for free surface flow},
  volume    = {235},
  ISSN      = {0001-5970},
  number    = {12},
  journal   = {Acta Mechanica},
  publisher = {Springer Science and Business Media LLC},
  author    = {Ruan, Jianwu and Wang, Lihua and Zhou, Yueting},
  year      = {2024},
  month     = dec,
  pages     = {7351--7368},
  language  = {en}
}

@article{Sun2015,
  title   = {Numerical simulation of interactions between free surface and rigid body using a robust {SPH} method},
  volume  = {98},
  journal = {Ocean Engineering},
  author  = {Sun, P. and Ming, F. and Zhang, A.},
  year    = {2015},
  pages   = {32--49}
}

@article{Negishi2024,
  title     = {Fluid--rigid body coupling simulations with the passively moving solid model based on a physically consistent particle method},
  volume    = {36},
  ISSN      = {1070-6631},
  number    = {3},
  journal   = {Physics of Fluids},
  publisher = {AIP Publishing},
  author    = {Negishi, Hideyo and Kondo, Masahiro and Takahashi, Hidenao and Amakawa, Hiroaki and Obara, Shingo and Kurose, Ryoichi},
  year      = {2024},
  pages     = {033327}
}

@article{Zhang2024,
  title     = {Numerical investigations on water entry and/or exit problems using a multi-resolution {Delta-plus-SPH model with TIC}},
  volume    = {292},
  journal   = {Ocean Engineering},
  publisher = {Elsevier BV},
  author    = {Zhang, Guiyong and Liang, Guangqi and Yang, Xi and Zhang, Zhifan},
  year      = {2024},
  month     = jan,
  pages     = {116560},
  language  = {en}
}

@article{Sun2019,
  title   = {Study of a complex fluid-structure dam-breaking benchmark problem using a multi-phase {SPH} method with {APR}},
  volume  = {104},
  journal = {Engineering Analysis with Boundary Elements},
  author  = {Sun, P. N. and Le Touz\'{e}, D. and Zhang, A. M.},
  year    = {2019},
  pages   = {240--258}
}

@article{Gotoh2026b,
  title     = {An enhanced coupled {ISPH-TLSPH FSI} solver with a {Riemann} fluid-structure acceleration scheme},
  volume    = {16},
  ISSN      = {2196-4378},
  journal   = {Computational Particle Mechanics},
  publisher = {Elsevier BV},
  author    = {Gotoh, Takafumi and Khayyer, Abbas and Liang, Dongfang and Gotoh, Hitoshi},
  year      = {2026},
  month     = aug,
  pages     = {286--304},
  language  = {en}
}

@article{Chen2026,
  title     = {{GPU}-accelerated multi-material {FVM--SPH} with immersed boundary method for high-fidelity modeling of shock--structure interactions},
  volume    = {77},
  ISSN      = {0178-7675},
  number    = {5},
  journal   = {Computational Mechanics},
  publisher = {Springer Science and Business Media LLC},
  author    = {Chen, Jian-Yu and Zhao, Jidong and Feng, Ruo-Feng and Ni, Rui-Chen and Peng, Chong},
  year      = {2026},
  month     = may,
  pages     = {1267--1289},
  language  = {en}
}
